# An Illustrated Introduction to the Basic Biological Principles

Simon Fu, PhD

*Department of Biochemistry and Molecular Biology, Keck School of Medicine, University of Southern California, Los Angeles, CA 90033, USA*

Correspondence: fyla686@hotmail.com

## Abstract

Both external environmental selection and internal lower-level evolution are essential for an integral picture of evolution. This paper proposes that the division of internal evolution into DNA/RNA pattern formation (genotype) and protein functional action (phenotype) resolves a universal conflict between fitness and evolvability. Specifically, this paper explains how this universal conflict drove the emergence of genotype-phenotype division, why this labor division is responsible for the extraordinary complexity of life, and how the specific ways of genotype-phenotype mapping in the labor division determine the paths and forms of evolution and development.

# Table of Contents

## List of Figures

## I. A Rough Sketch

Evolution is the sampling of configuration space* by physical entities under environmental constraint. Evolution as a whole is not biased to fitness or complexity increase. However, why and how some specific types of evolutionary entities have become more complex than others is the focus of evolutionary biology. As an evolutionary entity with extraordinary complexity and fitness, terrestrial life occupies only extremely small regions in the enormous configuration space[1]. The major theme of evolution is how blind evolution finds its way to these small regions under environmental constraint. Although the blindness of evolution has been widely accepted as a rebuttal to creationism and intelligent design[4-9], the profound influence of blindness on evolution is not fully appreciated. ***The key to understand the role of blindness is to recognize that because evolution is blind and purposeless, incomplete sampling of the possible configurations of the next step is unfavorable to the increase of fitness and complexity***. Due to its blindness and purposelessness, evolution does not "know" in advance which path will lead to the increase of fitness or complexity or how fluctuating environment will change. Therefore, retrospectively, the best "strategy" to increase complexity and fitness is to take every possible path at every next step. As a result, no complex or fit configurations will be missed. In contrast, if the evolutionary entity only takes a part of paths at the current location, only configurations downstream of these paths can be reached: all other configurations will be missed. From the angle of blind evolution rather than intelligent humans, the greater the incompleteness in configuration sampling, the smaller the probability for blind evolution to increase complexity or fitness is. Even when intelligent humans (re)design enzymes, blind random mutagenesis followed by screening is more successful than "rational" approaches using computational predictions based on chemical principles[10]. There is no better "strategy" for blind evolution, especially during the origin and early evolution of life. It must be emphasized there is no purposeful pursue for the "strategy": some branches blindly acquired the mechanisms of such a "strategy" and consequently had high evolvability and complexity.

Incomplete sampling can be due to the insufficient number of individuals (parallel sampler) for sampling all possible configurations at every step. In this case, giving sufficient time, evolution can still locate extremely small regions of the vast configuration space of life (repeated sampling). In other words, if the missed configurations are distributed randomly, namely unbiased, the incompleteness of sampling can be remedied by parallel or repeated sampling. However, if the incompleteness is biased to certain configurations, parallel or repeated sampling cannot remedy it. ***The bias in configuration sampling is harmful for blind evolution to increase fitness and complexity, because the bias prevents complete configuration sampling.*** The conventional concept of diversity is actually the degree of the unbiasedness in configuration sampling. The blindness of evolution answers why diversity, a property without involving net fitness gain, is beneficial to life. ***The degree of the unbiasedness in configuration space sampling is one of the two essentials of evolvability;*** the other one is the resolution and efficiency of natural selection (chapter V).

---

* Configuration is a general term introduced from physics, which represents all interrelationships of constituent elements of an entity. Configuration space is the collection of all possible configurations of an entity, either abiotic or biotic. In terrestrial life with genotype-phenotype division, the genotype space is the sequence space of genetic material, which is a subset of the configuration space of genetic material. Gene frequency is only a very small subset of the genotype space of a population. The phenotype is determined by the configuration of the functional components and the environment. If the environment remains constant, the phenotype space can be considered as a subset of the configuration space of functional components. Both taxonomic space[1] and morphospace[2,3] are a subset of configuration space.

It has already been noticed that evolutionary biases, such as mutational robustness and thermostability, constrain evolution, as the ridges on the landscape blocks evolutionary exploration[11-16]. However, those findings are not linked to the blindness of evolution to explicitly conclude that bias is harmful to the increase of the fitness and complexity of evolution. As a result, the understanding of evolutionary bias is unclear. In the conventional view, a process of evolution is not clearly distinguished from the associated bias. For example, mutational bias is often confused with mutation *per se* and hence is consider introducing novelty for evolution[17]. Mutation generates novelty/diversity, but mutational bias only makes the available paths less than the possible paths and thus reduces the novelty/diversity.

It is understandable that completely unbiased sampling is only an unattainable ideal in real condition. Can evolutionary sampling approach unbiasedness close enough that sampling bias is reduced to a harmless level? The answer is no, because there is an intrinsic and universal conflict between the unbiased sampling and the functional activity required by terrestrial life.

The existing configurations of macromolecules and their organization are not random samples of the configuration space. Sampling configuration space by an evolutionary entity is performed by its internal physicochemical processes under environmental constraint, namely its internal pattern formation* under environmental constraint. Sampling must have the bias of underlying processes. All physicochemical processes of evolution are biased more or less, because there is an intrinsic and universal conflict between the unbiased pattern formation and vigorous activity. An active entity or system must be strongly biased to one direction in its reaction and evolution, while a weakly biased entity must be inert

* The abstract relationship of a configuration is the pattern in a general sense. Specifically, the relationship which is extracted from the configuration of an entity and segregated from its original physical embodiment is the pattern. For example, the distribution of soluble molecules in a reaction-diffusion system can specify the spatial arrangement of an organ during embryonic development[18]. The concept of pattern is especially important when a configuration does not have important function but provides relational templates for other types of evolution. Such examples of pattern are the sequence of DNA/RNA, intergenic relations, and the gene pool of a population.

in its activity[19-20]. Activity per se is bias. The vigorous activity results in the bias to a subset of configurations in evolution; evolution of active entities is constrained in the local optimums on the rugged landscape[11-12, 15-16] (Fig. 1). The stronger the activity, the greater the bias and the constraint are. However, on the other hand, vigorous activity is required to realize the function and fitness of life.

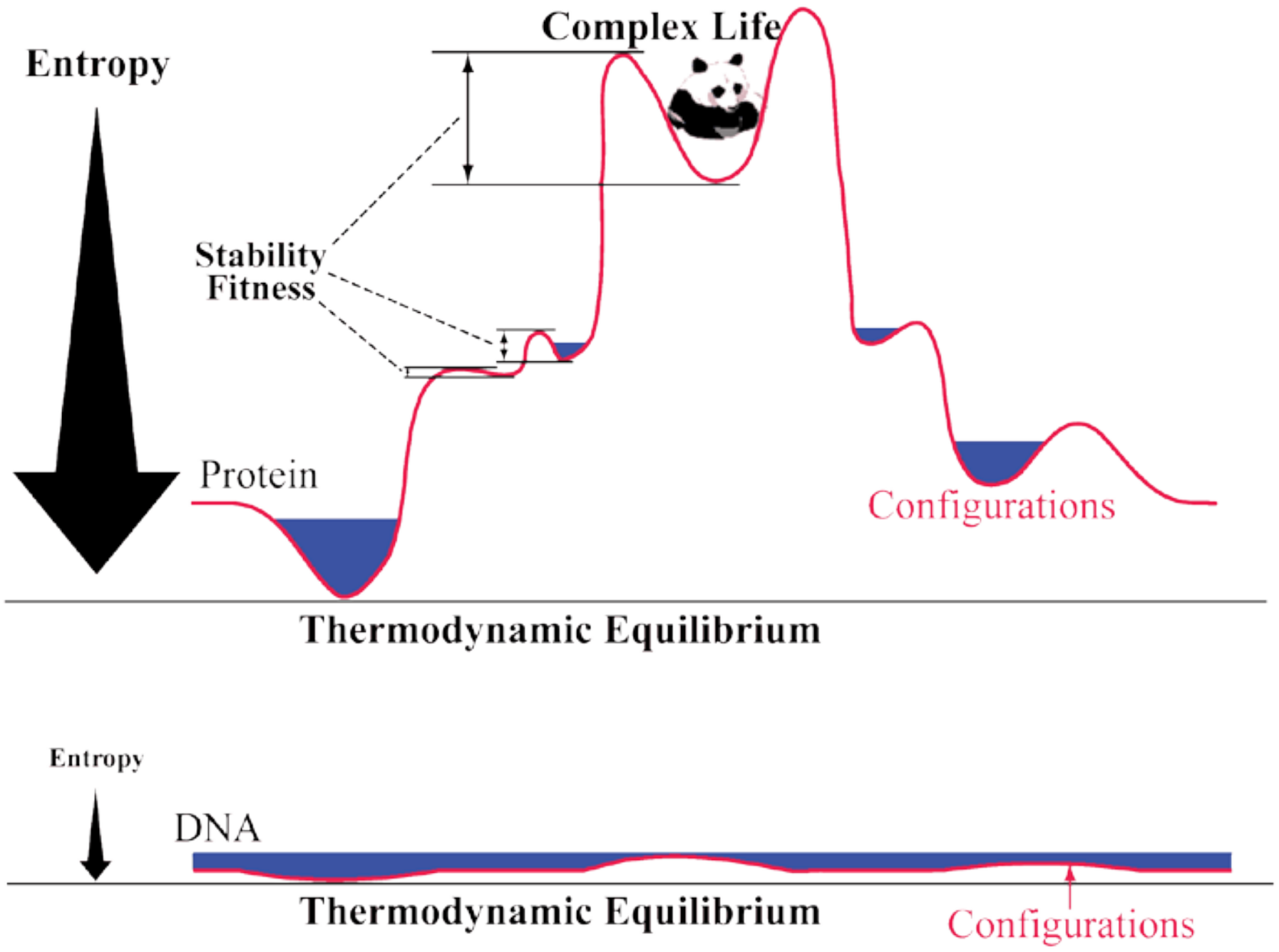

**Fig. 1. The energy landscape of evolution and the universal polarity of evolution.** The red line represents the configurations of the evolutionary entity. The ground state of the landscape is the thermodynamic equilibrium, and the altitude is the deviation from equilibrium. According to the second law of thermodynamics, the general trend of evolution is downward on the energy landscape, unless external energy is effectively utilized. Stability/robustness/fitness is the accumulative height of all ridges surrounding a local minimum (valley). A complex state must deviate from equilibrium, although complexity is not proportional to the deviation. A rugged landscape, such as that of protein, has peaks (local maximum) and valleys (local minimum), but the evolution is constrained in the valleys of low altitude (in blue). The evolution on a smooth landscape, such as that of DNA, is not constrained but has neither deep valleys (stability/fitness) nor evolutionary altitude (complexity). This conflict reflects a universal polarity of evolution: active entities have potential complexity but their pattern formation is biased to local minimums, and configuration sampling is thus incomplete and biased. In contrast, functionally inert entities have less biased pattern formation and thus more complete sampling of configurations, but lack functional and complex states that highly deviate from equilibrium. The real energy landscape and configuration space are much more complicated than this illustration.

As stable complexity, life requires biases, namely valleys, ridges, and various barriers on the landscape of functional domain, to sustain its configurations to achieve stability/robustness. Inert entities with a relative smooth landscape do not have vigorous motion on their landscape to strongly bias to certain configurations, let alone they cannot highly deviate from equilibrium. In conventional words, inert components cannot constitute an evolutionary entity with sufficient functional activity to support the extraordinary complexity and high fitness of life. As the stability/robustness of a lineage of biotic entities, fitness must be realized through various concrete functional activities. As the physical basis of phenotype, those functional activities are actually the embodiment of the biases and constraints on the rugged landscape which sustain the configurations of life.

At molecular level, on the one hand, life requires active components to realize its extraordinary complexity and high fitness. That is why active proteins, rather than the relatively inert DNA, are the major structural block and functional performer of terrestrial life. On the other hand, life uses relatively inert DNA as the principle pattern generator which guides the evolution of life.

At organismal level, on the one hand, a local minimum is actually a configuration or configurations which all surrounding configurations are biased to; the local minimums provide a stable existence for life (fitness); on the other hand, the stable existence at a local minimum (fitness) is unfavorable to further exploration of other regions of the configuration space.

***The opposition at both levels is the manifestation of a universal polarity of evolution: unbiased sampling of configuration space and vigorous functional activity are opposite to each other*** (chapter II). This conflict is intrinsic and universal so the author names it a 'polarity'. In conventional terms, there is an intrinsic and universal conflict between diversity generation and vigorous activity. Constraining the increase of complexity and stability/robustness/fitness, this intrinsic polarity applies to all evolutionary entities, for example atoms, molecules, and cells. To terrestrial life, active proteins, as a functional performer, are advantaged in functional action but disadvantaged in configuration sampling: the configurations of protein are biased to the lowlands (stable configurations) of its rugged energy landscape, as a result of its diverse and vigorous activity (Fig. 1). In contrast, inert entities such as DNA have a flat landscape and generate configurations that are less biased than proteins, but lack vigorous activity and useful function except for pattern formation (Fig. 1). Compared to DNA, unstable protein biased to degraded configurations is only one of important biases. Another is the sequence-dependent stability variations, which bias polymers to stable sequences. The stability of an evolutionary entity is not only determined by its thermodynamic property, although thermostability is a principal factor in the transition from nonlife to life. Instead, the stability of an entity is an all-inclusive stability attributed to its all interactions under a specific environment. Therefore, the stability of an entity can vary in different environments. For example, the GC stability in genomic evolution varies significant in different species[21]. All biases and constraints are caused by the differential stability of configurations, which, on the other hand, drives the running of reaction and evolution.

The universal polarity of evolution is actually the conflict between evolvability and fitness in the conventional view of evolution[22-23]. Evolvability is the degree of unbiasedness in the sampling of configuration space. As explained above, functional activities, as the physical basis of phenotype, are the embodiment of the biases and constraints which sustain the configurations of life to achieve stability/robustness. As the stability/robustness of a lineage of biotic entities, fitness provides an opportunity of sustained existence for some configurations of life, but constraints further exploration of other configurations. The major difference between the conventional view and the present theory is that this conflict is extended here from biotic evolution to abiotic evolution.

In abiotic evolution, functional action and configuration sampling/pattern formation are inherently bonded: there is no specialized and separated pattern generator or functional performer. Because both are required for complexity and fitness increase, the conflict caused by the universal polarity leads to the low complexity of abiotic evolution. Either configuration sampling through pattern formation is constrained by the strong bias of vigorous functional action, or the activity of the functional performer is inhibited by the unbiased pattern formation, or both. Therefore, the complexity of abiotic evolution is very low.

***The only solution for this universal polarity is a labor division of internal lower-level evolution to configuration sampling/pattern formation and functional action, which is actually the genotype-phenotype division*** (Fig. 2 & 4). Such a labor division is an essential characteristic of biotic evolution. In terrestrial life, sampling of configuration space is mainly guided by a relatively inert pattern generator - DNA (RNA in some viruses), whose patterns (including both genes and intergenic relations) are more diverse and less biased than protein patterns (evolvability). Meanwhile, active proteins provide the organism with various vigorous functional action to support the stable existence of its lineage (fitness). Through labor division, the evolutionary constraint imposed by the universal polarity is broken. ***Known as the genotype-phenotype division, the division of biotic evolution to pattern formation and functional action is responsible for the miraculous complexity of life and thus is the essence of life.*** The bifurcations of genetic vs. epigenetic heredity and Darwinian vs. Lamarckian evolution* are the manifestation of this division of labor. Moreover, this division of labor is actually a molecular cooperation/altruism (functional performers give up descendibility), which is a necessary basis for the altruism at organismal level (chapter VI).

Because of the separation of DNA/RNA pattern formation and protein functional action, namely the genotype-phenotype division, divided biotic evolution requires links to connect DNA/RNA pattern formation and protein functional action. Genotype-phenotype mapping emerges as a result. In one direction, translation, a biological heterodomain mapping (heteromapping)* between two distinct evolutionary domains, maps the pattern in the DNA/RNA domain to the protein functional domain with energy dissipation (Fig. 2) (chapter III). Heteromapping is the first step of genotype-phenotype mapping. Inert DNA provides less biased/constrained pattern formation than active proteins, and results in less biased/constrained sampling of configuration space. Moreover, because DNA and protein are different chemicals, their configurations near thermodynamic equilibrium do not correspond to each other in translation. In other words, the degeneration of DNA does not result in degenerated proteins in translation. That is why a biological lineage (continuance through germline information) is much more stable than a biological individual (continuance through somatic information and non-informational proteins). In this way, the thermodynamic bias in evolution is avoided (Fig. 2). The advantage of heteromapping is further enhanced by coarse graining/canalizing genetic code under the selective pressure for unbiased pattern formation (4th section of chapter III) (Fig. 5&6). Not only the sequence of proteins but also the relations between proteins, namely the spatial and temporal organization of proteins and other components, can be mapped from the source domain of heteromapping. As a result, configuration space sampling is not blocked or biased by the roughness of the landscape of the target functional domain. The heteromapping from DNA/RNA to protein is the most important step of the mapping from genotype to phenotype and also the only unique characteristic of life; the others steps are protein folding and organization/hierarchization (chapter V and VI), which are only more complex than their counterparts in abiotic evolution.

---

* Lamarckian evolution is a form of evolution by passing characteristics that the organism acquires during its lifetime to its offspring. Epigenetic heredity is a type of Lamarckian evolution. In contrast, Darwinian evolution is another form of evolution by natural selection of pre-existing inheritable variations. The environmental action in Lamarckian evolution is transformation, while that in Darwinian evolution is natural selection. In the following chapters, this article explains why Darwinian evolution, in stead of Lamarckian evolution, is the principal contributor to the complexity of life.

* Extraction of pattern from one type of evolutionary entity for different entities is named as heterodomain mapping, in contrast to the homodomain mapping in which pattern extraction and application occurs in the same type of entities. For example, translation is a biological form of heterodomain mapping while replication of DNA/RNA is homodomain mapping.

In the other direction of genotype-phenotype mapping, coupled selection of DNA/RNA to the host organism feeds back the fitness of protein function to the corresponding DNA/RNA pattern domain (Fig. 4, 11, and 14): DNA/RNA corresponding to low-fitness is eliminated together with its host by selection (chapter IV). Herein, coupled selection indicates that the two evolutionary entities always have the same fate in natural selection. Coupled selection is usually achieved by the dependence of one entity on the integrity of its host to survive or exist. For example, DNA/RNA and the translation system require the integrity of its host cell to sustain their existence and function. Only after the fitness of protein function is coupled to the corresponding DNA/RNA patterns during selection, can the patterns be tuned for the fitness of host organism. This self-evident and apparently trivial principle is very important for understanding evolution. In multilevel hierarchies, for example multicellular organisms, genetic information can couple to all hierarchical levels; the way of coupling influences the evolution and development of multicellular organisms (germline and the dichotomy of animals and plants, chapter VII). Heteromapping and coupled selection are actually the bidirectional mapping between genotype and phenotype, which is usually considered as unidirectional from genotype to phenotype in the conventional view of evolution.

The phenomena of heteromapping and coupled selection are obvious, but their role in evolution has not been fully understood because the universal polarity of evolution has not been appreciated. Herein, the author specially names, defines, and briefly describes these two mechanisms, and will explain their origin, evolution, and crucial importance in the origin and evolution of life. When combined with ***coarse graining/canalization***, heteromapping and coupled selection can explain many fundamental phenomena of life unitarily and parsimoniously. Some of them are not satisfactorily explained by the conventional theories. For example, the Eigen's paradox of the error threshold in the origin of replication[24], the advantage of sex, and the emergence of altruism. Others are not even questioned because of the limited angle of the conventional view. For example, why genetic information is linear despite that all other entities are 3-D? What is the consequence of that linearity? What is the reason behind the central dogma? What are the cause and the role of canalization? What is the role of germline in the dichotomy of animals and plants? The key of the answers is to understand heteromapping and coupled selection under various conditions in the history of life.

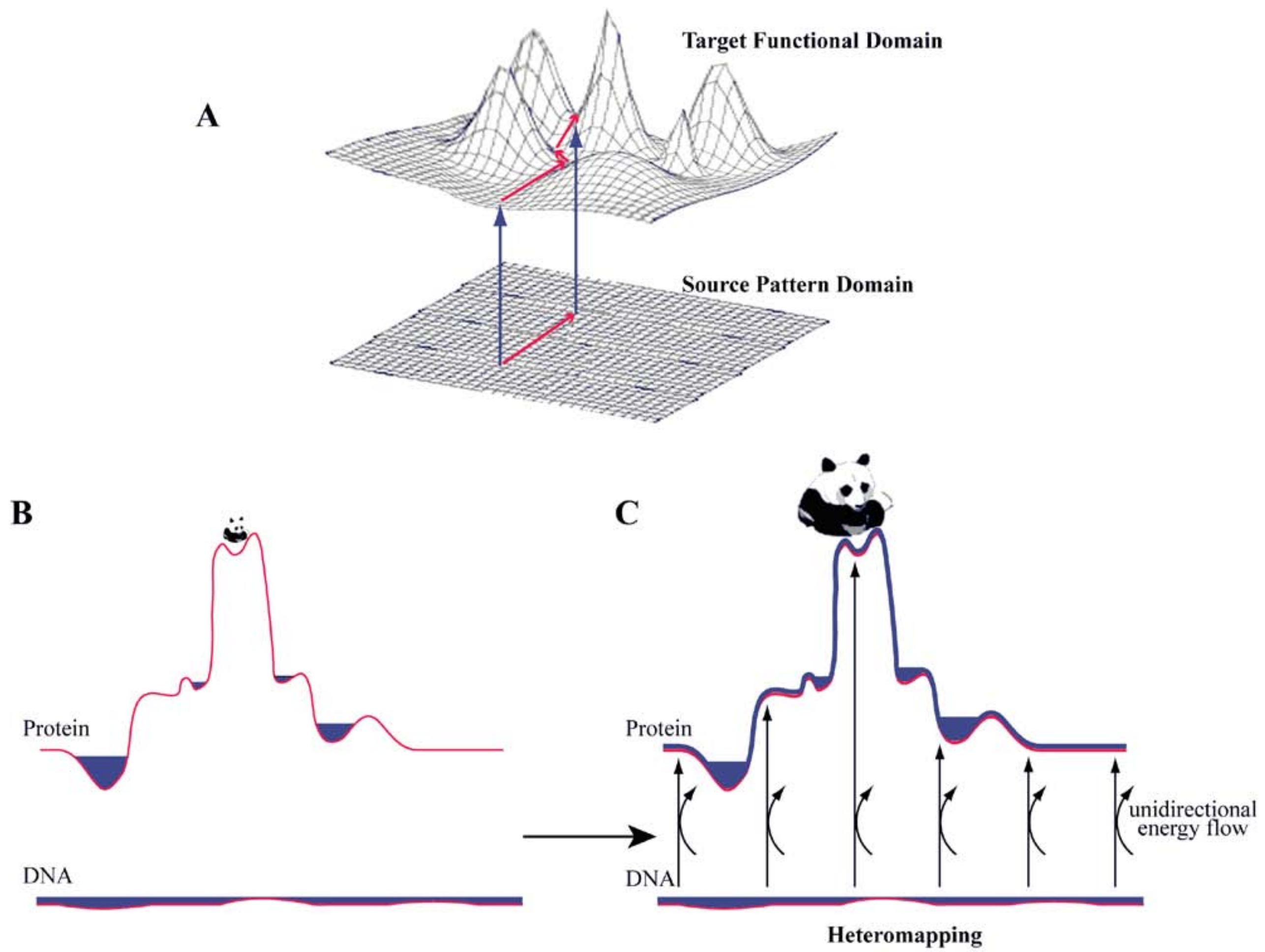

**Fig. 2. What is heterodomain mapping? A.** The lower smooth landscape represents the evolution of a source pattern domain; the upper rugged landscape represents the evolution of a target functional domain. Red arrows on a landscape represent the evolution of the corresponding domain. Blue arrows represent the mapping from the smooth source domain to the rugged target domain. A spontaneous upward evolution from a low-altitude local minimum to a high-altitude local minimum is rare on the target functional domain, but it can be achieved through heteromapping from a smooth evolution on the source domain. **B.** The protein has advantage in functional action, but has a rugged landscape with high altitude. The configuration of the protein is restricted in the valleys of low altitude (blue area). It is difficult for protein evolution to reach valleys of high altitude, which represent complex and stable configurations such as terrestrial life. DNA is inert in functional activity compared to protein. However, due to this inertness, DNA has a flat evolutionary landscape and thus the patterns generated in DNA

evolution are less biased. **C.** Translation, as a unidirectional biological heterodomain mapping with energy dissipation, generates proteins in the configuration specified by genetic coding according to DNA patterns. Because of the smoothness of the DNA landscape, the configuration of the proteins generated by translation is not restricted by its landscape anymore: it combines both DNA patterns and spontaneous protein patterns. Moreover, because DNA and protein are very different chemicals, their regions near thermodynamic equilibrium do not correspond to each other in translation. In other words, the degeneration of DNA does not result in degenerated proteins through translation. In this way, heterodomain mapping has the advantages of both protein and DNA: vigorous functional action and unbiased pattern formation. It must be emphasized that the real heteromapping and the source and target domains are much more complicated than this illustration.

The labor division does not directly output any advantageous function in the struggle for survival. Instead, it only permits the acquisition and development of those advantageous functions by eliminating various barriers in configuration space exploration. It is permissive rather than endowing. In conventional words, genotype-phenotype division only directly increases evolvability rather than fitness or complexity: heteromapping enhances diversity generation to approach the ideal complete sampling of configuration space; coupled selection feeds back the evolutionary evaluation of functions to the corresponding patterns. As the result of the labor division, the bidirectional genotype-phenotype mapping is the center of biological evolution. ***On the one hand, the principal content of biotic evolution is the consolidation and improvement of genotype-phenotype mapping in the labor division by various mechanisms, which emerge under the selective pressure for both unbiased configuration sampling through pattern formation (evolvability) and vigorous functional activity (fitness).*** The unidirectionality of translation, the linearity of genetic information, genetic code, and nuclear compartmentation are all the mechanism to preserve and further decrease the reduced bias in the pattern formation by DNA/RNA (chapter III & IV). As the last step of genotype-phenotype mapping, the coarse graining/canalization in hierarchization transform macromolecules to organisms and significantly improve the evolvability of the functional domain (chapter V & VI). However, hierarchization complicates biotic evolution with inter-level conflict and cooperation. Sex, altruism, and germline are all the mechanism to improve the labor division in the hierarchy (chapter V - VII). For example, the germ-soma division is a cellular labor division to pattern formation (germ) and functional action (soma) in multicellular animals, because the early-specified germline is sequestered from functional differentiation and selection (chapter VII). ***On the other hand, the specific ways of genotype-phenotype mapping under different situations determine the paths and forms of evolution and development.*** For example, animals have the early-specified germline, but plants have germ cells derived from vegetative (somatic) cells and hence do not have a specified germline; the early-specified germline couples genetic information to multicellular organism rather than to individual cells, and accounts for not only the much greater complexity of animals as compared to that of plants, but also the differences between them in nourishment, motility, cell fate, plasticity, development, and oncogenesis. In this way, evolution is united with the ecology as well as the development of life. Thus, the evo-devo framework is extended to the evo-devo-eco framework (chapter VII). This article presents this novel theory of evolution in detail and propounds it as a unitary and parsimonious explanation for the origin and evolution of life.

# II. A Universal Constraint on Evolution – the Universal Polarity of Unbiased Configuration Sampling and Vigorous Functional Activity

## The molecular view of general evolution: conceptual preparations

The modern evolutionary synthesis is mainly a union of Darwinism and Mendelian genetics at the populational level. Although the modern synthesis is the current mainstream theory of evolution, its framework was constructed in the 1930s when the molecular mechanism of heredity was completely unknown. Since biological evolution is a temporal and spatial continuum from molecules to cells, organisms, species, and biospheres, incorporation of molecular and cellular biology into the theory of evolution can bring a better understanding of life. Specifically, analyzing biological evolution from the molecular angle helps elucidating the origin and early evolution of life when molecules are the leading actor of evolution, and the long-term evolution of life in which the subtle effect of the bias in molecular evolution is no longer negligible. Before doing that, it is necessary to re-examine the fundamental concepts of evolution from the angle of molecules.

Any entity subject to change is an evolutionary entity, whether an elementary particle, a hypercycle of Eigen's type[25], or a human. A change from one state to another is the evolution in a general sense. This definition covers all forms of evolution and thus is compatible with the concept of Darwinian selection. For example, $H_2O$ changes from liquid state to solid state under freezing point. This phenomenon is actually an environmental action against unstable liquid configuration and for stable solid configuration. Here, stability is an abiotic extension of fitness: in abiotic evolution, the most stable is the fittest. The "stability" in this paper is a comprehensive stability which results from various environmental actions, not only thermostability. For example, the degradation of a protein may decrease the fitness of the host organism; however, to the protein molecule, degraded configuration is thermodynamically and evolutionarily more stable than the polymer state; therefore, it still holds that the "fittest" (or, more precisely, the sufficiently fit) survives. Therefore, the fitness of molecules is stability. Correspondingly, Darwinian selection, for example the birth and death of a life, is also the assembly and disassembly of a complex entity, still a configuration change rather than the beginning and end of existence. Evolution is the temporal extension of the form of existence. It is neither purposeful nor foresighted[4-9]. Complexity does not necessarily increase in evolution. However, why and how some specific types of evolutionary entities have become more complex than others is the theme of this paper.

The property of an evolutionary entity is the manifestation of its constituent elements and configuration. All evolutionary entities, either abiotic or biotic, use the same set of chemical elements. Form this angle, evolution is mainly the configurational change under environmental constraint. To a specific type of entity, evolution is the exploration of its configuration space under environmental constraint. To understand evolution from this angle, the conventional fitness landscape needs to be extended (Fig. 1).

The conventional evolutionary landscape is a fitness landscape of organisms[26-27]. The altitude in this landscape represents fitness of organisms. Climbing a peak stands for the fitness increase in evolution. However, the fitness landscape does not represent the real general process of evolution: fitness increase only occurs in a part of branches of evolution. Fitness landscape describes the macroscopic increase in fitness from human angle, which is useful in visualizing the relationship between genotype/phenotype and fitness. However, the conventional fitness landscape does not include the underlying microscopic changes into its analysis. In order to elucidate the origin and the essence of genetic life which consist in bottom-level processes, energy landscape, an extension of conventional fitness landscape to molecules, is introduced here[*].

The energy landscape of evolution represents the real physical process underlying evolution. Also known as the potential landscape or entropy landscape, the energy landscape has been broadly used in physics and chemistry to illustrate the relationship between energy change and physicochemical reactions[28-29], such as the catalysis by an enzyme[19-20]. Based on the thermodynamic aspect of life[30-32], the energy landscape has also been used to analyze biotic evolution[33-35]. All landscapes in this paper refer to energy landscape if not specified otherwise. Here, the thermodynamically equilibrium state is set as the ground state of the landscape, and the deviation from equilibrium as the ordinate axis (Y axis). The abscissa (X axis) represents the configuration space of the evolutionary entity (Fig. 1). Therefore, low altitude stands for close-to-equilibrium, while high altitude stands for far-from-equilibrium. Energy landscape uses a local minimum (valley) to represent the local optimum of fitness/stability/robustness, while fitness landscape uses a local maximum (peak) to represent the local optimum of fitness/stability/robustness. However, their opposite representation is only a difference in the angle of analysis, not a fundamental difference in the understanding of evolution. It is must be emphasized that the real energy landscape and configuration space of life are very complicated[33, 36-39]; all relevant figures in this paper are schematic drawing only for illustration (Fig. 1 & 2).

According to the second law of thermodynamics that a closed system tends to approach equilibrium, the general trend of evolution is downward on the energy landscape, unless external energy is effectively utilized to drive evolution upward[32]. Energy dissipation is the only driving or maintaining force for the upward evolution which deviates from equilibrium[30-32]. Although life, as an open system, can stably deviate from equilibrium through energy dissipation, the trend of entropy increase is still important in the origin and evolution of life (Suppl. Text 1, *The trend of entropy increase in the origin and evolution of life*).

A property opposite to thermodynamic equilibrium is complexity. As a crucial aspect of evolution, complexity is a measurement of the internal interactions of an entity. Complexity can be measured as the number of interactions per entity[40]. Complexity increase has two ways: one is the increase of horizontal interactions, and the other is the increase of vertical interactions[41], namely multilevel hierarchization, which will be discussed in chapter V and VI. Many characteristic properties of complex systems, such as unpredictability, circular causality, feedback loops, and emergence[42], are actually the consequence of an enormous number of interactions that goes beyond current human capacity of reduction, rather than qualitatively distinct properties.

[*] A fitness or energy landscape must be dynamic. The environment and organisms influence each other. The landscape of an evolutionary entity is changed by both environmental factors and the evolution of that entity *per se*.

Thermodynamic equilibrium represents the randomness in its components' evolution, the independence among components, and thus the lack of interactions among components[43-45]. Therefore, a complex state must deviate from thermodynamic equilibrium. An entity's deviation from thermodynamic equilibrium is the order of that entity. ***Although complex state must deviate from the equilibrium, complexity is not proportional to the degree of order***, because completely ordered entities, for example crystals, are highly deviated but only mildly complex[46]. Therefore, simple input of energy does not necessarily increase complexity, although it may increase order. Biotic entities are both complex and highly deviated from equilibrium. Moreover, their complexity is nontrivial, namely it is organized rather than disorganized complexity. The organization of biotic entities is the configuration underlying the mechanism that utilizes energy to maintain their deviation from equilibrium.

An entity in disequilibrium, such as life, can be stable if and only if it is in a local minimum (a valley) where the surrounding energy ridge prevents the entity from regressing toward equilibrium (Fig. 1). The stability/robustness relative to a reference state is determined by the accumulative height of all ridges that the entity has to cross to reach the reference state (usually the ground state)[*]. The altitude of a local minimum is irrelevant to the height of surrounding ridges. Therefore, complexity is not necessarily related to stability/robustness. A stable, complex, and organized entity, such as life, must be highly deviated from equilibrium[30-32], and that is represented as a high-altitude local minimum on the energy landscape, like the crater of a volcano (Fig. 1).

According to the above generalization of evolution, biological evolution is only a special type of general evolution. Compared to abiotic entities, biotic entities is special in their discontinuous spatial extension but continuous temporal extension through reproduction/replication (continuous germline). What matters in the continuous temporal extension is the continuance of abstract information as a lineage through reproduction/replication (continuance through germline genetic information). The lineage is as physical as the ordinary entity whose spatial extension and temporal extension are both continuous (continuance through somatic information and non-informational proteins). The fitness in biology is the ability of a biotic entity to sustain and maximize its lineage under the environmental constraint. Fitness is just a special form of stability/robustness of a lineage of entities, namely the accumulative height of all ridges surrounding the local minimum (valley) of that lineage. However, fitness or the stability/robustness of lineage is different from that of individual entities, because the ability to sustain and maximize its lineage, namely replication/reproduction, is not required for the stability/robustness of individuals. Therefore, replicate/reproduction may be selected for at the cost of the stability/robustness of individuals. The number and survival rate of offspring are only two of the specific parameters of the fitness of reproductive life. The biological embodiment of the surrounding energy barriers is the functional activities of life, which realize the stability/robustness/fitness of life by sustaining the configurations of life through effective energy consumption (for why a biological lineage is more stable than a biological individual, please see the 2nd section of chapter III). The above definition of fitness interprets biological fitness from the angle of general evolution, including both abiotic and biotic evolution. This angle of interpretation is necessary for understanding the transition from abiotic evolution to biotic evolution.

---

[*] If there are multiple paths to the reference state, then the least accumulative height of all barriers is the stability.

Both abiotic and biotic evolution is the change in the configuration of an evolutionary entity by environmental action. The "natural selection" in Darwinism is only a special type of configuration change, namely destruction of a lineage together with its genetic information from a complex configuration to very simple elements. In contrast, abiotic evolution can have a relatively continuous and gradual configuration change without destruction, namely the transformation in Lamarckism[47-48]. Biotic molecules such as DNA undergo both types of environmental action: mutation is Lamarckian transformation while destruction with the host organism is natural selection. Why natural selection rather than Lamarckian transformation is the principal contributor to the complexity of life is the major topic of this article. Since environmental action is universal, environmental action alone cannot explain why only a specific type of evolutionary entity has extraordinary complexity[7]. As a purely eliminative process, selection at one level only eliminates the existing individuals at that level and never generates new substrates for selection at that level. This understanding is the semantic and physical meaning of selection, and also the essence of Darwinism vs. Lamarckism. However, selection at lower level, for example molecules and cells, changes the configuration of higher level and thus generates new phenotypes and genotypes as the substrate for higher-level selection. Internal evolution of lower levels is the source of new configurations and thus the substrate of natural selection[7-9, 49-51] (Suppl. Text 2, *The generalization of Darwinian selection*).

The viewpoint that natural selection of organisms can explain everything is misleading[6-8, 49, 52-54]. Although natural selection is an essential of evolution, the universal selection of organisms alone cannot explain why some organisms are more complex than others. Natural selection can only see the immediate fitness. ***As in the general evolution, the complexity of a form of life does not necessarily correlate with its fitness.*** For instance, although the panda is much more complex than *E. coli*, *E. coli* could have much higher fitness, namely the stability/robustness of its lineage. Therefore, natural selection for stability/robustness/fitness alone cannot result in the complexity of life. Why there is a large-scale trend of complexity increase in the history of life[55-58]? Like diffusion from the bottom to the top, sampling configuration space starts from the state near equilibrium (the simplest configurations): configurations of low complexity are reached before those of high complexity[55-58]. Meanwhile, backward evolution (decrease in complexity) occurs equally, if not more. The destruction of a lineage in Darwinian selection is actually the backward evolution of that lineage. The bottom of diffusion is the thermodynamic equilibrium with maximal entropy and minimal complexity. Although there is a universal trend of entropy increase, energy/negative entropy counteracts the trend of entropy increase and drives the deviation of life from equilibrium. As a result, both the maximal and average complexity of evolution are increasing despite the blindness of evolution, like diffusion goes upward despite the randomness of molecular motion[55-58].

This widely accepted diffusion model has a hidden assumption: evolutionary sampling of configuration space (diffusion) is random[57], which means that the configuration sampling is neither constrained nor biased. However, there is a universal polarity that constrains evolutionary sampling: the landscape of active entities, for example proteins, is rugged, so the evolution of active entity is trapped in local maximums of fitness/stability/robustness. In biotic evolution, the nearly unbiased sampling of configuration space is realized through inert entities, for example DNA, whose landscape is relatively smooth. The relatively unconstrained evolution of inert entities guides the evolution of active entities through genotype-phenotype mapping. The whole history of life is the emergence and improvement of genotype-phenotype mapping. As the selective pressure for genotype-phenotype dichotomy, the universal polarity of evolution is crucial for understanding of life and thus deserves detailed examination.

## The universal polarity of unbiased configuration sampling and vigorous functional action is the intrinsic conflict between evolvability and fitness

***Because evolution is blind and purposeless[4-8], the bias in sampling of configuration space is unfavorable to the increase of complexity and fitness***. What is the scientific meaning of the term "blind" here? The meaning is that no process can necessarily lead to the increase in complexity or fitness. When evolution is examined from the angle of molecules, natural selection only concerns the immediate fitness/stability/robustness; therefore, natural selection cannot locate configurations of higher fitness from the current local optimum of fitness. Simple energy/negative entropy input only increases thermodynamic order; as explained above, neither fitness/stability/robustness nor order correlates with complexity. "Prediction" of complexity, fitness, or environmental change by a physical process is impossible in blind evolution. Therefore, the best strategy for blind evolution to increase complexity or fitness is to sample configurations as complete as possible at every step of configuration space exploration, although perfectly complete sampling is an unattainable ideal. If the missed configurations are distributed randomly, namely the incompleteness in sampling is unbiased, the incompleteness can be remedied by repeated or parallel sampling. However, if the incompleteness is biased, then there is no remedy. The bias in sampling will make the evolutionary entity miss the paths leading to rare complex and stable/fit configurations.

All existing configurations of macromolecules are not random samples of the configuration space. The bias is caused by the molecular selection of the fitness of macromolecules, namely the stability/robustness of macromolecules under the specific environment[33, 36-39]. Some configurations are unstable and thus do not have sustained existence. Some configurations are stable but are inaccessible to blind evolution because they are isolated by unstable configurations on the landscape. Such molecular selection constrains the sampling of configuration space by blind evolution[1, 16, 33, 36-39] and thus restricts the accessibility of some advantageous configurations/phenotypes.

From the researches on RNA evolution it also concludes that evolutionary biases, such as mutational robustness and thermostability, constrain evolution, like the ridges on the landscape blocks evolutionary exploration[11-12, 15-16]. However, above findings are not linked to the blindness of evolution to explicitly conclude that bias is harmful to the increase of fitness and complexity of evolution. As a result, the understanding of evolutionary bias is unclear. In the conventional view, a process of evolution is not clearly distinguished from the associated bias. For example, mutational bias is often confused with mutation *per se* and hence is consider introducing novelty for evolution[17]. Mutation generates novelty/diversity, but mutational bias only makes the available paths less than the possible paths and thus reduces the novelty/diversity. Internal bias of lower-level evolution is generally considered too weak compared to natural selection, and thus unimportant in evolution. However, in the transition from non-life to life and the long-term evolution of life, molecular and other lower-level biases play an important role and cannot be safely ignored.

A smooth landscape without bias is favorable to configuration sampling. However, the same evolutionary landscape topology can have completely different effect between genotype and phenotype[11]. Failure to distinguish the effects of a smooth landscape in genotype and phenotype may lead to the confusion in understanding evolvability. When the difference between genotype and phenotype in the effect of landscape topology is extended to all types of evolutionary entities from molecules through cells to organisms, an intrinsic and universal constraint for evolutionary exploration surfaces. Specifically, every entity has a definite landscape in an environment; the roughness of the landscape represents the activity of the entity and the degree of the bias in its evolution. An active entity or system must be biased to one direction in its bidirectional or multidirectional reaction and evolution, while a weakly biased entity must be inert in its activity[19-20]. The bias of evolution reduces the availability and accessibility of configurations; in conventional words, it reduces diversity. Because of the bias in the evolution of active entities and systems, their energy landscape of evolution is rugged and full of steep slopes, which represent the strong biases in evolution (Fig. 1). The evolution of active entities is trapped in local minimums. It is difficult for a highly active entity to escape from a valley (local minimum), or climb over a ridge (local maximum). For example, highly active sodium readily reacts with other elements in the environment, such as oxygen, and form stable compounds, which fixes the relation of sodium to its surrounding entities; namely, the configuration is highly biased. Accordingly, the evolution of highly active entities, either small inorganic molecules or high molecular weight biotic polymers, is the ineffective motion constrained in low-altitude local minimums which represent stable configurations. The configuration space of the active entity has high-altitude peaks and valleys that represent complex and far-from-equilibrium states, but it is very difficult for the active entity to reach high altitude configurations (Fig. 1). In other words, sampling the configuration space by active entities is biased to the local minimums of low altitude, opposite to the high altitude positions of life.

If the activity of constitutive entities is inert, the landscape will be relatively smooth and the switch between different configurations will be easy. For example, iron is less active than sodium; as a result, in the same environment, iron has more types of relation to its surrounding entities than sodium does: iron can have both free and several compound configurations while sodium cannot. The configurations of inert entities are less biased and more diverse than those of active entities. The less active, the more diverse the configurations are, and the less bias in sampling configuration space. However, an inactive entity is functionally inadequate: it cannot have any practically useful function (except for approaching unbiased sampling configuration space) due to its inertness. An extreme is the inert gaseous element helium, which is completely useless in the function of life. In other words, the whole configuration space of inert entities is close to equilibrium: the entity cannot acquire significant complexity, which must deviate from equilibrium (Fig. 1).

***In short, evolution has a universal polarity: active entities have potential complexity but their configuration space sampling is biased to local minimums of low altitude; functionally inert entities have less biased sampling of configuration space, but have very limited function and complexity.*** Unbiased configuration sampling and vigorous activity are the two opposites of the intrinsic property of all entities. Therefore, this conflict is named as the universal polarity which applies to both abiotic and biotic entities. This universal polarity can be understood intuitively: a rugged landscape has peaks but the motion is trapped in the valleys of low altitude, while the motion on a flat landscape is not constrained but has neither evolutionary altitude nor stability/robustness (Fig. 1). The unbiased configuration sampling by blind evolution is just the conventional concepts of diversity, which are the key of evolvability[11, 22-23, 59]. Although the conventional view recognizes the general importance of diversity, it considers the intrinsic bias or barrier in diversity generation trivial or inexistent: only environmental selection of diversity at organismal and higher levels is important. As a result, the conventional view overlooks why diversity is important at root, how diversity is compromised in real situations, and what is the consequence of compromised diversity.

The universal polarity of evolution is actually the conflict between evolvability and fitness in the conventional view of evolution[22-23]. Life is stable complexity, which requires biases, namely valleys, ridges, and various barriers, on the landscape of functional domain to sustain its configurations. Inert entities with a smooth landscape cannot have stable configurations even if they can highly deviate from equilibrium. As the stability/robustness of a lineage of biotic entities, fitness provides an opportunity of sustained existence for some configurations of life, but constrains further exploration of other configurations. As the physical basis of phenotype, functional activities are just the embodiment of the biases and constraints on the rugged landscape of functional domain which actualize the stability/robustness/fitness of life. Therefore, the universal polarity of unbiased sampling and functional action is actually the intrinsic conflict between evolvability and stability/robustness/fitness. The major difference of the present theory from the conventional view is that the fitness and phenotype of life are extended here to the comprehensive stability and the configuration of evolutionary lineages, respectively. As a result, the conflict between evolvability and fitness is extended from biotic evolution to abiotic evolution. The polarity between unbiased sampling and functional action has many specific manifestations, among which the spatial constraint on the *en bloc* evolution is the most important.

## The spatial constraint on the *en bloc* evolution: pattern duplication require a spare spatial dimension

Most evolutionary entities are 3-dimensional (3-D) because the present space is 3-D. If not specifically constrained, all entities will use all dimensions of the space. It is peculiar that genetic information, whether in DNA or RNA, is always 1-dimensional (1-D), no matter in working state or in packaged state. Although proteins are synthesized as 1-D product according to the 1-D DNA/RNA template, proteins work in 3-D form. The reason behind the linearity of genetic information is the conflict between the efficiency of configuration sampling and the vigorousness of functional action, which is one of the specific manifestations of the universal polarity of evolution. What is *en bloc* evolution? Briefly, *en bloc* evolution is the preservation of configuration/pattern during evolution. Like a point mutation, point evolution is the change of a single element of the configuration/pattern. Although resulting in pattern change, point evolution *per se* does not contain any pattern. Therefore, point evolution cannot make a change in the form of pattern. In other words, point evolution always alters the original pattern to a new one, and cannot preserve or transfer an already existing pattern in evolution. In contrast, the change in the form of an existing pattern is *en bloc* pattern evolution, where an existing pattern is the unit of change. For example, replication, gene duplication, recombination, segregation, incorporation, and transposition are all *en bloc* evolution of DNA. Similar to modular evolution[14, 60], *en bloc* evolution has a special importance in evolution. ***When an existing configuration/pattern is a beneficial product of previous evolution, preservation of this configuration/pattern as a whole from disintegration is important. In other words, existing beneficial configurations/patterns need to evolve en bloc.*** In this way, the achievement of previous evolution becomes accumulable. If the structural complexity can accumulate through *en bloc* evolution, evolution will finally achieve remarkable complexity given sufficient time and unbiased sampling. *En bloc* evolution cannot be replaced by a group of point evolutions, for example a group of point mutations, because the pattern of this group of point changes is not the consequence of previous natural selection and thus is not beneficial in most cases.

*En bloc* pattern evolution is crucial in the origin and evolution of life. Replication is a typical example. The importance of replication consists in the branching of evolution. Natural selection would be meaningless without variation[14]. However, branching of evolution is as important as variation in promoting evolution. ***Variation without branching is only serial fluctuation rather than parallel diversification, which is the substrate of biological selection. Selection of serial fluctuation resets the evolution to the abiotic evolution and thus destroys the achievement of the previous evolution**.** In contrast, selection of parallel branches eliminates

---

* Individual evolutionary entities have only limited lifetime because of the degeneration resulting from the universal trend of entropy increasing; although life as a whole consumes energy to drive and maintain its deviation from equilibrium, the components of life are always under the erosion of entropy increasing. During the origin and early evolution of life when a repair mechanism is not available, the influence of such erosion is especially strong. Even when a repair mechanism is

branches of low fitness and keeps branches of high fitness; therefore, biological evolution preserves the previous achievement and incorporates the improvement. The substrate of selection in branching evolution includes both the differential reproductive capability in the conventional theory and other qualities related to fitness. Even if all novel branches are eliminated, the complexity and fitness of the original branching point are preserved. Parallel diversity plus natural selection improves fitness and complexity step by step. Although the complexity increase in every step may be very small, the branched evolution will acquire significant fitness and complexity given sufficient time. Without branching, blind pinpointing extremely small regions of life in the vast configuration space by a single entity is prohibitively improbable, even if the landscape is perfectly smooth. Branching is not essential to evolution, but is required for efficient fitness and complexity increase, and that is the reason why all forms of life are reproductive. ***Evolutionary branching is actually a parallel sampling of configuration space, which is a breakthrough in the efficiency of configuration sampling.*** In conventional words, evolutionary branching contributes to genetic novelty[61]. Biological branching, namely reproduction, is very complicated, but the key to branching is the replication of the patterns and configurations of life. In all forms of *en bloc* evolution, replication is the best example of the advantage of *en bloc* evolution.

However, there is a universal spatial barrier to the *en bloc* evolution of patterns. All configurations/patterns have dimensions, which are the physical degree of freedom. ***If the dimensionality of configuration/pattern is the same as the space, then there is no spare degree of freedom for en bloc evolution that keeps the integrity of the pattern.*** In 3-D space, it is impossible to separate *intertwined* 3-D configurations/patterns or combined separated 3-D configurations/patterns without changing them. The reason is that all spatial degrees of freedom are used in the construction of patterns, and thus no spare degree is available for innocuous separation/incorporation. This phenomenon is consistent with the physical meaning of dimension that a dimension is a degree of freedom. As illustrated in Fig. 3A, in a 2-dimensional (2-D) plane with 2-D concentric circles, the interior circle cannot be separated from the exterior circle without breaking the exterior circle. However, in the 3-D space, the 2-D interior circle can be separated from the 2-D exterior circle through the third dimension (Fig. 3A). Similarly, separating *intertwined* 3-D patterns in 3-D space must destroy the patterns, as undoing a knot in 3-D space requires the knot to be cut[*]. If the dimensionality of intertwined patterns is lower than the dimensionality of space, the separation can be fulfilled without changing the patterns, as undoing a 3-D knot in 4-dimensional space without cutting it[62]. Simple 3-D structures, for example relatively homogeneous fluid coacervates and microspheres, can divide without disintegration. However, such division is only a physical split, not a true replication of the parental 3-D structures[63], moreover, the parental 3-D structure is indeed changed during split, but the change is insufficient to disrupt the very simple structure.

---

available at late stage, not all erosions can be repaired with energy utilization. Therefore, the death/disintegration of individual entities is inevitable.

[*] A curved line, for example a knot, is a curved 1-D space, although a knot looks like a 3-D object. Similarly, a curved surface, for example, the surface of a ball, is a curved 2-D space, although it envelops a 3-D space. The curvature, or nontechnically, the shape, of a space does not change the dimensionality of the space.

Any *en bloc* operation on the 3-D structure, such as replication (vertical transfer), segregation, or incorporation (horizontal transfer), involves separating *intertwined* 3-D patterns or combining originally separated 3-D patterns to form novel *intertwined* 3-D patterns, both of which destroy the 3-D patterns. Restoring these destroyed 3-D patterns requires pattern storage of lower dimensionality, which can be replicated in 3-D space without damage. When the 3-D configuration/pattern is the principal form of complexity, the extensive destruction of the 3-D pattern during replication is irreversible, because the patterns of lower dimensionality are insufficient to restore the 3-D patterns.

If not specifically guided or restricted, all evolutionary entities or systems tend to fill the space and thus occupy as many dimensions as possible. 3-D structures have more and better functional activities than those of fewer dimensions. Therefore, under selective pressure, all evolutionary entities use all available physical degrees of freedom. For example, internal compartmentation is functionally advantageous compared to the diffusive reaction system; therefore, internal compartmentation is required for all entities whose complexity is beyond a certain level, for example the terrestrial life. However, the 3-D internal compartments block the replication of their host entity. All 3-D structures prevents *en bloc* evolution, and thus blocks evolutionary branching and complexity accumulation. The complexity of protocells must be limited by this spatial constraint until protocells acquire a mechanism to solve this problem.

This barrier to *en bloc* evolution constrains the accumulation of complexity through configuration/pattern preservation. As *en bloc* pattern evolution is a special form of pattern formation, such spatial constraint is a special and important manifestation of the universal polarity of evolution: functional activity seeks as high dimensionality as possible, while efficient configuration sampling and preservation through *en bloc* evolution seeks as low dimensionality as possible. The universal polarity of configuration sampling and functional activity is the universal barrier to life. Life is just the evolutionary entity that has crossed the barrier and thus has acquired extraordinary complexity in long-term evolution. What mechanism does the terrestrial life use to break the universal polarity of evolution?

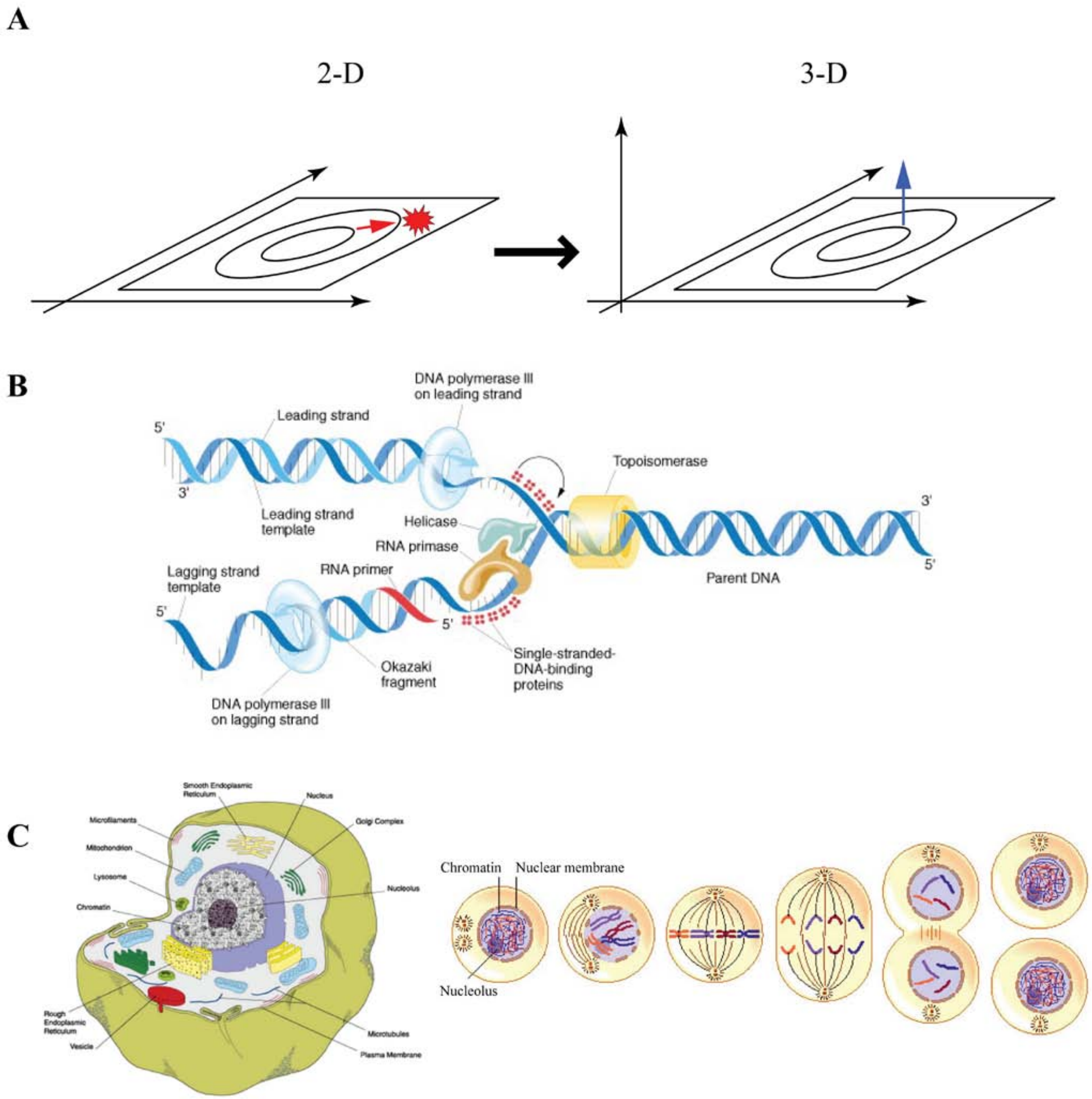

**Fig. 3. Replication requires spare spatial dimensions. A.** Separation of intertwined patterns, an essential step in pattern replication, requires at least one spare spatial dimension, i.e. one spare degree of freedom. As illustrated, separating concentric circles without breaking the large circle requires at least one spare dimension. In a 2-D plane with 2-D concentric circles, the interior circle cannot be separated from the exterior circle without breaking the exterior circle. When 2-D concentric circles are placed in a 3-D space, the interior circle can be separated from the exterior circle through the third dimension. **B.** Although DNA is 3-D, the DNA patterns that are used in life are 1-D. Therefore, the DNA patterns are not changed during replication and separation in 3-D space. Even when DNA is severed by topoisomerase, the pattern can be maintained by proteins using other two spare dimensions. **C.** The patterns of the cell are 3-D, such as folded proteins, subcellular

compartments, and organelles. Cells cannot be replicated and separated without destroying these intertwined 3-D patterns. In cell division, most organelles, such as the nuclear envelop, are destroyed to allow the separation of two 3-D daughter cells. The information for restoring the compartments and organelles is mainly stored in DNA as 1-D patterns, which are intact during division. The pattern with the lower dimensionality than the space is the only escape from the destruction during *en bloc* evolution.

## III. The Molecular Interpretation of Genotype-phenotype Mapping

The only solution to the universal polarity of evolution is the labor division of internal evolution to pattern formation (configuration sampling) and functional action (Fig. 4), which is the threshold mechanism of life and thus the fundamental difference between nonlife and life. This division of labor requires bidirectional links between the separated pattern formation and functional action. One direction is that the specialized pattern generator, namely DNA/RNA sequence space, guides the evolution of proteins through heterodomain mapping, which is actually the genotype-phenotype mapping. The other direction is the coupling of the specialized DNA/RNA pattern domain to the protein functional domain in natural selection, which feeds back the fitness of functional domain to the corresponding pattern domain. The bidirectional genotype-phenotype mapping only increases evolvability: heteromapping enhances the diversity generation by eliminating the barriers in the configuration space sampling, and coupled selection feeds back the fitness of the functional output to the pattern domain. In conventional words, the genotype-phenotype division does not directly produce any functional output in the struggle for survival; rather, it is a permissive mechanism that eliminates the barriers in the acquisition and development of those functions.

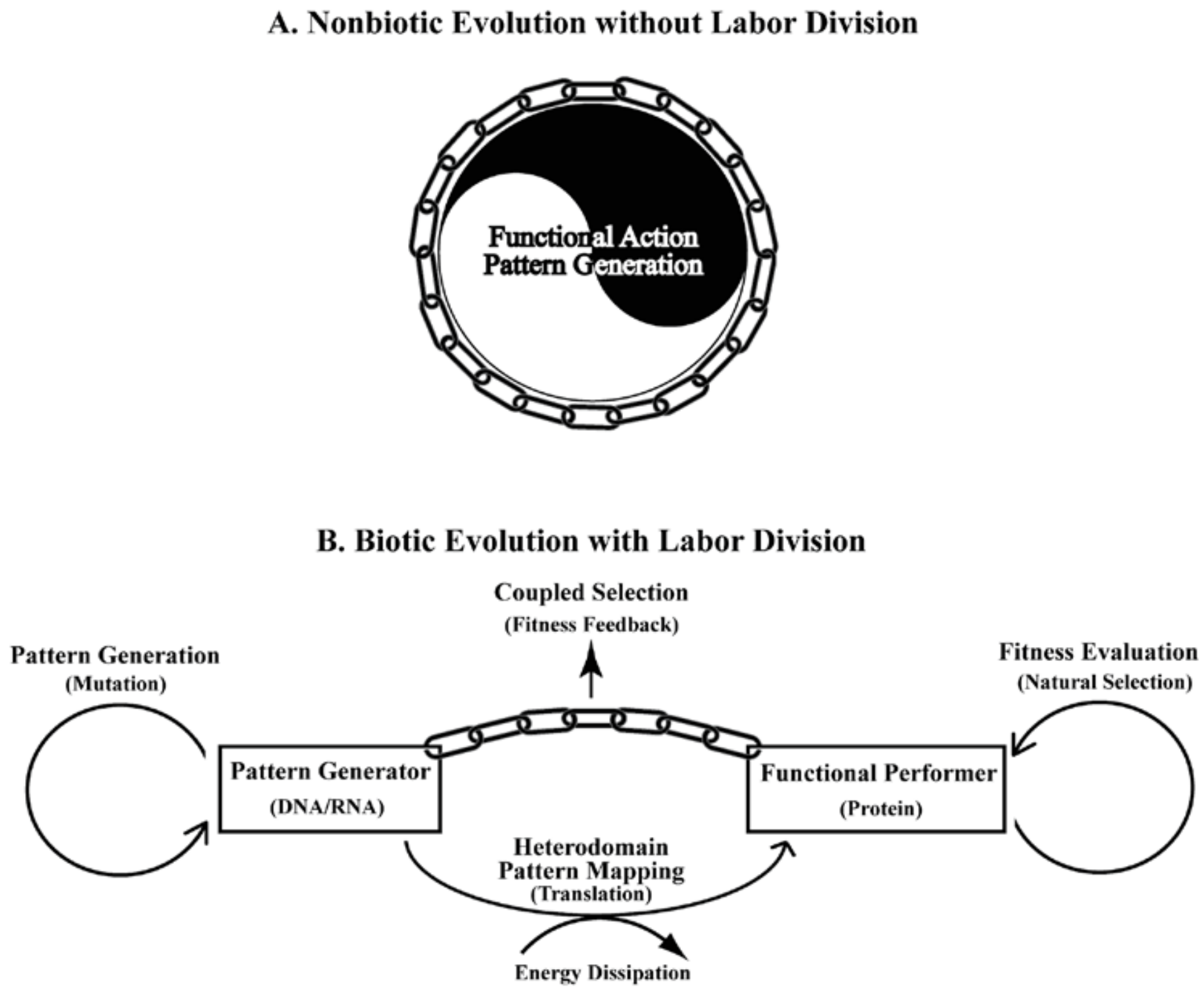

**Fig. 4. Labor division is the essence of life.** Both unbiased pattern formation and vigorous functional activity are required for complexity and fitness to increase. However, unbiased pattern formation and vigorous functional activity are opposite to each other and forms a universal polarity. In abiotic evolution, pattern formation and functional action are inherently bonded (**A**). Biotic evolution breaks the universal polarity through a division of labor to pattern formation and functional action (**B**). The separated DNA/RNA pattern generator and protein functional performer are linked by heterodomain mapping and coupled selection: translation, a biological heterodomain mapping, maps the patterns in DNA/RNA domain to the functional protein domain with unidirectional energy flow, while coupled selection feeds back the fitness of the functional action to the pattern generator. As the essence of life, such labor division not only accounts for the tremendous complexity of life, but also explains the paths and forms of biotic evolution and development.

## The watershed between life and nonlife

Many phenomena are considered as key features of life, such as metabolism, replication, compartmentation, adaptation, growth, homeostasis, hierarchization/organization, *etc*. However, all these phenomena have abiotic counterparts, although they are much more complex than their abiotic counterparts. So there is no unbridgeable gap between life and nonlife[24]. The essence of life should be a threshold mechanism that breaks the limits to the complexity increase in abiotic evolution. The internal division of evolution to pattern generation and functional action is the threshold mechanism of life. The pattern domain has a relatively smooth landscape with advantage in pattern formation and preservation, while the target domain has a rugged landscape with advantage in functional action. Through mapping from the pattern domain to the functional domain, namely heterodomain mapping from genotype to phenotype, the evolution of protein functional domain is guided by the nearly unbiased evolution of genetic information to realize the nearly unbiased sampling of configuration space. The labor division combines the advantages of both domains, and thus breaks the mechanistic limit to the complexity increase of a single domain due to the universal polarity (Fig. 2). The labor division of internal evolution to pattern formation and functional action is the distinction between life and nonlife.

The labor division has four components: the DNA/RNA pattern domain, the protein functional domain, mapping from genotype to phenotype, and the coupled selection of the pattern domain to the functional domain. The precursors of pattern and functional domains exist as isolated evolutionary entities before the labor division. The coupled selection of pattern domain with functional domain also precedes the labor division, because the integrity of DNA/RNA patterns requires the integrity of the host organism. Therefore, among the four components of labor division, genotype to phenotype mapping, the biological heteromapping, is the decisive one. The mapping has three steps: genetic translation, namely the heteromapping from linear DNA/RNA to linear protein, protein folding, and hierarchization/organization of proteins and other macromolecules to form cellular organisms. Only translation (heteromapping) is the unique characteristic of life; protein folding and hierarchization are homodomain transformation and have counterparts in abiotic and prebiotic evolution, although protein folding and hierarchization are much more complex than their counterparts in abiotic and prebiotic evolution. Therefore, the emergence of translation is a decisive step in the origin of life.

## Translation breaks the universal polarity

Genetic translation breaks the polarity constraint on configuration sampling (pattern formation) and functional activity. Translation is a unidirectional heteromapping. The source domain is nucleic acid and the target domain is protein. The amino acid residues of the target domain are more active than the nucleotides of the source domain, so proteins are relatively active and have diverse functions. Therefore, the landscape of amino acids and proteins is rugged. The configuration and sequence of proteins are restricted in the low-altitude valleys, which represent spontaneously formed small peptides and degraded products. Although proteins have vast configuration space, evolution from primitive peptides to long and complex proteins is blocked by the rugged landscape, let alone the complex configurations underlying the fitness and function of life (Fig. 2).

In the source domain, the chemical activity of nucleic acids, particularly DNA, is much weaker than that of proteins. The landscape of nucleic acids is smoother than that of amino acids and proteins. The landscape of the linear DNA, namely the sequence space of DNA, is especially smooth. The smoother landscape of DNA has two important consequences. First, the difference in stability between monomers and linear DNA polymers is smaller than the difference between amino acids and proteins. Therefore, nucleic acids can form longer and more stable polymer than amino acids. This property enables DNA to hold more patterns than protein. Second, the stability of linear polymer of nucleic acids is only weakly affected by the sequence when compared to proteins. The strongly sequence-specific evolution of DNA/RNA occurs only after the emergence of proteinaceous enzyme. Because proteinaceous enzyme is the translational product of DNA, tightly regulated specific action on DNA by enzymes is actually an extended form of the internal interaction and organization of genome (3rd section of chapter IV). Therefore, the landscape of nucleic acid sequences, particularly DNA, is much smoother than that of protein sequences. The pattern generated by DNA evolution is less biased because its landscape is smoother than that of protein. In short, the pattern of DNA is greater and less biased than the pattern of protein.

A protein with a desired function may be at a high-altitude position or isolated by barriers on its landscape. Nevertheless, its corresponding DNA sequence is on a relatively smooth landscape. Through transcription and translation, the protein can be synthesized according to the DNA template. The relatively smooth evolution of source domain is a blueprint of the rugged evolution of target domain (Fig. 2). Even if there are weak barriers on the landscape of DNA, the distribution of the barriers is different from those in the protein domain, because they are heterogeneous. The smooth area of DNA landscape maps to the peaks on the rugged protein landscape; therefore, peaks on protein landscape can be reached through heteromapping.

Meanwhile, the distributions of valleys are different between source and target domains. In other words, the degeneration of DNA (mutation) does not result in degenerated proteins in translation. As an active functional performer, proteins tend to degenerate, which is a principal factor limiting the life of individual organisms. In contrast, as a pattern generator, relatively inert DNA does not tend to degenerate as much as proteins; at the late stage of life when novelty is not needed as much as in the early stage, various mechanisms emerge to increase the stability of DNA, for example nucleosomes. Moreover, because of the heterogeneity in the genotype-phenotype mapping, degeneration of DNA does not result in degenerated proteins in translation; the effect of DNA degeneration is equivalent to the almost random change in the protein domain, not the degeneration of the protein domain. That is why a biological lineage (continuance through germline information) is much more stable than a biological individual (continuance through somatic information and non-informational proteins).

Not only the sequence of proteins but also the organization of proteins, namely the relationship between proteins, can be mapped from linear patterns of DNA (3rd section of chapter IV). The advantage of heterodomain is further enhanced by the tuned genetic code (4th section of this chapter). Therefore, through heteromapping from DNA, the evolution of protein can go through barriers to reach peaks, valleys, or regions isolated by peaks or valleys (Fig. 2). Another result is the flexibility of biotic evolution: it is not limited by the stability of its performer - proteins. The cause of such flexibility of biotic evolution is the labor division: the pattern generator can command the functional domain to overcome immediate disadvantages and go beyond immediate advantages, because the pattern domain, the planner of evolution, has a smoother landscape than the functional domain, the performer of evolution. In contrast, abiotic evolution is rigid because of the intrinsic unity of pattern domain and functional domain. The difference between the pattern domain and the functional domain is the cause of altruism and intelligence, as will be explained in chapter VI.

Energy is the only force driving evolutionary entities to climb peaks and ridges on the energy landscape. However, complexity gained by energy input is unstable even if it climbs to a high-altitude local minimum, because energy is a double-edged sword: it equally accelerates the disintegration of complex entity at a high-altitude local minimum through climbing the surrounding ridges; therefore, energy input raises the altitude but lowers or removes the surrounding ridge at the same time; as a result, the topology of the landscape is changed. As in chemical reactions, energy influences all directions of the reaction equally[19-20]. In other words, the effect of energy is bidirectional in nature. Energy flow is an intrinsic part of evolution. The diversity of the patterns generated in energy flow is still limited by the form and nature of a specific type of evolution; namely, energy flow is a part of the landscape and thus changes the topology of landscape (Fig. 2). In order to drive the upward evolution effectively, energy flow must be unidirectional, and that in turn requires a corresponding change in the internal configuration of evolutionary entities.

The heterodomain mapping from DNA/RNA to proteins requires energy. The key driving step is the dissipation of energy to ***blindly*** activate and link individual amino acids to form a protein chain[19-20], which represents the climb from the bottom of equilibrium to a high-altitude terrain on the landscape of protein. The blindness of energy dissipation is essential for blind evolution to utilize energy. Here, energy flows in a fixed unidirectional way and does not affect the pattern generated in translation. The only determinant of the generated pattern is the DNA/RNA template. In other words, the pattern of energy flow is separated from the generated pattern because of the labor division in biotic evolution: the unidirectional flow is not a part of the landscape and thus does not affect the topology of the landscape. In contrast, abiotic energy flow is an intrinsic part of abiotic evolution and cannot be separated from the pattern generated during evolution. ***Because energy dissipation is the only way to drive upward evolution, the separation of the pattern of energy flow from the generated pattern is a breakthrough in evolution.*** The complex biotic metabolism, including cellular respiration and the subsequent energy consumption, is an advanced extension of the unidirectional energy flow in translation, because it develops from the output of unidirectional translation and also forces the proteins and other molecules to the configurations encoded in the genomic patterns.

## Translation breaks the spatial barrier to *en bloc* evolution

The spatial barrier to *en bloc* evolution is an important manifestation of the universal polarity. If the source domain of heteromapping has lower dimensionality than the space, then the spatial barrier is broken naturally. In genetic translation, the source domain, DNA, is 1-D. Although its physical structure is still 3-D, the organization of genetic material is the 1-D linkage of elements, so the pattern of DNA/RNA used in translation is 1-D too. Theoretically, any configuration/pattern of dimensionality lower than three can break the barrier to replication. However, the fewer the dimensions of the structure, the more stable the structure is during replication, because more degrees of freedom are available for the replicating operation. In 3-D space, *en bloc* evolution of the 1-D genome, including replication, segregation, and incorporation, does not interfere with the patterns stored in DNA (Fig. 3B&C). For example, the division of biotic cells undergoes a extensive reorganization: chromosomes change from extended form for replication/transcription to condensed form for segregation, and many 3-D subcellular organelles, such as nuclear envelope, are destroyed to fulfill separation[64]. The regulation of the destruction, division, and restoration of intertwined 3-D structures heavily depends on the patterns in 1-D DNA sequences[64], which is intact during cell division (Fig. 3C). This is a mechanism specific to life. The individual life is mortal, but the genetic information is perpetual because of replication and transmission of genetic material. In this way, the complexity of life, in the form of DNA pattern, can increase without limit. Given sufficient time, life can acquire extraordinary complexity through genetic heteromapping and natural selection. Because of this property of linear genetic information, the higher percentage of genetic information in the total evolving and transmitted patterns in an evolutionary entity, the higher evolvability the entity has and the higher complexity will be achieved in evolution. The higher complexity of eukaryotes than that of prokaryotes can be explained by the advantage of high percentage of genetic information in heredity (3rd section of chapter IV).

Replication is often considered as the central part of life. It accelerates complexity increase in evolution by generating varied branches for natural selection. However, as explained above, replication of a 3-D pattern is not allowed in 3-D space. Only after the emergence of heterodomain mapping, does the branching of biotic evolution become feasible through the replication of 1-D DNA/RNA. Branching/replication is an important component of genetic heredity (also the horizontal transfer in the origin and early evolution of life), but it is the consequence of translation, the heterodomain mapping from 1-D DNA/RNA to 3-D protein. Moreover, separated from translation, DNA/RNA replication is fundamentally similar to abiotic replication, as in the growth of a crystal, and cannot promote complexity increase. Crossing the barrier to replication is an important step logically and temporally following the emergence of translation. In the emergence and evolution of genetic information system, there is a clear trend of dimensionality decrease for pattern generator and dimensionality increase for functional performer: in the prebiotic RNA world, RNA can have primary, secondary, and tertiary structures, but only the secondary structure is stable. So 2-D structure is the basis of RNA's intermediate role between pattern generator and functional performer[65]. In contrast, protein tertiary structures are much more stable than secondary structure, which is consistent with its enhanced functional action[65]. As a specialized pattern generator/storage, DNA mainly works at stable primary structure, namely at 1-D state (Fig. 7B).

The dichotomy of genotype and phenotype brings at least five other advantages. First, 1-D genetic pattern makes the double strand structure of DNA possible, which not only adds redundancy to genetic domain but also reduces the mutation rate and the attendant mutational bias through pairing of the bases on DNA double strands[66-73]. Second, the carrier of genetic information can be segregated, transmitted, and incorporated both horizontally and vertically; this can expedite evolution, particularly at the early stage of evolution. Third, the source domain evolution produces not only coding sequences, but also non-coding sequences, such as pseudogenes. The non-coding sequence can provide an information reservoir for host evolution[74]. Fourth, the patterns provided by DNA are not limited to the coding sequences, as the regulatory function of non-coding sequence provides additional patterns for the complexity increase[75-78]. The relational patterns performed by the regulatory sequences are mapped to the relations between proteins and thus increase the complexity and fitness of the host, although regulatory sequences do not have direct translational product. As a result, the other two steps of genotype-phenotype mapping, namely protein folding and organization/hierarchization, are also encoded as genetic information. Fifth, the convergent mapping, namely multiple configurations/patterns in pattern domain mapping to one configuration of functional domain, can bring robustness to biotic evolution, for example the mutational robustness resulting from the neutral network[11, 22]. Such robustness is a property of the mapping, not the stability of either domain alone, but it results in stable phenotypes in biotic evolution. Because genotype-phenotype mapping can be fine-tuned in biotic evolution to gain advantages without affecting the pattern and functional domains, the universal conflict of evolution can be further relieved through fine-tuning the rule of mapping.

### The rule of translation further reduces biases but keeps the dynamics of evolution

Although the inert property and the attendant smooth landscape are beneficial for unbiased pattern formation, a certain level of activity is required for pattern formation. First, complete inert entities, for example as inert as helium, evolve extremely slowly; namely, no activity drives evolution. Although slow evolution can preserve patterns, it is adverse to the generation of novel patterns. It must be emphasized that configuration/pattern preservation, namely stability/robustness, is only one side of evolution; the other side is novelty, which is essential to evolution. Stability/robustness of genetic material/code is a double-edged sword to blind evolution, especially in the origin and early stage of life and in the evolution in a changing environment when novelty is needed. Second, the pattern domain requires a certain level of activity for operation, for example the recognition of patterns during evolution. As a result, the pattern generator must have a certain level of activity and the consequent biases in pattern formation. Therefore, although the landscape of DNA/RNA is smoother than that of proteins, it still has rugged terrains due to various biases. For example, in animal nuclear genomes, transitional mutations, i.e. purine to purine or pyrimidine to pyrimidine, occur twice as frequently as transversion, i.e. pyrimidine to purine or vice-versa[21]. The GC mutational pressure is another example[21]. The GC mutational pressure also exists in the origin of life as well as at the late stage of evolution: cytosine is less stable than other nucleotides, which makes the first genetic material bias to AU[79]. Such biases cannot be eliminated by decreasing the bias in the activity of pattern generator without impairing the running of evolution, because biases drive evolution. One subtle but important role of genetic mapping is to convert the landscape of DNA/RNA to a smoother landscape of genetic information without affecting the dynamics of evolution.

Different nucleotides and sequences have different stability due to their physicochemical property, and thus different appearance rates in the genome. Here, the stability is a comprehensive stability which results from various environmental actions, not only a thermostability. On the one hand, the differential stability of nucleotides and sequences makes the landscape of genetic materials rough and thus biases the patterns generated by DNA evolution. On the other hand, such a difference in stability is a prerequisite for DNA evolution, because a completely smooth landscape does not have evolution, as water does not flow on a flat surface. A reaction must be biased to one direction to produce effect. This conflict between unbiased pattern formation and the driving force of evolution is also a manifestation of the universal polarity of evolution. Triplet genetic code of translation solves this conflict by converting (coarse graining/canalizing, see chapter V) the landscape of DNA to a smoother landscape of genetic information without impairing the driving force of genetic evolution.

First, the degeneracy of the genetic code makes the coding of most amino acids redundant. As a result, many mutations are synonymous. The so-called fault-tolerance is actually a conversion of a rough terrain on the landscape of DNA/RNA to a flat terrain on the landscape of genetic information (Fig. 5). For example, CUG and CUC are more stable than CUU; CUU tends to mutate to CUG or CUC, represented by a slope on the DNA landscape. However, to genetic information, that represents a relatively flat region even if the codon usage bias is included, because CUC, CUG, and CUU all map to leucine (Fig. 5). Different amino acids, for example leucine and phenylalanine, may have genetic codons of similar stability, which converts the rough protein landscape to a relatively smooth DNA landscape. Meanwhile, such a conversion does not affect the rate of DNA mutation, which is the underlying driving force of informational evolution. Other types of fault-tolerance of genetic code, such as translational errors[80-82], have the same effect in smoothing over the rugged landscape.

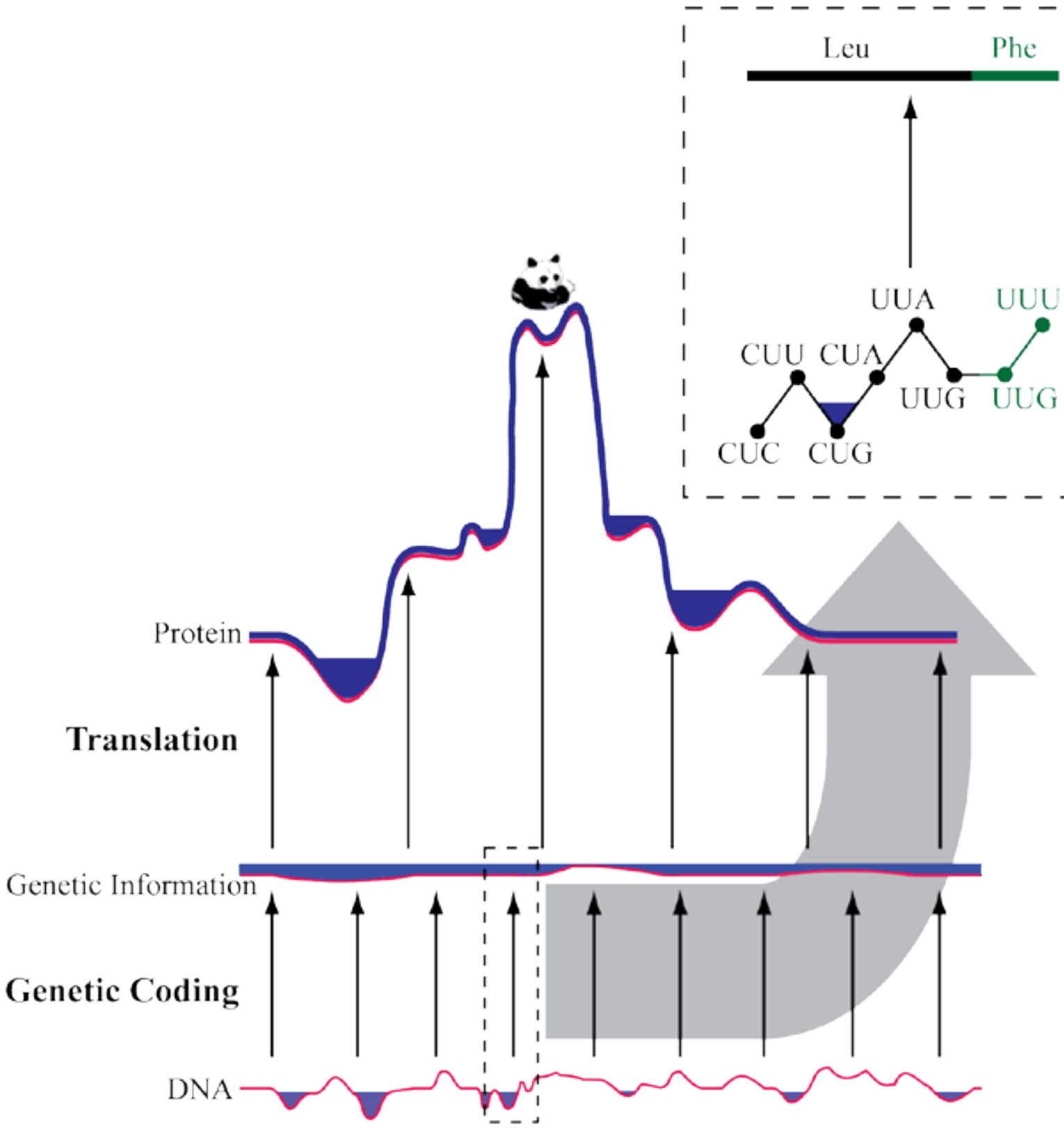

**Fig. 5. The advantage of genetic coding.** The differential stability of nucleotides makes the DNA landscape rough: mutations are biased to stable nucleotides. Because of the degeneracy and fault-tolerance of the triplet genetic code, some mutations are synonymous. In this sense, degeneracy and fault-tolerance smooth out the roughness of the DNA landscape partially. Thus, the evolution of genetic information is less biased than that of DNA. The direction of GC pressure varies in different species and at different stages of life history.

Second, the composition of genetic codons is tuned to smooth out the differential stability of nucleotides in genetic mapping. The GC ratio of the genome varies greatly both among different species and within the individual genome. It is demonstrated to be the consequence of mutational bias[83-84]. Interpretations based on selectionism have been refuted[85-87]. Because of the nature of the triplet genetic code, the GC ratio of the codons of different amino acids varies greatly (Fig. 6A). Therefore, the GC mutational pressure can influence the composition of proteins: the ratio of the amino acids with high and low GC ratio in their codons correlates with the GC ratio of the genome positively and negatively, respectively[83]. However, the genetic code is tuned that GC ratio is similar in various functional groups of amino acids. Specifically, the basic, acidic, polar, and nonpolar groups of amino acids all have a GC ratio that is very close to 50% (Fig. 6B). It is very unlikely that this phenomenon occurs by chance. In view of its effect on protein function, the GC ratio very close to 50% in all groups should be shaped by the selective pressure on the unbiased pattern formation for protein synthesis.

Consequently, although GC mutational pressure influences the ratio of individual amino acids, the functional composition of proteins is not affected because of the similar GC ratio in all functional groups of amino acids. In this way, the roughness of DNA/RNA landscape is smoothed over by the fine-tuned mapping rule. The differential stability of nucleotides in RNA also exists in the origin of life[79]. Under the selective pressure for unbiased pattern formation, genetic code is tuned to smooth out all strong mutational biases in the origin and early evolution of life, such as the bias to transition and the GC mutational bias. Actually, the genetic mapping rule is a form of coarse graining/canalization, another type of pattern transformation (chapter V). Other shaping forces of genetic code include the modulation to expand the genetic code to synthesize more amino acids[88], and the tuning for splicing, localization, folding, and regulation[82]. At later stages, the genetic code becomes fixed gradually and its tuning becomes very weak or halted. ***According to this explanation, genetic information is fundamentally different from the pattern of DNA, the carrier of genetic information, because of the function of mapping rule.***

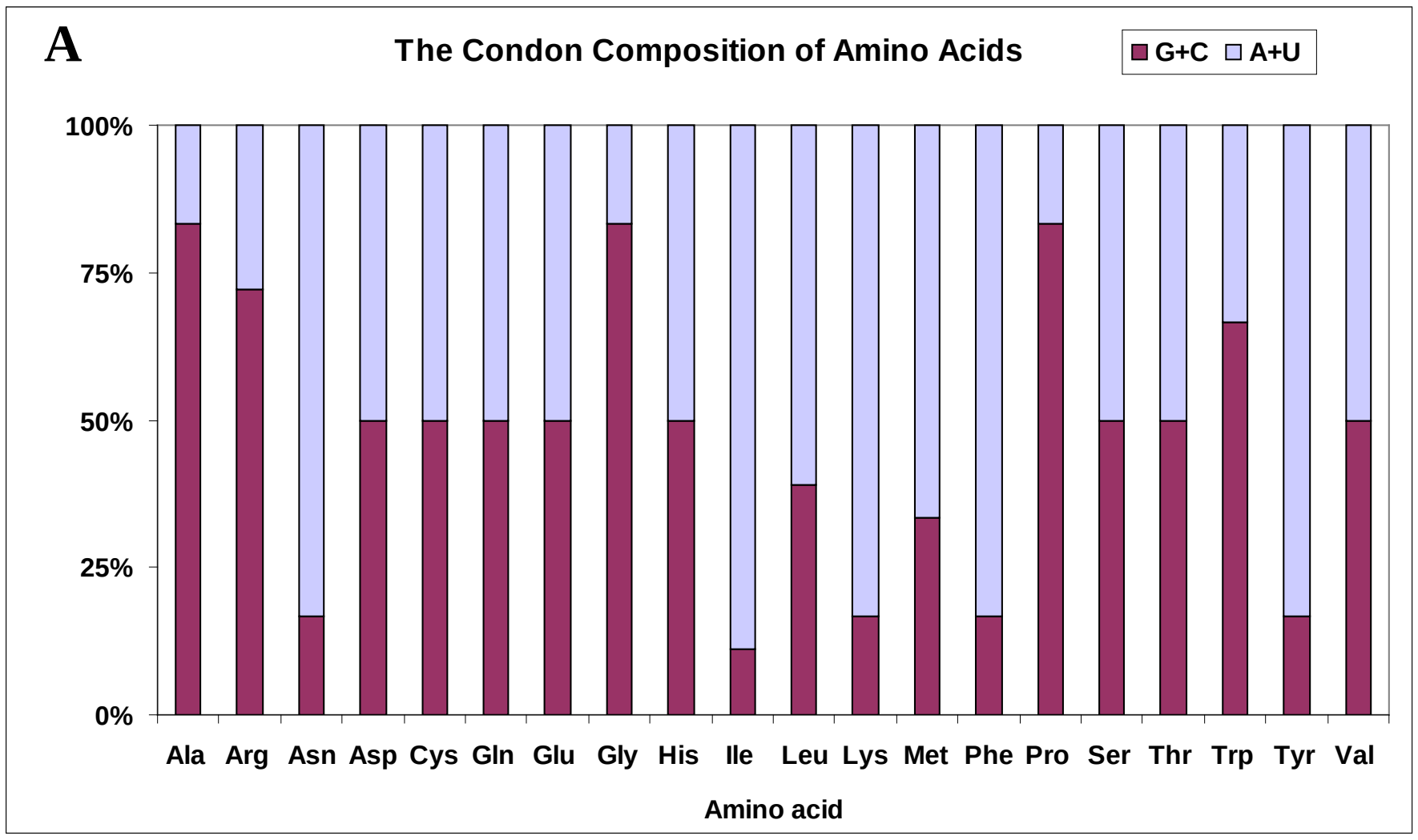

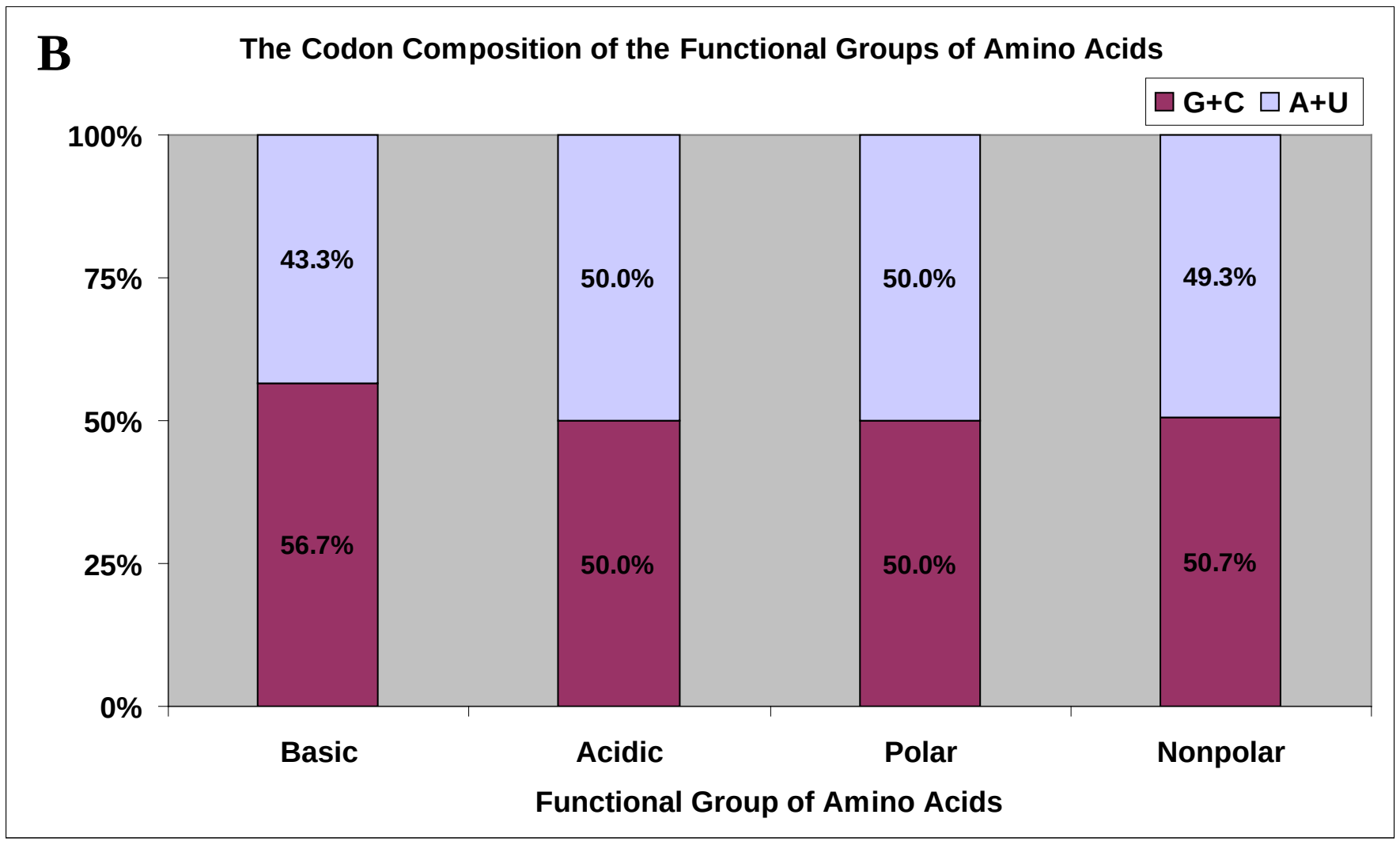

**Fig. 6. Genetic code is tuned to smooth out the differential stability of nucleotides. A.** Due to the nature of the triplet genetic code, the GC ratio in the codons of amino acids is not evenly distributed. Therefore, the GC mutational pressure biases genetic information to a specific group of amino acids. **B.** However, the GC ratio is almost evenly distributed in the codons of four functional groups of amino acids. Therefore, the influence of the bias on protein function is reduced by the fine-tuned genetic code.

## The essentials of genotype-phenotype mapping

The real genotype and phenotype spaces and the mapping between them are much more complicated than those illustrated in the figures of this paper[33, 36-39]. Moreover, some factors in evolution are out of the control of genetic mechanism: phenotype is determined by genotype and epigenotype, while fitness is determined by phenotype and environment (Suppl. Text 3, *Genotype, phenotype, and epigenotype*). The sequence space of DNA or protein has a tremendous number of dimensions and each dimension has a very small number of variables. The phenotype space, namely the configuration space of life, is even more complicated than the sequence space. Genotype-phenotype mapping has three steps, translation, folding, and organization/hierarchization, each of which has different complicated properties. However, in spite of the extreme complicatedness, heterogeneous genotype-phenotype mapping has some common characteristics.

Generally, the essence of heterodomain mapping is a stable causal chain that transforms the patterns in one type of evolution to the patterns in another type of evolution. All interactions can be viewed as a mapping event, but most of them are transient: they do not have fixed source and target domains or stable mapping rule. Only a **stable causal chain** can serve the function of mapping. Besides stability, the **heterogeneity** in evolutionary landscape between two ends of causal chain affects its effectiveness: the source domain serves as a pattern generator, so its landscape should be as smooth as possible, while the landscape of the target functional domain should be as rough as possible. ***Even if the pattern and functional domains are not as ideal as above, heteromapping still provides advantage as long as the two domains are different: the greater the heterogeneity between two domains, the greater the advantage is***. Unidirectionality is not an essential for heteromapping. Under some conditions, heteromapping can be bidirectional, especially in a heteromapping which is based on another unidirectional heteromapping. Lower-level heteromapping can calibrate higher-level heteromapping. Therefore, only the bottom heteromapping has to be unidirectional. For example, perception and motion are the bidirectional neural mapping between neural system and environment, which is based on the genetic mapping. Although heteromapping can be bidirectional, the energy flow must be consistent with the direction of mapping: the energy flow within either direction must be unidirectional. **Unidirectional energy flow** thermodynamically stabilizes the direction of causal chain.

The stability of the causal chain underlying heteromapping can be fine-drawn to four essentials. First, heteromapping must be unambiguous. Specifically, the units in the source domain have at most one correspondence in the target domain. Some patterns in the source domain play a role in the regulation and organization of the source domain, and thus may not have direct translational product in the target domain; however, these patterns still have corresponding patterns in the target domain: these corresponding patterns are the relations between proteins, not protein sequences. In real condition, genotype-phenotype mapping includes translation of linear proteins (heteromapping), folding of linear proteins, and organization/hierarchization of folded proteins to generate complex functions. In order to prevent multiple mapping, not only heteromapping and protein folding are unambiguous but also the noise in organization/hierarchization is smoothed out by coarse graining/canalization to achieve unambiguousness (2nd section of chapter V, *Coarse graining/canalization reduces the noise in genotype-phenotype mapping for unambiguousness*).

Second, the code of heteromapping must be temporally stable. Any change in the code will result in the global loss of accumulated complexity. The code can only be optimized at the very early stage of evolution when the heterodomain mapping is still very crude, and then the code becomes fixed rapidly.

Third, the code of heterodomain mapping must be spatially uniform to the whole source domain and target domain. Otherwise, different parts of life using different codes will evolve separately and will be in conflict to one another. Such conflict either inhibits the complexity increase of all parts or finally leads to one dominant part enslaving other parts.

Fourth, consistent with the above essentials, all entities in the source domain, either elementary or compound, must have a unique and defined relation with all other entities during heteromapping. Although the relations can be changed, the relations in one event of translation or regulation must be fixed. Otherwise, uniform and stable heteromapping cannot be achieved. This characteristic ensures that reading the source domain by the heteromapping machine is determinate rather than haphazard. In the translation and genetic regulation of genes, the determinate reading is achieved by sequential scanning of linear DNA. Determinate reading is an essential characteristic of information. In other words, information, irrespective of its form, must be indexed or addressed in order to be used to increase the fitness and complexity of its user in evolution; all informational entities, either elementary or compound, have only one unique position in the index, although this index is alterable. Transient and regional errors in the index are possible and their effect is similar to mutation, but the index must be basically unambiguous and stable in long term; otherwise, any persistent or large-scale error in the index will severely jeopardize the integrity of the host entity, let alone the fitness or complexity. The collection of informational entities using the same index system forms an evolutionarily and functionally distinct source domain, for example the genome of terrestrial life.

These essentials seem self-evident and trivial. However, their effects are fundamental and far-reaching and can be found in various forms of heterodomain mapping.

## The origin of translation and transcription

In the origin of life, there has been a puzzle due to the replication error. The non-enzymatic replication of nucleic acids has a certain rate of error, and that limits the length of the whole genome to 100 or less nucleotides. To increase the genome size, a proteinaceous replicase is required. However, a genome coding for such an enzyme would be much more than 100 nucleotides. This “catch-22 of prebiotic evolution (Eigen's paradox)” is considered as an inevitable problem in all early replication systems irrespective of the form of template[24].

The cause of this “catch-22” is the misunderstanding of what a gene actually is in the conventional replication-first theory. Without a translation system, replication of nucleic acids or any other templates is only an abiotic replication, which is fundamentally the same as crystal growth. Translation is the basis of genetic heredity. The nucleic acids become a gene because of translation rather than replication. Emergence of any complexity beyond the limit of abiotic evolution, for instance, a high fidelity proteinaceous replicase, needs the participation of the translation system, while primitive translation does not require the replication system[80, 88]. Although it has been found that translation precedes replication through phylogenetic analysis[89], current understanding of evolution cannot explain that. Here, the order of emergence of the components in genetic information processing is explained unitarily and parsimoniously by the present theory.

Before the emergence of translation, RNA, as a pre-gene, plays a similar role as the protein - a functional performer. In the prebiotic RNA world, RNA and small peptides bind together to obtain more structural and functional capabilities than each alone[24, 65, 90]. The primitive translation mechanism develops from this functional association, which does not require complex proteins[88, 91]. Since the essence of translation is to generate pattern through heteromapping, the mapping rule does not have to be fixed and the mapping does not have to be precise at very early stage. Fidelity of translation is not important in the origin and early evolution of life when all proteins are crude and thus many translational errors are beneficial. Even a very primitive translation system provides significant selective advantage by producing larger and less biased proteins than non-translated primitive proteins, and that is the basis of better function for proteins. The protocells with functionally improved proteins are selected for, while those protocells without improved proteins are selected against. This improvement in protein function could feed back to the translation system, which then produces second-generation proteins with further improved function. Finally, the translation system could produce proteins whose functions are sufficient to resolve the “catch-22 of prebiotic evolution” in the emergence of replication. Primitive translation is a bootstrap in this process and thus avoids the deadlock of the replication-first theory (Fig. 7). This scenario is consistent with the RNA world theory[65, 80, 88].

Although an RNA genome and replication might be common in the origin and early evolution of life, the DNA genome and replication is the principal form of current life. The step after translation is the transition from RNA to DNA. The existence of this transition is supported by many reports and is widely accepted[24, 89, 92]. Why does DNA emerge at first place? It is argued that the enhanced chemical stability of deoxyribose and thymine in DNA, as opposed to the ribose and uracil in RNA, is the main selective force for the RNA to DNA transition. Unlike RNA, stable DNA can be replicated more faithfully and open up the possibility of large genomes[24, 92]. However, the size of genome is only one of the factors. Another more important reason is that the evolutionary landscape of DNA is smoother than that of RNA; thus, the patterns generated by DNA evolution are less biased than the patterns of RNA evolution. At the very early stage after the emergence of translation, DNA emerges and generates patterns for protein synthesis through bidirectional transcription between DNA and RNA. The complexity and function of proteins are further improved because of the enhanced diversity of DNA patterns compared with the RNA pattern, and the less biased sampling of genotype-phenotype space (Fig. 7). This benefit is immediate, while the benefit of large genome is realized slowly.

Because reverse transcription transforms the RNA pattern to the DNA pattern, the diversity of the DNA pattern is impaired by the incorporation of more biased RNA patterns. Therefore, after the DNA genome emerges, organisms with reverse transcription are eliminated by the selective pressure on unbiased pattern formation (Fig. 7). That is why only primitive viruses use reverse transcription as a principal component of information processing, as only some viruses use relic U-DNA in their genome[92-93]. In contrast, retrotranscription plays a very minor role in the evolution of cellular organisms. Derived from retrovirus, retrotranscription in cellular organisms only occurs inside the virus-like particle produced by retrotransposons, and serves mainly as a mechanism of transposition and duplication of DNA entities inside DNA genome. Functional gene duplication through retrotransposition is much less than the “normal” gene duplication[94]. Most products of retrotransposition have no function. Therefore, the role of retrotransposition in evolution is very similar to that of mutation. Moreover, the evolutionary benefits of retrotransposition are mainly due to the structure of mRNA[94], rather than the relatively rough landscape of RNA. For example, retrotransposons do not have regulatory elements and have to recruit new regulatory elements; this character makes retrogenes more likely to gain new expression pattern and thus new function; the retrogenes also evolve more efficiently, because the evolutionary constraint imposed by splicing signals is relaxed by the intronless structure of mRNA[94]. ***The paucity of retrotranscription cannot be explained merely by the higher stability of DNA than that of RNA, because retrotranscription does not affect the stability of DNA. A plausible explanation is that pattern formation by DNA the less biased than that of RNA.***

The complexity and function of proteins are progressively improved by these transitions. This improvement prepares for the emergence of proteinaceous enzymes that are capable of accurate replication of DNA. Providing time and the selective pressure on protocells, genetic heredity would finally emerge and reach its current state. Studies of the universal phylogenetic tree showed that the order of emergence of the components in genetic information processing is translation first, then transcription, and finally replication[89]. This investigation supports the present theory of the universal polarity and the consequent division of internal evolution to pattern formation and functional action (Fig. 7).

The emergence of asymmetrical unidirectional translation from the symmetrical RNA-protein association is actually the symmetry breaking in physics, which is fundamentally similar to the differentiation in biology. This symmetry breaking at the molecular level is the core of the central dogma. The understanding of the horizontal symmetry breaking in the central dogma can be extended to the vertical symmetry breaking in hierarchical life.

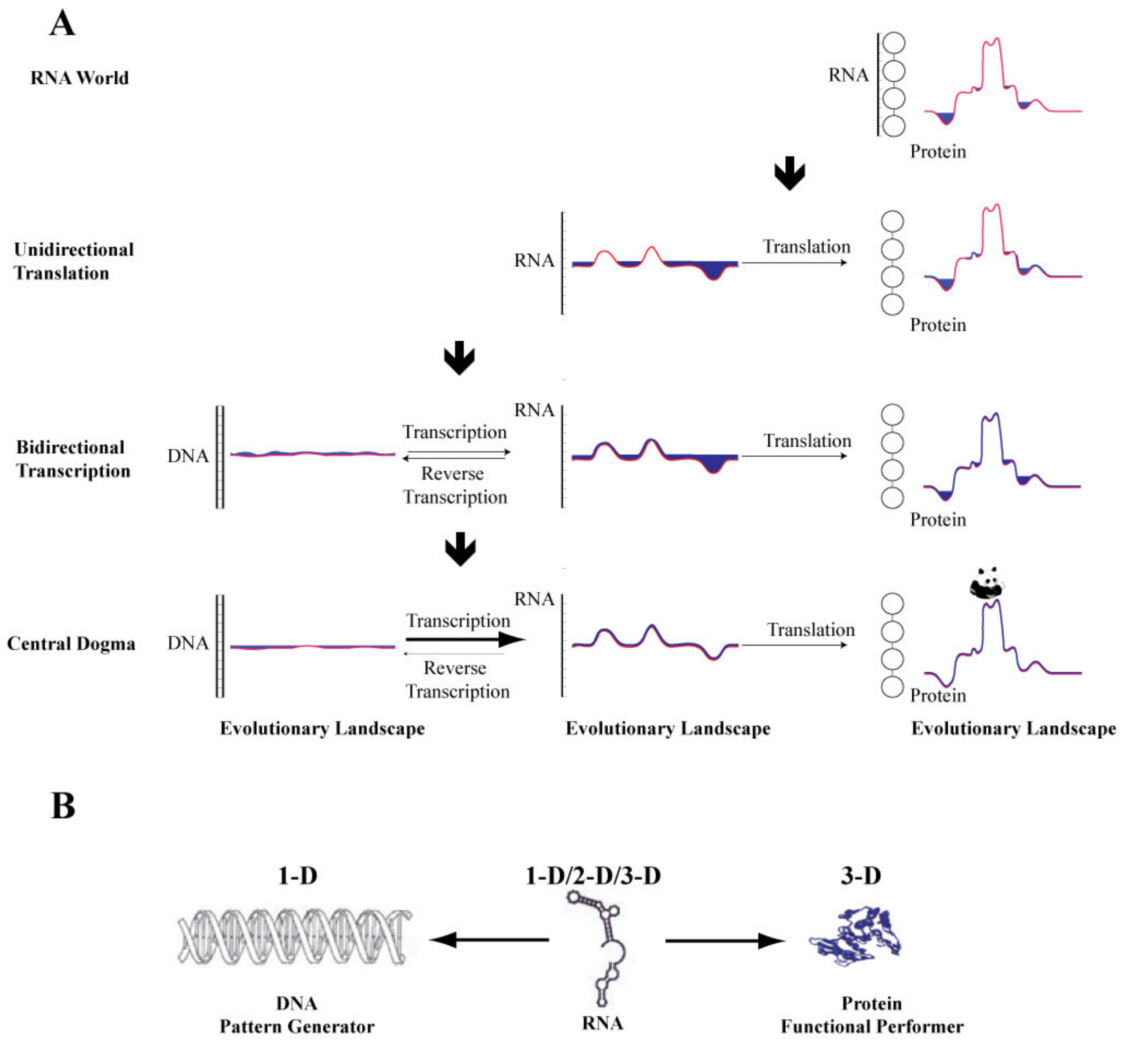

**Fig. 7. The origin of translation and transcription. A.** The pattern generator of life shifts from protein to RNA and finally to DNA in the origin and early evolution of life. The shifting reflects that the evolutionary landscape of pattern generator becomes more and more flat, and that is driven by the selective pressure for unbiased pattern formation and the consequent unbiased sampling of the phenotype space of life, i.e. the configuration space of life. In the RNA world, RNA/protein complexes are both the functional performer and pattern generator. In the origin of life, the RNA/protein complex differentiates to the RNA as a pattern generator and the protein as a functional performer. The mapping between the RNA patterns and protein products is unidirectional in order to block highly biased protein patterns interfering with RNA pattern formation through retrotranslation.

The constraint in pattern formation is further reduced when pattern formation is shifted from RNA to DNA through bidirectional transcription. Reverse transcription maps the more biased RNA patterns to the less biased DNA domain and thus is harmful to the unbiased phenotype space exploration. Therefore, reverse transcription as a principal flow of information is only present in some borderline evolutionary entities between nonlife and life, for example some viruses. **B.** The change in dimensionality during the origin of translation and transcription. As a primitive functional performer and pattern generator, RNA works at 1-D, 2-D, and 3-D states. The specialized pattern generator, DNA, works at 1-D state, while the specialized functional performer, protein, works at 3-D state. Such change in dimensionality is consistent with the theory of the spatial constraint on *en bloc* evolution.

## The essence and the extension of the central dogma

The central dogma of molecular biology (Fig. 8) was proposed in 1958[95] and restated in 1970[96] by Francis Crick. It states that "once information has got into a protein it can't get out again[95]" or "information cannot be transferred from protein to either protein or nucleic acid[96]". In addition to this explicit meaning, the central dogma implies that the transfers from RNA to DNA, from RNA to RNA, and from DNA to protein are minor while only the transfer from DNA to RNA to protein is principal[96](Fig. 8). Although the central dogma is one of the keystones of molecular biology, it has only a definition but no explanation so far. Therefore, there are some confusions about the central dogma because of the misunderstanding of the central dogma[97]. Some biologists consider the central dogma no longer valid after some phenomena, for example RNA editing and prion, have been discovered[97]; some consider that the central dogma is needless to explain and/or unexplainable because it is a starting point of molecular biology, like a mathematical axiom that is neither deducible nor demonstrable. Here, the author explains the cause of the central dogma and extends the central dogma from molecular biology to evolutionary biology.

In cellular organisms, the mainstream of informational flow is the unidirectional flow of genetic information from DNA to RNA to protein. Only the retrovirus uses the flow from RNA to RNA or DNA as the mainstream of informational processing. The flow from DNA to protein is only able to be performed in the *in vitro* cell-free system. The specific direction of informational flow is not a frozen accident. In stead, it is a necessity of the labor division to pattern formation and functional action in life.

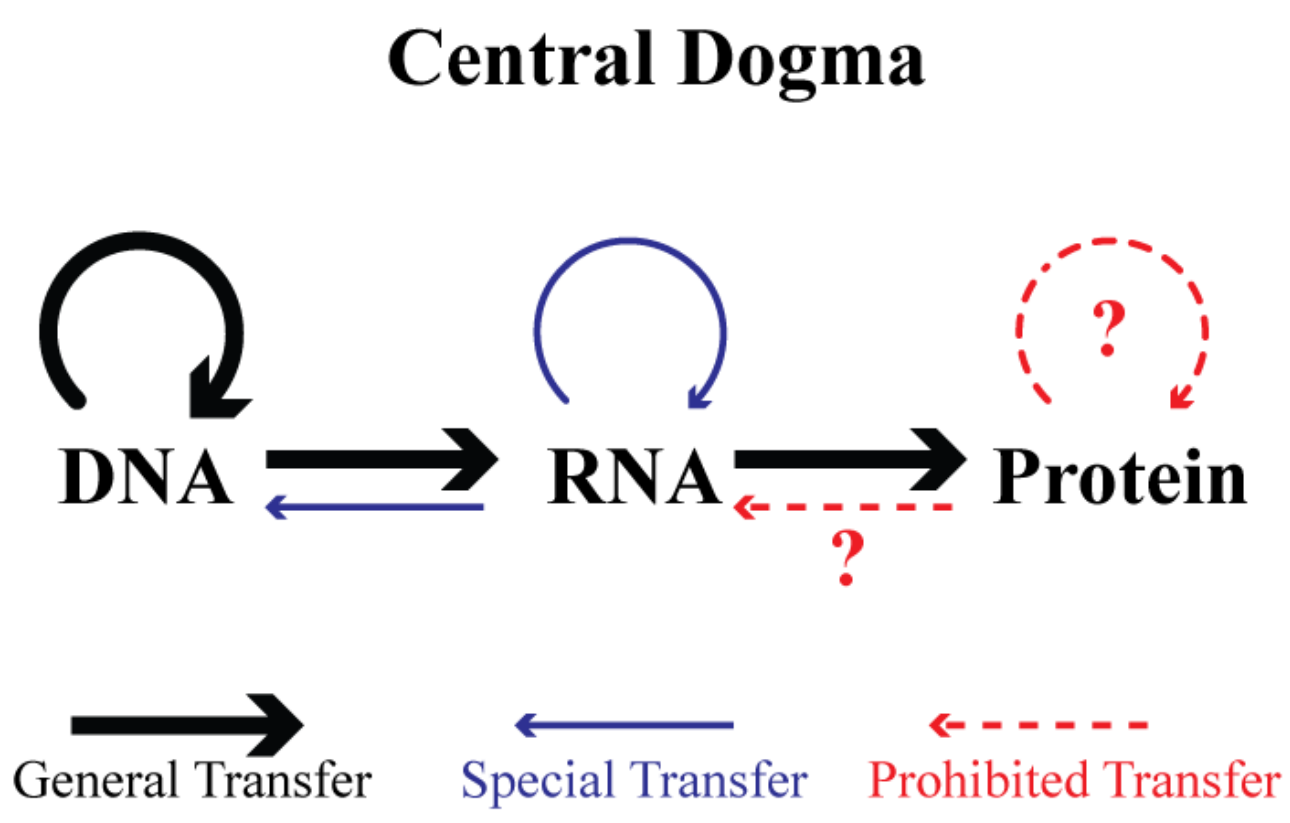

**Fig. 8. Why central dogma?** The transfer of genetic information from DNA to RNA to protein is the principal informational flow in life. The special transfer from RNA to DNA or RNA as a mainstream of information flow only occurs in the primitive forms of life, such as retroviruses. The absence of the information flow out of protein is a necessity rather than a frozen accident of biological evolution. This necessity reflects the importance of unbiased pattern

formation in the complexity increase in evolution. Drawn according to Crick F. (1970): Central Dogma of Molecular Biology. Nature 227, 561.

Heteromapping can be bidirectional. Why is genetic translation unidirectional? The precursor of translation, the association between RNA and proteins, is symmetrical[24, 65, 90]. During the origin of translation, bidirectional mapping may briefly exist. Even in the modern cell, it is possible that a protein is unfolded to a linear state and then retrotranslated to RNA with the aid of enzymes. The universal unidirectionality of translation in terrestrial life is not accidental. Instead, it protects the patterns in DNA/RNA from the influence of protein evolution. Therefore, the advantage of DNA/RNA in unbiased pattern formation is preserved (Fig. 9).

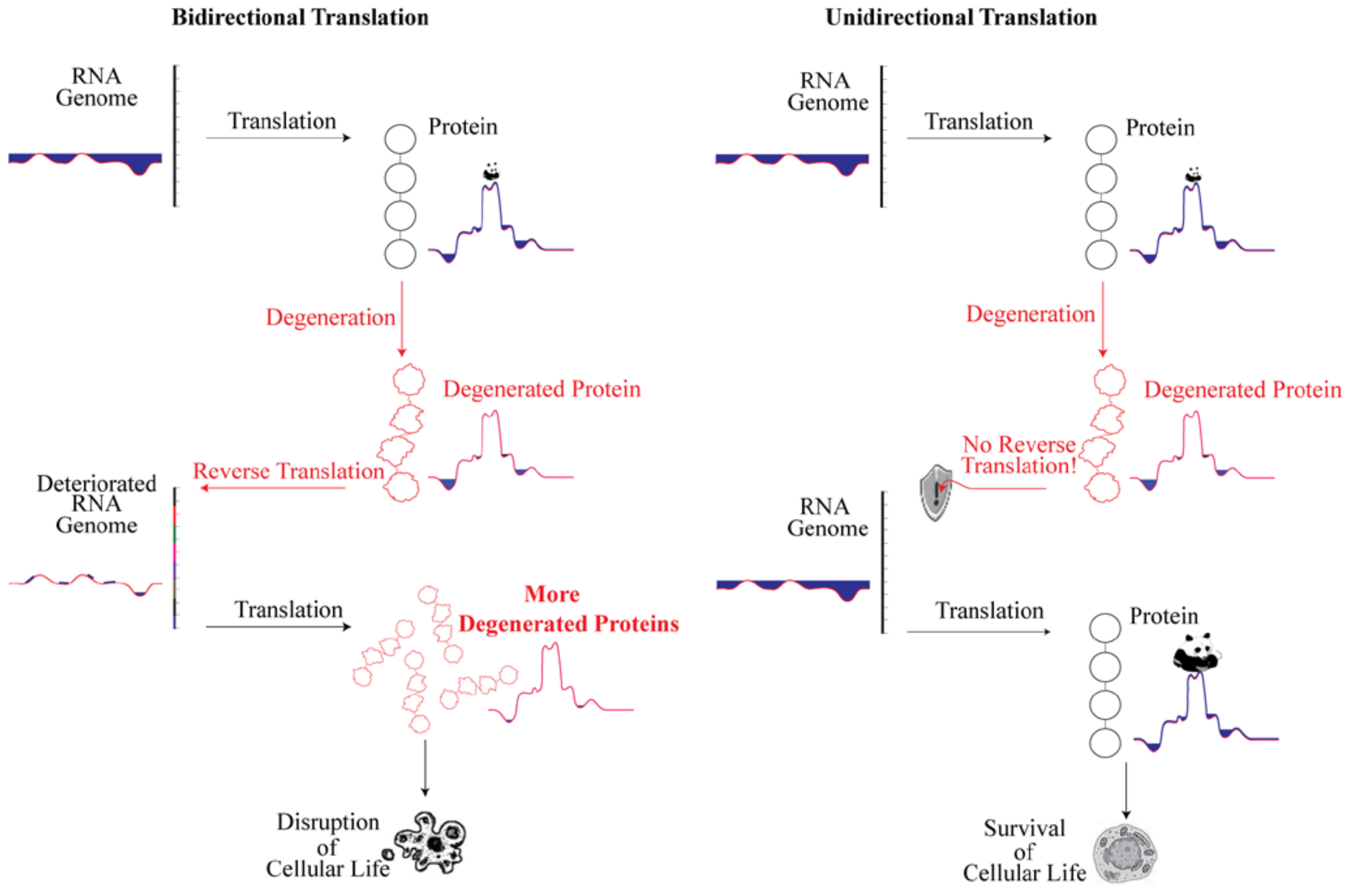

**Fig. 9. The unidirectionality of translation.** Protein is more active than DNA. Therefore, protein is less stable and more susceptible to degeneration, which is represented as the low-altitude valleys on the evolutionary landscape (blue area). Reverse translation feeds back the sequence information of degenerated proteins to genome. Such deleterious genetic information is then translated to degenerated proteins. In this way, protein degeneration will be magnified and the inundation of degenerated proteins will destroy cellular life. On the one hand, the unidirectionality of translation ensures the separation of pattern formation and functional action. On the other hand, the unidirectionality decouples genetic information from proteins in natural selection, and thus consolidates the coupling of genetic information to the host cell, which sustains the integrity of cell and increase the complexity of cellular life.

Specifically, retrotranslation brings many serious disadvantages. First, according to the second law of thermodynamics, everything tends to go towards the equilibrium state unless negative entropy is consumed. Therefore, protein degeneration is inevitable. Because proteins are more active than DNA, and DNA information has fault-tolerance, redundant duplex structure, and a repairing system, protein degeneration is much severer than DNA degeneration. If degenerated proteins are retrotranslated into genetic information, the genome will incorporate the sequence information of degenerated proteins, and, in turn, the genome will be translated to produce more degenerated proteins, and so on. In this way, both genome and proteome will deteriorate quickly (Fig. 9). Bidirectional mapping connects two domains of evolution, which affect and incorporate each other. As a result, the degeneration of functional domain would ruin both domains and consequently the whole cell. Second, there is competition among individual proteins. This internal competition would be magnified through retrotranslation and translation, and finally become out of control. For example, a proteinase can degrade other proteins, and that will be magnified by retrotranslation and translation. Because protein competition is also a manifestation of thermodynamics in biochemistry, inundation with proteinase or advantaged proteins is a special type of degeneration. Since retrotranslation is lethal to cellular life, it must be transient if it ever naturally occurs (Fig. 9).

The unidirectionality of translation protects the flat landscape of the RNA pattern domain from the erosion of the rugged landscape of the protein functional domain. In this way, unbiased pattern formation and the consequent unbiased sampling of genotype-phenotype space are approached as closely as possible. Retrotranslation is so harmful that it is lethal to all early forms of life. That is why retrotranslation is absent in all existing forms of life. Similarly, retrotranscription is harmful to the patterns in the DNA domain, because DNA has a smoother landscape than that of RNA. That is why only primitive viruses use retrotranscription as a principal information flow. In the evolution of cellular organisms, retrotranscription plays a very minor role as compared to the normal transcription in cells or retrotranscription in viruses. The absence of protein-to-protein transfer is due to the spatial constraint on the replication of 3-D patterns in 3-D space, as explained in chapter II.

Understanding the cause of the flow of genetic information answers the challenges to the central dogma[97]. Proteins can specifically modify DNA/RNA and thus influence the evolution of genetic information, for example RNA editing and intein homing. However, unlike reverse mapping, such modification does not follow the genetic code and thus is meaningless to the genetic information in DNA/RNA. Although proteins may produce sequence-specific changes on DNA, the specificity is not based on genetic code and the consequent changes are not the mapping between pattern domain and functional domain. Such modification never builds a stable and universal mapping between protein and DNA/RNA, as the genetic translation does. The nature and effect of the transformation/selection of DNA/RNA by proteins either is similar to those of physicochemically induced mutations of DNA/RNA or is a proteinaceous extension of the interaction between genes (3rd section of chapter IV). The prion only induces conformational changes in other proteins of the same type; the nature of this phenomenon is the same as a proteinaceous enzyme inducing conformational changes in its substrates or abiotic crystal growth. The central dogma is about the flow of genetic information, so heritable epigenetic patterns, for example DNA methylation and protein conformation, are not in the reach of the central dogma. If we adhere to the original statement of the central dogma, "once information has got into a protein it can't get out again[95]" or "information cannot be transferred from protein to either protein or nucleic acid[96]", no exception to the central dogma has been found so far.

***The essence of the central dogma can be understood from two different angles. First, the unidirectionality of translation is to ensure the separation of pattern formation and functional action.*** The universal unidirectionality of translation in all forms of life proves that the division of labor to pattern formation and functional action is essential to life. The paucity of reverse transcription further shifts and concentrates pattern formation to DNA.

***Second, reverse transcription/translation is a type of coupled selection of DNA/RNA patterns to proteins, which feeds back the fitness of protein to the corresponding DNA/RNA pattern.*** Such feedback at the molecular level is beneficial to proteins at the cost of the host cell. When both translation and transcription are unidirectional, the survival of genetic information couples to that of the host cell rather than the individual RNA or protein. As a result, the complexity of the whole cell instead of individual molecules is increased in natural selection.

***The second understanding of the central dogma can be extended to that the hierarchical level coupled to the genetic information is coupled to different levels in the natural selection of multilevel hierarchies*** (Fig. 10). In a hierarchical entity, when the information couples to a specific level, the complexity of this level is increased at the cost of other levels in evolution. The underlying reason is that the property and fitness of a hierarchical level must be different from those of other levels; configurations/patterns beneficial to one level are usually harmful to other levels (chapter V & VI); for example, the fitness of molecules is different from that of cells, which is the cause of the unidirectionality of translation. Reverse translation/transcription is a type of horizontally coupled selection, because proteins are at the same level as DNA/RNA. Since all forms of life are composed of cells, which is a level higher than molecules, all types of horizontal and downward coupled selection are harmful to cellular life. Only the upward coupled selection can be beneficial.

The suppression of downward coupled selection is the downward extension of the central dogma. As explained in previous sections, genetic information is different from DNA, the carrier of information, because genetic coding can smooth out the rough terrain of the DNA landscape. Therefore, in the upward genotype-phenotype mapping, the carrier of information is lower than genetic information. The fitness of the informational carrier, namely the stability of DNA, must feed back to the carried information. In other words, the mutational bias of the underlying carrier of information must impair the evolution of information: the selection of information couples more or less to the transformation/selection of the carrier of information. In addition to genetic coding, nuclear compartmentation diminishes harmful downward coupling and is responsible for the emergence of multicellularity. Similarly, the central dogma can be upwardly extended in a multilevel hierarchy: coupling genetic information to the top level increases the complexity of the whole organism, while coupling to the intermediate cellular level decreases the complexity of the organism. This upward extension of the central dogma explains the bifurcation of animal and plant through the early specification of germline (Fig. 10). The downward and upward extensions are the topic of chapter IV and VII, respectively.

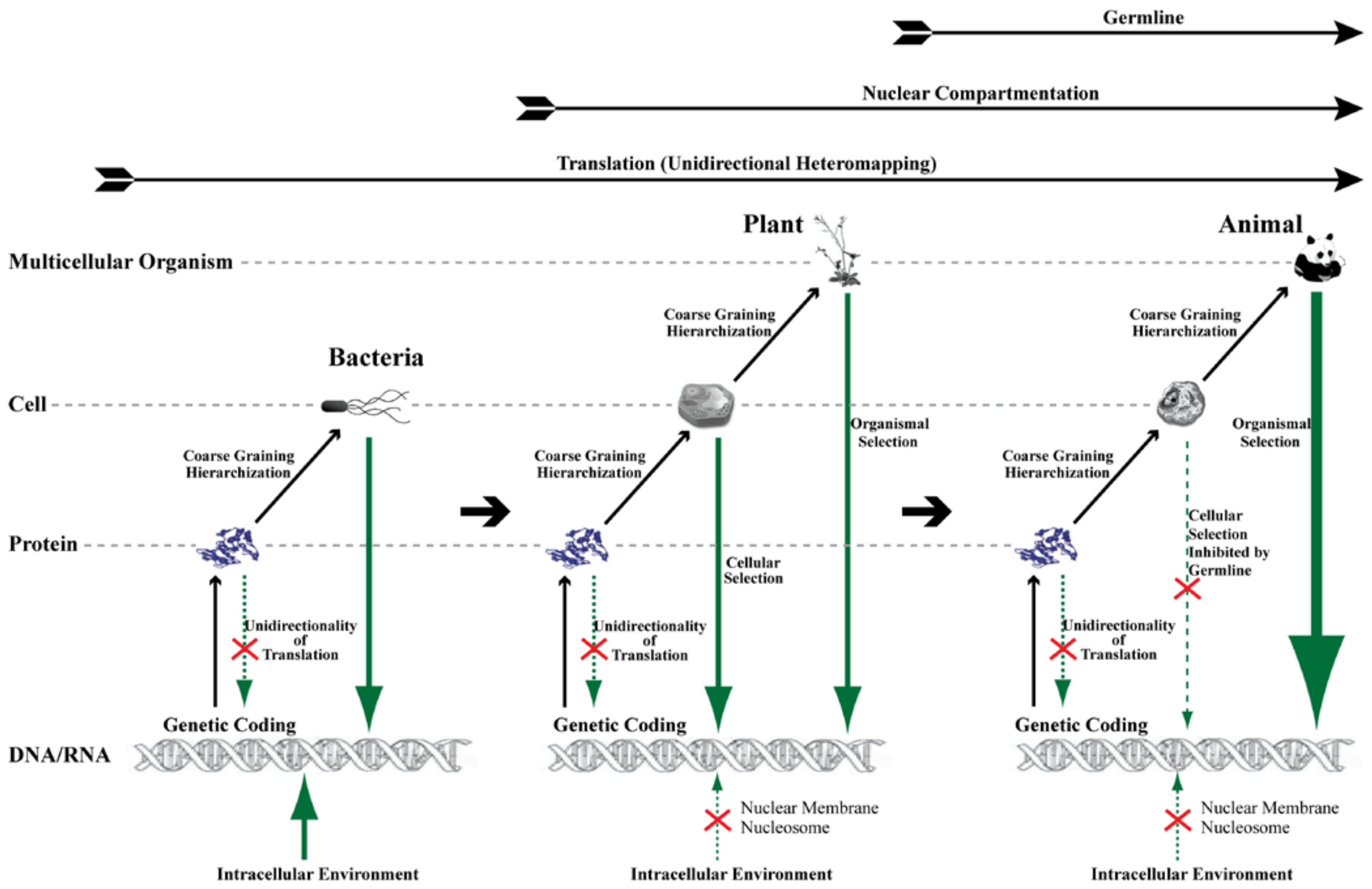

**Fig. 10. The extended central dogma.** As the argument of the central dogma, the unidirectionality of translation is to protect the patterns of DNA from harmful feedbacks of protein evolution through reverse translation. The central dogma can be extended to multilevel hierarchy: coupling of genetic information to one level increases the complexity of this level at the cost of other levels. Therefore, in order to maximize the complexity of the organism, coupling of genetic information to all lower levels should be inhibited. Unidirectionality decouples genetic information from protein evolution; nuclear compartmentation decouples genetic information from the DNA/RNA carrier of information; and, the early specified germline of animals decouples genetic information from the selection at the cellular level. All these mechanisms are crucial in the complexity increase of life. On the other hand, the extended central dogma can be explained by the universal labor division of internal evolution to DNA/RNA pattern formation and protein functional action at different levels of hierarchy.

## IV. The Molecular Interpretation of Darwinism

Natural selection is an essential component of genotype-phenotype mapping, but in a direction reverse to heterodomain mapping. Despite its importance, heterodomain mapping alone cannot increase the complexity of evolution. Functional action needs to feed back its fitness to the corresponding pattern because of the separation of the pattern and functional domains; similar to heteromapping, fitness feedback can be various forms of causal chain. However, in biotic evolution, the fitness of the functional action cannot directly and horizontally feed back to the pattern, because the labor division will be ruined by this horizontal feedback (reverse translation), as explained in last chapter. In this situation, a specialized vertical feedback mechanism has emerged – coupled selection. Briefly, in coupled selection, the integrity of the molecular pattern couples to the survival of the hierarchy of corresponding functional output, i.e. the host organism. In a multilevel hierarchy, the source domain can couple to several hierarchical hosts at different levels. For example, the mitochondrial genome couples to the host mitochondrion, cell, and organism simultaneously. Here, the author only analyze the case of the one coupling with the cell to elucidate the essence of coupled selection; the multiple coupling is discussed in chapter VI & VII on hierarchical evolution. ***Because of the division of internal evolution, the integration of coupled selection with heterodomain mapping is required to link the separated domains. The pattern produced in such a labor division is genetic information, which has greater evolvability than the pattern generated through the undivided pattern formation and functional action in abiotic evolution.***

Generation of information has two integrated components: the heterodomain mapping from the source pattern domain to the target functional domain, and the coupled selection of the source pattern with the functional output (Fig. 11). Without heteromapping, the pattern in an isolated domain does not have functional output, such as the pattern in the isolated DNA. Without coupled selection, the patterns in the source domain are not selected according to the fitness of its user and thus are evolutionarily valueless. Moreover, heterodomain mapping is not only a passive translator. Instead, mapping can modulate the source domain pattern to generate additional advantages for the target domain, as explained in chapter III. It must be emphasized that genetic information is fundamentally different from the carrier of information, i.e. the source domain. Genetic information is the integration of the source domain of informational carrier, mapping, the target domain of output, and coupled selection (Fig. 11). Actually, all forms of information have these four essentials. Genetic information is a primitive form of information, so its origin and composition are more evident that other forms of information which have more complicated origin and composition.

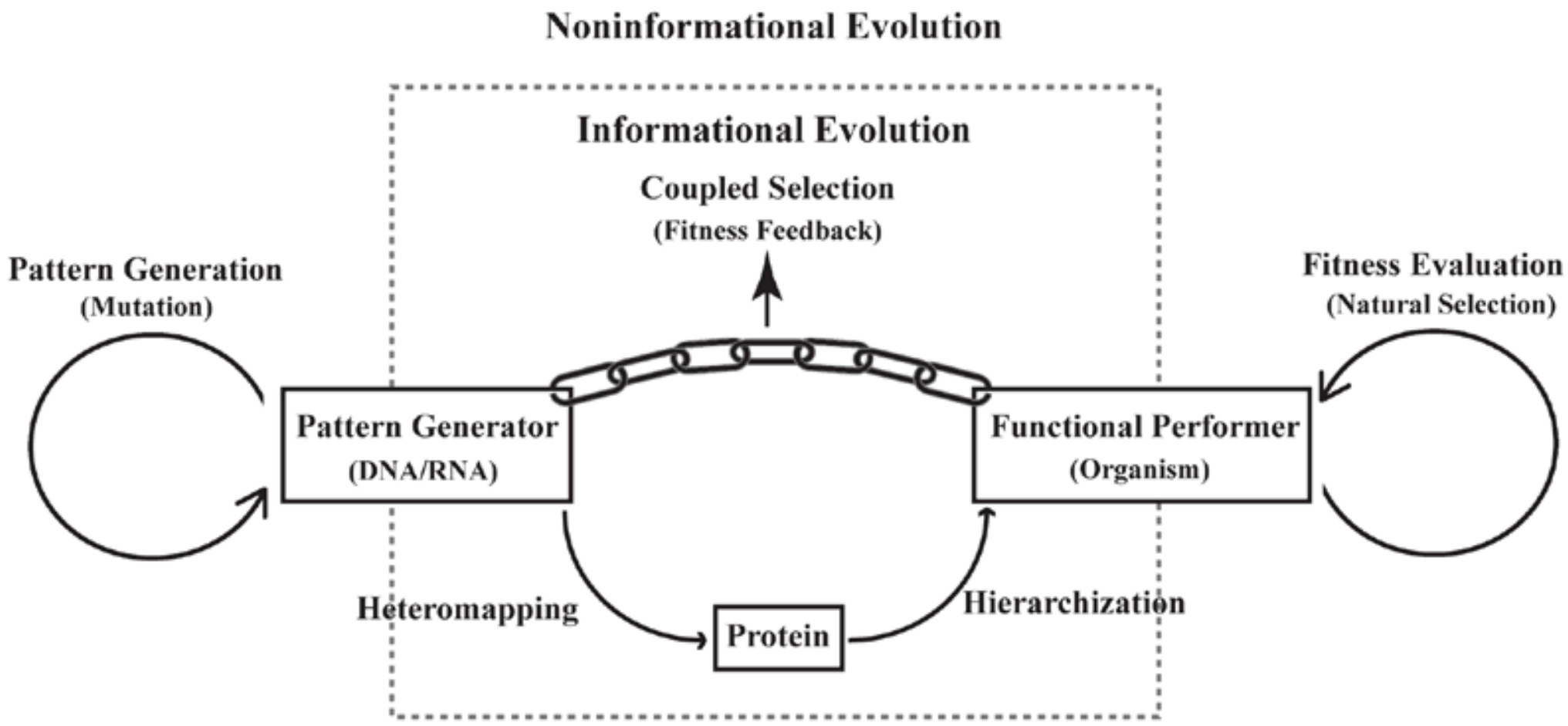

**Fig. 11. What is genetic information and informational evolution?** Genetic information is the patterns generated by the labor division of internal evolution to pattern formation and functional action. Informational evolution is the integration of pattern formation and functional action through heteromapping and coupled selection. DNA is the specialized pattern generator while protein is the specialized functional performer. Such separation requires a bidirectional link between them: heteromapping transforms patterns generated by DNA to protein patterns, while coupled selection of DNA patterns to the host organism feeds back the fitness of functional action to the corresponding pattern. Heteromapping and coupled selection constitute an integral cycle of informational evolution. Isolated evolution of both domains is noninformational. The noninformational evolution of DNA/RNA domain is the source of the patterns, while the noninformational evolution of functional domain is the criterion for the selection of information. Not restricted in the DNA-protein based biotic evolution, informational evolution is widely used to increase fitness and complexity.

## The molecular view of Darwinian selection

As the primitive form of information, genetic information is an illustrative norm of information: the isolated evolution of both or either of two domains of genetic translation is noninformational. Specifically, the mutation of isolated DNA is not informational evolution, because it is the evolution of DNA alone without host complexity or fitness increase. In contrast, the mutation fixed by the selection of organisms during the alternation of generations is informational evolution, because fixation of mutation results from the selection of the whole organism, which includes both the source domain, DNA, the genetic mapping, and the target domain, proteins. Such a distinction is important because ***informational evolution, rather than noninformational evolution, is the major contributor to the fitness and complexity of life***.

The evolution of DNA and protein domains has different roles in genetic information. The noninformational evolution of DNA source domain generates patterns for functional action, while the noninformational evolution of protein target domain is the criterion for the selection of patterns in the source domain. Between the two domains, heteromapping projects the noninformational evolution of the source domain to the target domain, while coupled selection projects the noninformational evolution of target domain back to the source domain. The net result is that the pattern in the source domain is tuned for the fitness of target functional domain (Fig. 11). Coupled selection is the scientific meaning of natural selection in Darwinism, and explains why natural selection, rather than Lamarckian transformation[47-48], is the principal form of environmental action on life.

In terrestrial life, coupled selection is enforced through the dependent internal genome: the internal genome cannot maintain its integrity when the host organism is eliminated in natural selection. In principle, genetic materials can be independent of their users. For instance, an organism can store stable genetic materials outside in another organism or in the environment and access it when necessary. Alternatively, genetic materials inside the host can keep their integrity independently after the host's death and continue to provide information to other organisms. There are many reasons why such mechanisms are disadvantaged, but the most important reason is that even if these mechanisms work very well, the organisms using these mechanisms cannot acquire complexity and fitness through the selection of genetic information. Dependent genetic materials are inside the host only because this is the most reliable way with the least cost to ensure the coupled selection. One consequence of coupled selection is that all genes are selected together as a part of the organism; therefore, Darwinian selection cannot recognize individual genes. In order to increase the resolution and efficiency of Darwinian selection, sex emerges to fine-grain the coupled selection of organisms. Because coupled selection results from the labor division of internal evolution to pattern formation and functional action, sex is a necessary consequence of the genotype-phenotype division. This topic will be discussed in detail in chapter V.

***The informational evolution is Darwinian, while the noninformational evolution is Lamarckian.*** The evolution of the source and the target domains is the superposition of both noninformational evolution and informational evolution. Because one cycle of informational evolution has more steps than one cycle of noninformational evolution, informational evolution and noninformational evolution of the same domain have different constituents and temporal scales. For example, the noninformational evolution of DNA is intrageneration mutations; the lifespan of such noninformational evolution is one generation of the host organism. The informational evolution of DNA is the intergeneration selection of DNA patterns; one generation is only a minimal step of the informational evolution. Informational evolution extends beyond one generation through reproduction. Therefore, noninformational evolution and informational evolution of the same domain are completely different and thus have distinct roles in evolution. Informational evolution has much greater evolvability than noninformational evolution. ***In an evolutionary entity, the higher percentage of information in the total heritable configurations/patterns, the higher evolvability the entity has and the higher complexity will be achieved in evolution.*** Such discrimination between noninformational evolution and informational evolution is crucial to the evolution in abiotic domains as well as biotic domains. ***Informational evolution is a universal mechanism to increase complexity and fitness. Biotic evolution is only a specific form of informational evolution based on DNA/RNA and protein.***

## Why natural selection? - the essence of Darwinism

What is the meaning of selection in Darwin's magnificent theory of natural selection? Natural selection is only one way of organism-environment interaction which drives the evolution of life. Why does selection, rather than other more common ways of organism-environment interaction such as modification or transformation[48], become the principal driving force of biotic evolution? These questions touch the core of life. Evolution is the configuration change of an entity under environmental constraint. However, there is a crucial difference between abiotic and biotic evolution. In abiotic evolution, for example the abiotic evolution of DNA, there is no division to pattern formation and functional action. Any configuration change of an abiotic entity, either small conformational change or eliminative destruction, reflects that that entity evolves to fit the environment; moreover, any configuration change brings novelty to abiotic evolution. In biotic evolution, because of the unidirectionality of translation and other insulations to ensure the internal labor division, configuration changes in functional protein domain cannot change the pattern of DNA, the principal pattern generator/storage at the bottom level. Changes at organismal level, even beneficial changes in response to the environment, are in the functional domain rather than in DNA, and thus cannot be passed to descendants. Only genetic or epigenetic changes in gametes are heritable. However, such gametic changes below the level of organism are tuned for the fitness/stability of lower levels and thus are irrelevant to the fitness of host organism. Therefore, the only way to tune genetic information for the fitness of organism is to eliminate genetic information through the complete destruction of its host. In other words, genetic information is coupled to the death and survival of its user and functional embodiment. Similarly, coupled selection is also the only way to tune genetic information for the fitness of other biotic levels higher than the bottom, such as cells and species. That is why only purely eliminative selection is the principal environmental driving force for biotic evolution among diverse forms of organism-environment interaction.

As a fitness sieve, selection only eliminates biotic entities with low fitness and permits the survival and further evolution of entities with high fitness. The survival and further evolution are the consequence of selection, not the process of selection *per se*. Therefore, selection must be negative elimination, which is not only the scientific and semantic meaning of selection but also the essence of Darwinian selection vs. Lamarckian transformation[48]. Any form of positive "selection" that favors or promotes the survival/existence or reproduction/replication of evolutionary entities is either a metaphor or a mistake, not a concrete form of selection (Please see the case of kin selection in chapter VI). Eliminative selection at one level cannot generate novelty for that level, but may change the configuration of higher levels and thus generate novelty for higher levels. For example, somatic cell death in a multicellular organism may change the configuration and function of that organism, and organismal death may change the structure of the population. However, such configuration changes at higher levels are not encoded as genetic information and thus are not the principal pattern generator in the evolution of life.

Therefore, contrary to the conventional viewpoint that natural selection is creative, natural selection never produces any patterns that are different from the existing ones at the level of selection. This standpoint conforms to the essence of Darwinism vs. Lamarckism: evolution by selecting existing diversity vs. evolution by *de novo* producing diversity. The diversity or variation in the Darwinian theory of evolution is achieved through internal lower-level pattern formation by DNA/RNA. Internal pattern formation, including mutation, recombination, and genetic drift, explores the configuration space and blindly alters the complexity of the host evolutionary entity. Therefore, internal lower-level pattern formation is responsible for complexity change, either increase or decrease of complexity as determined by the specific physical process of pattern formation. However, complexity is only one side of evolution. The other side is stability/fitness: an unstable entity is eliminated and, hence, its further increase in complexity is blocked. In biotic evolution, internal pattern formation is responsible for complexity change, while external environmental action is only responsible for stability/fitness change.

If internal pattern formation is responsible for complexity, what is the role of external environmental selection? As a purely eliminative process, natural selection destroys unstable complexity and permits stable complexity to survive. This eliminative process provides a stability/fitness basis for the further complexity increase upon the survived complexity. If evolution is viewed as blind processes of configuration sampling, environmental transformation/selection is the foothold, but the sampler is the DNA/RNA pattern generator. In abiotic evolution, the trajectory of evolution is determined by the distribution of footholds in the configuration space. In biotic evolution, because of genotype-phenotype division, configuration sampling is performed by the protein domain but guided by the specialized DNA/RNA pattern generator; therefore, both of the distribution of organismal footholds (fitness) and the trajectory of sampling in the DNA/RNA configuration space (DNA/RNA foothold) are required to determine the trajectory of evolution; the trajectory of sampling by DNA/RNA is determined by the landscape of DNA/RNA configuration space: the more flat the landscape, the more even the distribution is and the less bias in sampling. Although environmental action is universal, only specific types of evolutionary entity have extraordinary complexity because they have smoother landscape of pattern generator and thus less biased sampling. Lives in other parts of universe may have very different building blocks and thus have completely different forms. However, the division of internal evolution to pattern formation and functional action must be an essential characteristic to all forms of life. Moreover, coupled selection should also be the principal means of evolution for extraterrestrial life.

## Nuclear compartmentation and gene regulation: the downward extension of the central dogma

Why and how prokaryotes differ greatly from eukaryotes in their evolution is an important but mysterious question to biologists, as Ernst Mayr emphasized in an interview[54]. The role of nucleus, or more precisely, nuclear compartmentation, in evolution is explained by the present theory satisfactorily. As explained in the section on the central dogma, when the genetic information couples to a hierarchical level in natural selection, the complexity of this level is increased at the cost of other levels during evolution. In order to increase the complexity of an organism, the coupling of genetic information to the top-level organism should be maximized while the coupling to lower levels should be inhibited. One of the themes of biotic evolution is the emergence of various mechanisms to reinforce the coupling of information to the host organism and weaken the coupling with lower levels. The central dogma is only a case embodying the principle of coupled selection at the level of molecules.

The central dogma can be extended downward in the hierarchy: any coupled selection below horizontal translation is harmful to life. For example, the triplet genetic code weakens the coupling of genetic information to the carrier of genetic information: genetic coding reduces the bias of DNA evolution to any specific functional group of amino acids. Nuclear compartmentation is another mechanism that decouples information from the environmental transformation/selection of the carrier of information.

As explained in the section on the central dogma, specific modification of DNA/RNA by protein is not a mapping between the DNA/RNA pattern domain and the protein functional domain. Actually, such proteinaceous modification, despite its specificity, acts on the carrier of genetic information rather than the information. In contrast, reverse mapping, such as retrotranslation, is the informational modification of genetic information. Because protein-mediated modification of DNA is noninformational, it does not couple genetic information to proteins in natural selection. Instead, it is the interaction between informational carrier and its environment, particularly its proteinaceous environment: it is a coupling of genetic information to the evolution of informational carrier. Such noninformational modification, like the background noise, weakens the couple of genetic information to the host organism.

Nuclear compartmentation, including nuclear membrane and nucleosome, protects genetic information from unsolicited modification by proteins[98-101] and other physicochemical damages[100, 102-105]. The constraint on the pattern formation by DNA is reduced because the mutation rate and the attendant mutational bias are decreased by the nuclear compartmentation. In other words, the coupling of genetic information to the organism is strengthened because the coupling of information to the DNA carrier is weakened.

Moreover, because of nuclear compartmentation, not only the production of proteins but also their actions on the genome are tightly controlled by the genetic information. ***Since all proteins are the translational output of genes, protein-DNA interaction is actually a proteinaceous extension of intergenic relations.*** Nuclear compartmentation, including both nuclear membrane and nucleosome, plays a central role in gene regulation. Because of the relatively inert activity of DNA, a gene seldom directly regulates another gene: a gene regulates another gene mainly through its proteinaceous outputs, for example transcription factors. Therefore, as the shield and gate of genome, nuclear compartmentation regulates the intergenic relations by controlling protein-DNA interaction. Nuclear compartmentation is an active regulation, not merely a passive separation. Without nuclear compartmentation, protein-DNA interaction is out of control and thus is a noninformational leakage of intergenic relation. ***When protein-DNA interaction is tightly regulated by nuclear compartmentation, the originally stochastic protein-DNA interaction is determined by the output of genetic information and thus is converted to an intrinsic part of intra-genomic relations. In this way, the interaction between the proteome and the genome is encoded as the linear DNA information in genome.*** Because of the advantages of one-dimensional genetic information, the relations between proteins and genes in the form of linear information can be mutated, recombined, selected, and preserved effectively in evolution. In contrast, in prokaryotes, the interaction between proteome and genome is noninformational; the evolvability of the relations in the form of noninformational pattern is much lower than in the form of one-dimensional information. ***This explains why the nuclear compartmentation is very important in the complexity increase of genetic information, which contributes to the emergence and evolution of multicellular life***[100, 104, 106]**.**

This theory explains why the prokaryotic genome is compact and simple. Without nuclear compartmentation, the prokaryotic genome is exposed to the noninformational erosion by proteins and other chemicals that are out of control. Because proteins are the functional embodiment of genes, such erosion actually results in disordered and noninformational intergenic relations. Therefore, the complexity of intra-genomic relations remains very low in prokaryotes. Nuclear compartmentation converts the disordered and noninformational protein-DNA interaction to purely informational patterns. Therefore, eukaryotic genome is large and has complicated internal relations that are finally mapped to the target protein domain. Various non-coding sequences in eukaryotic genome are the result of enhanced evolvability and complexity of genome.

In addition to nuclear compartmentation, sex increases the resolution of natural selection from organism/genome to individual nucleotides, which contributes to the complexity and fitness increase in the form of intergenic relations; otherwise, coarse-grained selection of whole organism/genome cannot shape delicate intergenic relations: it's impossible to draw a Persian miniature with a scrub brush (section II and III in chapter V). ***In prokaryotes, the absence of nuclear and organelle compartmentation suppresses the evolution of intergenic relations, and that limits the complexity increase of prokaryotes. Nuclear compartmentation and sex gradually make the relational changes between genes become the major theme of biotic evolution and the principal form of complexity and fitness increase of eukaryotes.*** The complex spatial and temporal intergenic relations are mapped to the functional domain and are responsible for the great complexity of eukaryotes. In the post-nucleus stage, the changes in the frequency of and the relation between old genetic elements, rather than the emergence of novel elements, are the major content of genomic evolution, especially in short-term evolution[60, 75-78, 107]. That is why the modern synthesis can use gene frequency to explain the short-term evolution at the post-nucleus stage without invoking the innovation of lower-level evolution, such as mutational bias. However, extrapolation of this special characteristic to pre-nucleus evolution or long-term evolution could be misleading[108]. Pre-nucleus evolution, or more generally, the origin and early evolution of life, does not have well-developed anti-bias mechanism; even with well-developed anti-bias mechanisms, the effect of weak residual biases accrues in long-term evolution. In both situations, the creativity of DNA evolution and the attendant biases are not negligible any longer.

### The evolution of organelle genome: enslavement through exclusive coupling

As nuclear compartmentation regulates protein-DNA interaction, organellar compartmentation regulates the interaction between proteins. However, organellar compartmentation is less important than nuclear compartmentation, because nuclear compartmentation influences genetic information more directly and tightly than organellar compartmentation. Nuclear compartmentation regulates protein-DNA interaction and thus converts protein-DNA interaction to intra-genomic relations; in contrast, organellar compartmentation mainly regulates protein-protein interaction and thus only indirectly affect genomic evolution. Although some organelles have their own genome, for example mitochondria and plastids, the size and the importance of organellar genome are decreasing.

The importance of organellar genome in evolution is determined by the position of organelle in hierarchical life. All organelles, including endosymbiotic organelles, are a lower level of cell. The reproduction of symbiotic organelles is relatively independent of the cellular host and hence organellar fitness is different from that of the host cell. Meanwhile, elimination of a cell destructs its organelles. Therefore, the fate of organelles in natural selection is determined dually by organellar fitness and cellular fitness. As a result, the evolution of organelles is balance between organelles and the host cell. The structure and function of organelles are not tuned for the host cell, and the fitness of the cell is compromised.

The only solution is the coupling of organellar genome and nuclear genome. As explained previously, genotype-phenotype division is the essence of life: genetic pattern, rather than epigenetic pattern, is the principal form of biological evolution. Moreover, genetic pattern is shaped by the targets which the pattern is coupled to, according to the principle of coupled selection (chapter IV). If organellar genome is decoupled from the organelle, the host cell will be the dominant determiner of organellar evolution, and, consequently, organellar evolution will be tuned for the fitness of the host cell. The principle of coupled selection can satisfactorily explain the transfer of organellar genes to nuclear genome.

Various hypotheses have been raised to explain the universal transfer of genes from endosymbiotic organelles to nuclear genome, but none of them can explain the universal transfer without exception. One hypothesis is that the transfer from a multiple-copy organellar genome to a single-copy nuclear genome is more frequent than the reverse; therefore, the biased transfer results in the migration of organellar genes to the nuclear genome. However, organisms with a single plastid also have organellar genome transfer[109-110]. The hypothesis that the prokaryotic property of endosymbionts cause unidirectional transfer is not viable, because eukaryotic endosymbionts (secondary endosymbiosis) also undergo genomic erosion[109, 111]. Muller's ratchet, a common cause of genetic reduction, cannot explain organellar genome erosion either because both mitochondria and chloroplast can be recombining[112-113]. It has been suggested that the internal environment of plastids and mitochondria is mutagenic because the redox reactions there produce oxygen free radicals; under the selective pressure for less mutation, organellar genes transfer to nuclear genome[114]. However, the free radical hypothesis fails to explain the gene transfer in secondary endosymbionts which do not produce free radicals[110]. The highly developed gene regulation in nuclear genome was once considered a cause for organellar gene transfer. However, the gene regulation in plastids is subordinated to the nuclear genome, so plastid genes don't have to transfer to the nuclear genome in order to be regulated by the host organism[115].

A hypothesis has been proposed to explain the universal organelle gene transfer: selection for small genome size drives the transfer of organelle genes[116]. However, neither the intracellular environment nor the life cycle of organelles apply strong selective pressure for fast division to organelles. The organellar gene transfers continuously at high rate through the whole history of eukaryotes[117-121]: the transfer is so strong that some mitochondria have lost their genome completely[122-123]. Even if the selective pressure for small size exists, the pressure will decrease with the reduction of organellar genome and can hardly drive the organellar genome to zero. The case of genome reduction in obligate intracellular parasites[116] is fundamentally different from the endosymbiotic organellar gene transfer: the loss of unnecessary genes in parasites is a specialization to a nutrition-rich intracellular environment, while the organellar gene reduction is a transfer of essential genes to the host genome[124]. The proposed selective pressure for fast division at some stages of host life[116], such as the zygotic bottleneck and gametogenesis, is actually a pressure on the host cell, not on the organelles. Such pressure on the cell only drives the cellular enslavement of organelles, which is realized through the transfer of organelle genes to the nuclear genome, as an embodiment of the principle of coupled selection. ***The difference between intracellular parasites and endosymbiotic organelles in genomic evolution just reveals that endosymbiotic organelles are enslaved by the host cell through genetic coupling, while intracellular parasites have successfully resist such enslavement. It further suggests that enslavement of invading organisms is initiated through genetic transfer and realized until genetic coupling is fulfilled; intracellular parasites should have a mechanism to resist genetic transfer.*** In brief, the small genome size of organellar genome is only the surface of the transfer of organelle genes to the host genome, which has abolished organelle autonomy and hence increased the fitness and complexity of the host organism[124].

In general, regulatory genes of organellar genome tend to transfer to the nuclear genome more readily than enzymatic and structural genes[115]. This trend is consistent with the functional enslavement of organelles for the fitness of the host organism. The remaining genes may have least conflict between organellar fitness and cellular fitness, so their selective pressure for transfer is very weak.

The degeneration of the Y chromosome is similar to the transfer of organellar genome: both are influenced by the conflict between hierarchical levels. However, because endosymbiotic organelles were autonomous and the organellar genes couple to the host organelle all the time, the conflict between the organelles and the cell is the primary force driving the decoupling of organellar genes from the host organelles. In contrast, the Y chromosome only couples to the host sperm transiently: most of the time, the Y couples to the organism. Moreover, the sperm selection is weakened through inhibiting post-meiotic gene expression and the intercellular bridges of spermatids[125]. Therefore, hierarchical conflict is only the secondary force for the degeneration of the Y chromosome; in stead, the asexuality of the Y is the primary force for its degeneration. Nevertheless, hierarchical conflict accounts for some gender-specific phenomena, which are analyzed in chapter VI.

## V. Canalization/Coarse Graining in Genotype-phenotype Mapping – the Basis of Hierarchization

As explained in previous chapters, the unidirectionality of translation couples DNA/RNA patterns to cells and decouples them from proteins, which is favorable to the increase of fitness and complexity. This reasoning has two hidden assumptions: as an organization of molecules, the cell has different evolutionary interest from that of its component molecules; moreover, the cell has higher evolvability than molecules. These two assumptions seem to be unrelated self-evident facts and hence are needless of explanation. However, analysis of these two simple and apparently unrelated facts uncovers a fundamental relation between them, which is crucial for understanding biological evolution.

As the last step of the mapping from genotype to phenotype, organization/hierarchization extends the output of translation from molecules through cells to organisms. DNA-protein based heterodomain mapping, coupled selection, and nuclear compartmentation bring great evolvability to life, but fulfilling this evolvability is constrained by the form of evolution, i.e. the physical property and evolvability of building blocks. Specifically, the evolvability of individual molecules is very limited, so even the simplest life is a cellular organization of molecules. However, because of the physicochemical property of molecules, the evolvability of unicellular life is still constrained[18]. A cell cannot have the size and complexity of human, because its form of evolution, such as intracellular transportation, cytoskeleton, and metabolism, sets a limit on the complexity of unicellular evolution[18]. As a result, the organization of cells further forms a novel type of evolutionary entity – multicellular life. Because cells have greater evolvability than proteins, the evolvability of multicellular life is much greater than that of unicellular life[126]. The extensive and relatively uniform organization is actually hierarchization, which forms a stable level; serial hierarchization forms a multilevel hierarchy. Functioning as an extension of translational output, organization/hierarchization gives full play to the evolvability of genetic information, i.e. the evolvability of protein-DNA based heterodomain mapping and coupled selection.

Organization/hierarchization is a very common mechanism to improve evolvability for both abiotic and biotic evolution. For example, elementary particles comprise atoms and atoms comprise molecules. Organization/hierarchization can occur in various forms. However, all forms of organization/hierarchization have a common characteristic: patterns at the lower level are coarse-grained/canalized to generate novel patterns for the higher level in the process of organization/hierarchization. Such coarse graining/canalization in pattern transformation is the basis of organization/hierarchization and explains some important properties of hierarchical life.

## What is coarse graining/canalization?

Coarse graining originally indicates a low-resolution imaging in which the fine details are smoothed out. Coarse graining is a way of modeling which is broadly used in physics and biochemistry[127-130]. However, no one has clearly pointed out that coarse graining is also a natural process, let alone the role of coarse graining in evolution. Coarse graining is that the details of an entity are masked in a process of interaction. If no detail is masked in this process, it is called fine graining. The details of an entity can be available in one process but unavailable in another process. Therefore, coarse graining is relative and process-specific. In biological evolution, coarse graining is actually canalization, which was proposed in 1940s as the reduction in the variability in the forms subject to natural selection[131-132] and has become a hot area of research[133-135]. A well-known example of canalization is the buffering of genetic changes by HSP90[136-138]. The robustness of neutral networks, or more generally, living systems, is also characterized by the reduction in variability[11, 23, 139-140]. It has been pointed out that robustness and canalization are an inevitable consequence of a complex network[133-134, 141], namely a property of complex systems. Coarse graining, robustness, canalization, degeneracy, and neutrality are all different descriptions of the same property of a system with internal interaction. Here, the author propose that, as an intrinsic property of organization/hierarchy, coarse graining/canalization increases evolvability by masking biases and transforming patterns.

The terms "robustness" and "canalization" emphasize the stabilization toward a specific state, which is currently considered as the cause of robustness/canalization. In contrast, the term "coarse graining" emphasizes its blindness to the effect of phenotype: coarse graining blindly smoothes over the rough landscape and thus transforms the pattern of the landscape. The coarse-grained patterns are selected according to the fitness and evolvability of the host entity. When the effect of coarse graining is extended from the robustness/stabilization of canalization to pattern transformation, not only more examples of coarse graining are discovered, but also the unnoticed important properties of coarse graining/canalization are revealed. One new example of coarse graining/canalization is Darwinian selection, which only recognizes the fitness of the organism as whole. The internal details of the organism, for example DNA sequence and protein conformation, cannot be directly recognized by Darwinian selection, as Ernst Mayr emphasized[54]. Under the pressure for high resolution and efficiency of selection, a special mechanism, sex, emerges to indirectly fine-grain Darwinian selection. In order to understand coarse graining, organization/hierarchization and the subsequent fine-graining mechanisms, the cause and the role of coarse graining in evolution need to be analyzed to show the unnoticed properties of coarse graining/canalization.

The cause of coarse graining is the interaction among the components of the coarse-grained entity. The reason why interaction causes coarse graining resides in the fundamental difference between an organized group and a simple aggregate. In other words, how is an organization of entities with complexity increase differentiated from a simple aggregate without complexity increase? In the simple aggregate, there is no interaction between entities and thus no change in their activity and properties, which are the same as when they are isolated. When there are interactions between entities, a part of the activity of the entities is used in the interactions, and is therefore unavailable to the external environment. Moreover, the activity remaining available to the environment is more or less changed compared to that of the isolated entities. The occupation of a part of activity masks the details of component entities, so the organized collection of these entities is coarse-grained.

The detail mask in coarse graining is fundamentally the same as the decrease in variability of a phenotype in canalization[131-134]. According to the laws of thermodynamics, when isolated entities become an interacting system, the entropy, namely the number of microstates the system can assume, decreases[32, 142]. If an organism has N genes and each gene has M alleles, then number of genotype is N x M. However, there are interactions between the products of these genes, so the number of the state of these products (configuration/organization) is less than N x M. Phenotype is a subset of the configuration of those products. Therefore, the number of phenotype must be less than genotype. ***Both the detail mask in coarse graining and the variability reduction in canalization are the manifestation of the entropy decrease of an organized system compared to its isolated components.*** Moreover, the more interaction, the more decrease in entropy, and the more reduction in the number of phenotype. Although phenotype is much more complex than the molecule in the usual thermodynamic models, it still abides the universal laws of thermodynamics. The key to link coarse graining, canalization, and entropy is interaction, which has already been suggested by some researches. It was pointed out that epistatic interactions rather than genetic redundancy causes the canalization in the phenotype of mutations[143]. It was further argued that canalization is an inevitable consequence of a complex network of interaction and the extent of canalization is determined by the complexity of the network[133, 141] (recall the measurement of complexity as the number of interactions per entity[40] in chapter II). However, those viewpoints have not had the attention they deserve. Canalization is often mistakenly understood as the result of long-term natural selection for optimal phenotypes[133]. This understanding has problem in explaining the property of phenotype canalization[133]: the selective pressure for canalization will decrease when the selected phenotype approaches or has reached the optimum[134]; it cannot explain why canalization can quickly evolve in a few generations[144-146, 147] and why mutational sensitivity is uncorrelated with fitness sensitivity[148-149] if canalization is shaped by selection. Moreover, it is hard to explain the universality of canalization in development[131] by stabilizing selection[133]. Canalization/coarse graining results from internal interaction rather than the natural selection for optimum, although some forms of canalization/coarse graining are indeed fine-tuned by natural selection (next section).

The functional consequence of coarse graining to the whole organization is determined by the functional contribution of the masked activity. The contributions of component activities to the host organization are not equal. Therefore, the effect of masking in interaction can range from strong activation or inhibition to no consequence. Because of coarse graining, not all changes in components can be transmitted to the external environment through the host organization. Namely, phenotype's norm of reaction is extended by coarse graining. In general, the extent of coarse graining is determined by the number of interactions, because coarse graining results from interaction. Complexity correlates with the number of interactions per entity. Therefore, the extent of coarse graining is determined by the complexity of the organization, as discovered by previous papers on canalization[133]. Regulatable coarse graining is the controlled change in the extent of coarse graining. Therefore, regulatable coarse graining only occurs in an organization which is complex enough to mask the change in the extent of coarse graining without disrupting its integrity. A typical example of regulatable coarse graining in evolution is that the effects of mutations is completely masked in a network that is regulated by HSP90: the function of the protein products of those mutations is completely masked by the coarse graining in an organization of proteins[136-137, 150-151]. Buffering stochastic perturbation by microRNAs is another example[152].

The organization, i.e. the coarse-grained collection, behaves as a whole, displays novel activity and property, and has a landscape different from that of the simple aggregate (Fig. 12). To a coarse-grained entity, the unavailable details are internal to the coarse-grained entity; the available details are altered and output to interact with external entities. Coarse graining is a necessary result of organization and hence an indicator of complexity increases. To an entity with defined components, complexity increase must result in coarse graining. Internal interaction is also an essential characteristic of group [153-154]. Failure to include internal reactions into the study of group selection leads to erroneous conclusions[155]. Actually, organization, hierarchy, group, and network, are different descriptions of a system with internal interaction.

Detail mask is only one aspect of coarse graining. Another aspect is that the remaining available details are more or less altered by the internal interaction. For example, a water molecule is an organization of hydrogen and oxygen molecules. Because of coarse graining, a water molecule doesn't exhibit the property of individual hydrogen or oxygen molecules; in contrast, the water molecule exhibits the altered property of hydrogen and oxygen molecules as a whole. Because of this nature, coarse graining transforms the property and pattern of individual components to a novel property and pattern of the organization. Evolutionary entities can form various new compound organizations that have properties and landscapes different from those of individual component entities. These compound organizations are selected as an indivisible unit according to their fitness and evolvability. In this novel unit of evolution, the patterns of component entities are transformed to the novel patterns of organizations (Fig. 12). For instance, unicellular organisms are a coarse graining of molecules; the patterns of molecules are coarse-grained to the patterns of cells and organisms, which are subject to natural selection as a whole. Darwinian selection is actually a form of organism-environment interaction. If the organism is coarse-grained relative to molecules, then the selection of organism must be coarse-grained relative to molecules. As a result, selection only differentiates the fitness of the organism as an indivisible whole and does not recognize its constitutive genes and proteins[54].

Coarse graining transforms one form of evolution to another with a different landscape (Fig. 12), and hence breaks the limit to complexity increase set by the form of evolution. This is why coarse graining is so common in both biotic and abiotic evolution. Coarse graining can be spatial, temporal, or, in most cases, both spatial and temporal. The amplitude modulation in telecommunication is a typical example. The individual oscillations of the carrier wave do not contain the sound signal we want to transmit. However, the amplitudes of individual oscillations are modulated to produce the signal. Because amplitude is only a part of an oscillation, the modulation consists of the coarse graining of a collection of oscillations to generate a new wave of much lower frequency to transmit sound signal (Fig. 12B). Although real amplitude modulation may contain many component processes, the most important one is coarse graining. This new wave generated by coarse gaining has new qualities, such as better robustness than the carrier wave, in addition to different frequency.

Coarse graining occurs at every level of the hierarchy. The cell in multicellular terrestrial life is a functional coarse graining of its components. Why does the cell rather than the subcellular or superacellular entity become the functional unit of coarse graining? The reason is that the cell was once the whole organism in the evolution to multicellular life. In other words, the cell is a unit of coarse graining in evolution. As a result, selection of unicellular organisms is coarse-grained: selection only differentiates the fitness of unicellular organisms as a unitary whole and does not recognize constitutive genes and proteins. Therefore, the fitness and complexity of the cell, rather than subcellular or superacellular entities, are increased during evolution at the age of unicellular life. Multicellular lives develop from unicellular lives and use the former unit of natural selection as their current unit of functional and developmental coarse graining. This phenomenon suggests that coarse graining, along with multilevel group selection[156], plays an important role in the history of life.

This transition from unicellular to multicellular life is a necessary result of evolution. Every type of evolution has a limit in complexity increase due to its physical form, mainly its function and evolvability of building blocks. Although heterodomain mapping, coupled selection, and nuclear compartmentation have great evolvability, effectuation of this evolvability is limited by the building blocks of cell. A cell cannot have the size and complexity of a human, because its form of evolution, such as intracellular transportation, cytoskeleton, and metabolism, limits the complexity of cellular evolution[18]. The great evolvability resulting from informational evolution is constrained by the form of the cell. The solution is the transformation of unicellular evolution to multicellular evolution by coarse graining. An assembly of cells can comprise a novel entity with greater evolvability than individual cells, because cells, as the building block of the novel multicellular entity, have much better function and evolvability than proteins and lipids, the major building block of the unicellular life. Therefore, multicellular life is much more complex than not only unicellular life but also the sum of individual component cells. Here, it must be emphasize that ***coarse graining/canalization per se is blind to fitness, complexity, or evolvability***; namely, the pattern transformation in coarse graining is blind; the coarse-grained patterns are selected according to the fitness and evolvability of the host entity. Therefore, the selected coarse-grained patterns are of high fitness and evolvability.

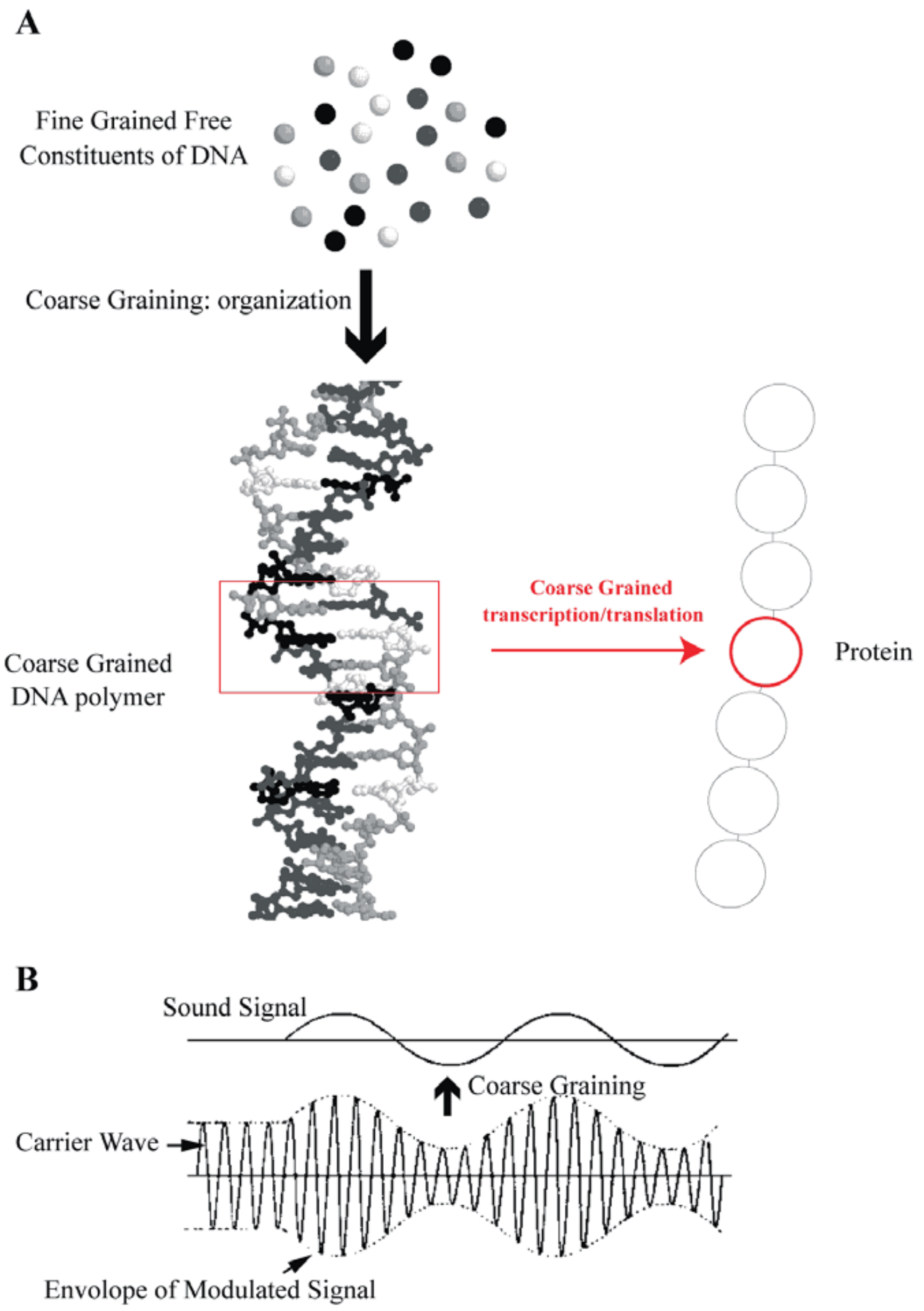

**Fig. 12. Coarse graining and pattern transformation. A.** A simple aggregation of the components of DNA is the fine graining of these components, while the DNA in translation is coarse-grained: three nucleotides of DNA participate in translation as an indivisible whole. Therefore, the components of DNA are masked in translation. In other words, the translational units of DNA are indivisible in translation but have internal complexity that is masked in translation. In this way, coarse graining transforms a compound evolutionary entity to a novel unitary evolutionary entity at a higher level. **B.** Coarse graining generates novel property and landscape. Extracting sound signal from an amplitude modulated carrier wave is an example of innovation by coarse graining. The sound signal is a type of coarse graining of carrier wave, because individual oscillations are required for generating sound signal, but those oscillations are masked in the modulated signal.

As introduced in chapter III, genetic mapping consists of coarse graining that smoothes out the rugged landscape of DNA without affecting the dynamics of DNA evolution. Although heteromapping and coarse graining are often integrated and hard to separate, they are distinct mechanisms. Coarse graining and heteromapping both are innovative, but coarse graining occurs within the same domain while heteromapping occurs between two different domains. Heterodomain mapping is a component of informational evolution, while coarse graining *per se* is noninformational. Coarse graining does not necessarily have a uniform code, as translation does. Therefore, heteromapping can have coarse graining as its inherent component, but coarse graining cannot have heteromapping as its inherent component. Moreover, coarse graining must be unidirectional in nature while heteromapping can be bidirectional. The upward unidirectionality of coarse graining prevents the feedback of patterns from the higher level to the lower level in the hierarchy. ***This natural unidirectionality of coarse graining blocks the feedback of patterns of higher-level organization in the functional domain downward to the bottom level of the hierarchy***, similar to the unidirectionality of translation that is produced and maintained under selective pressure. Therefore, coarse graining is fundamentally different from heteromapping. On the other hand, the coarse graining in translation must be universal and uniform because of the universal and uniform mapping rule.

## The profound role of coarse graining/canalization in evolution

The role of canalization in evolution is not only robustness/stability, which could be deleterious when novelty is needed in the origin and early evolution and/or in the presence of the changing environment. As explained in the previous section, coarse graining/canalization generates novel patterns for natural selection by transforming patterns, which breaks the physical limit to evolution. The transformed patterns have less variables and details than the fine-grained components. Therefore, the ruggedness of the landscape of the fine-grained components is reduced by coarse graining/canalization, very similar to smoothing the carrier wave to form a new wave of different quality (Fig. 12B). As a result, coarse graining/canalization has another fundamental role in evolution in addition to its innovative pattern transformation: reducing the bias in configuration space sampling by smoothing over the rugged landscape of evolution. For example, genetic mapping is coarse-grained/canalized: the molecular and submolecular changes that are not strong enough to change nucleotide recognition are all masked (Fig. 12A). The triplet genetic coding reduces the genetic bias resulting from differential stability of nucleotides, which is actually a form of canalization: it not only brings robustness to individual amino acids by degenerating the genetic codes of the same amino acids, but also smoothes over the rough landscape between the genetic codes of different amino acids (Fig. 5&6). Actually, canalization, coarse graining, neutrality, and degeneracy are all different names of the same phenomena, but with different emphasis. The blindness of evolution is the key in this beneficial effect of coarse graining/canalization. In conventional words, smoothing over the rough landscape provides "a molecular means by which adaptive peak shifts in large populations may occur without passing through an adaptive valley"[137]; namely, blind configuration space sampling is not constrained or biased by adaptive valleys and mountains.

Although internal interaction is necessary and sufficient for canalization/coarse graining[133], many forms of canalization have been fine-tuned by natural selection. However, the pressure in such selection is not the stabilization of phenotype or other variables to an optimum or a specific state. When the variation is stabilized to approach the target state, the variables of the substrate of coarse graining/canalization is reduced and the selective pressure is decreased[134]. If coarse graining/canalization is driven by the stabilizing selection of specific phenotypes to an optimum, it requires high mutation rates or the disruptive effects of mutations as the substrate for coarse graining/canalization to work on; however, neither is a plausible substrate for canalization[134]. The hypothesis of stabilizing selection toward a optimum has problem in explaining the characters of canalization[133], for example uncorrelated with fitness sensitivity[133, 148-149]. The selective pressure behind the fine-tuning of canalization is evolvability, not a specific phenotype or state. For example, the coarse graining in genetic coding makes translation not bias to any specific amino acids, which reduces sampling bias and is beneficial to blind evolution.

Similarly, in the canalization of phenotype, the selective pressure is the unambiguousness and stability in the mapping from genotype to phenotype. In translation, one genetic code stably maps to only one fixed amino acid. However, translation is only the first step in the mapping from genotype to phenotype. Other steps include protein folding and organization/hierarchization, both of which are not fixed and the mapping can be changed by epigenetic and environmental noise and disturbance. Noises and disturbance can change the mapping in a short period, convert the unique mapping to multiple mapping, or reversely, overlap two originally different genotypes in the phenotype space. For example, in a transcriptional regulation system, high-noise conditions convert a unimodal distribution of protein production to a bimodal distribution[157]. Both multiple mapping and varying mapping are deleterious or fatal to life because they globally destroy the accumulated fitness and complexity, as explained in chapter III. The noise and disturbance, either environmental or epigenetic, are caused by the processes that are out of the control of the genetic information, for example the thermal motion of molecules. The nature of noise and disturbance are stochastic biases with small effect. Coarse graining/canalization can smoothing out such small stochastic biases. As long as there is a unique and stable mapping between genotype and phenotype, the specific trajectory in heteromapping is unimportant. This noise abatement to stabilize heteromapping, neither stabilizing phenotype to a specific state nor masking a specific variable, accounts for the role of and the selective pressure for the coarse graining/canalization of phenotypes.

If coarse graining/canalization is fixed to reduce bias and stabilize heteromapping, then what is the role of reversible canalization - such as the buffering by HSP90? In other words, why some forms of canalization are reversible if canalization is advantageous? One reason could be that those forms of canalization cannot be fixed as a genetic mechanism. However, at least some forms of reversible canalization are well regulated and coordinated with the physiology of the host organism. For example, the genetic buffering regulated by HSP90 can be turned down by heat shock[136-138, 150]. It is unlikely that the reversibility of such highly regulatable canalizations is due to the organismal incapability to fix them.

The buffering by chaperone is the epigenetic[136-138, 158] version of the "adaptive" mutation[66-71, 73] in response to stress: their mechanisms and roles are fundamentally similar. Most spontaneous lower-level patterns, either genetic mutation or epigenetic noise, are deleterious to the stability/fitness of higher levels. Blind evolution cannot judge the usefulness of a lower-level pattern for organisms, so a feasible strategy is to mask lower-level evolution. However, the mask also reduces the genetic and epigenetic mutability, and thus decreases the evolvability. Therefore, the best strategy is a balance between two extremes: zero evolvability with neither deleterious change nor beneficial change to preserve the current configuration, or high evolvability with deleterious and beneficial changes to alter the current configuration. The determinant of the balance point is the organismal fitness to the environment. If the organism fits well to the environment, then the best strategy is to tune down mutability and evolvability to avoid deleterious changes. If the organism does not fit well, then the best strategy is to tune up mutability and evolvability to find an escape in new changes. In the buffering regulated by HSP90, the buffered/masked genetic changes are released by heat shock, similar to the new mutations generated in other stresses[66-71, 73]. Such neutral to non-neutral switch provides evolutionary innovations[22]. The transient competence for DNA uptake in response to stress is another manifestation of this strategy, although its mechanism is poorly understood[69]. Those new changes are blind to the stressful situation, but Darwinian selection will pick out beneficial changes. Therefore, the changes in response to stress, either genetic mutations or epigenetic changes, are apparently adaptive. The "adaptive mutation" is realized through tuning the fidelity of genetic information processing system[66-71, 73], while the epigenetic buffering is realized through turning down nongenetic coarse graining by chaperones and other noninformational components[136-138, 152]. HSP90 is the key factor that links the internal coarse graining/canalization and the external environmental stress[136-138]. The strategy of "adaptive mutation" or epigenetic buffering is a desperate struggle for survival in grave crises. The complex mechanisms behind this strategy has been selected for and fixed in evolution, although these mechanisms *per se* are not any direct functional output.

There is a major difference between the "adaptive mutation" and the buffering by HSP90: the former is only seen in unicellular microorganisms[69], while the latter is seen in both unicellular and multicellular organisms[136-137, 159]. Because mutation is blind, component cells of multicellular organisms will acquire different mutations; consequently, the response of the multicellular organism to stress through "adaptive mutation" will be heterogeneous and thus uncoordinated: the probability of fitness increase through "adaptive mutation" will be significantly decreased. In contrast, HSP90 regulated responses to stress either expose existing masked germline genetic changes[136-137] or generate heritable epigenetic changes[138]. The phenotype changes are relatively uniform in the multicellular organisms experiencing heat shock[136-137, 159]. Moreover, such epigenetic changes are more rapid and flexible than *de novo* "adaptive mutations"; these characteristics of epigenetic changes are beneficial for multicellular organisms with long life span to handle transient difficult situations. Such coordination in the utilization of Darwinian and Lamarckian mechanisms in different organisms strongly suggests that the regulatable canalization is highly fine-tuned by natural selection, not merely a byproduct of organization/hierarchization/network.

## Coarse-grained/canalized selection: the limit to natural selection

As explained above, natural selection of organisms is coarse-grained/canalized, because all component molecules are masked in the selection of host organism. Coarse-grained selection cannot differentiate the molecular details of an organism. Directly fine-graining a coarse-grained entity may change the way of organization/coarse graining and more or less affect its function and complexity, because fine graining needs to interact with the internal components to probe the inside and thus interferes with the internal interactions between its components. To the unintelligent environment and life, direct fine-graining is always destructive. Moreover, the degree of coarse graining is process-specific. Some processes cannot probe low-level details due to its coarse-grained nature. Darwinian selection of organism is such a process and thus cannot resolve the details of molecules - it's impossible to draw a Persian miniature with a scrub brush.

The fundamental reason of the coarse graining of Darwinian selection is that genetic evolution integrates both domains of heteromapping, the mapping rule, and the coupled selection, as explained in chapter IV; therefore, the whole source pattern domain must couple to the whole target functional domain to increase fitness and complexity of the functional domain; the coupled selection of the whole machinery of heteromapping, not individual pattern in the source domain, is the essence of biotic evolution. Integration of the pattern domain and the function domain is a form of organization, so both domains are inevitably coarse-grained. That is why genes and genome are never the direct substrate of Darwinian selection in biotic evolution, as Ernst Mayr has emphasized that "a gene is never visible to natural selection"[54]. In contrast, DNA is the substrate of abiotic molecular selection or transformation, for example mutations; however, such lower-level selection is the noninformational evolution of the vehicle of genes, not the informational evolution of genes, as explained in chapter IV; abiotic evolution of DNA, namely mutation, increase the stability of DNA rather than the fitness of the host organism. In brief, ***the coarse graining of Darwinian selection is the consequence of the coupled selection of heterodomains, which in turn results from the labor division of internal evolution to pattern formation and functional action.***

At the early stage of evolution, translated proteins and relations between proteins are still crude, so many blind mutations are beneficial. With the improvement of life, the percentage of beneficial mutations decreases. Finally, the genome reaches an equilibrium in which the number and influence of harmful mutations overwhelm those of beneficial mutations. Blind evolution cannot improve the whole genome although there is still considerable potential for individual genes to improve. Some genes may improve through spontaneous mutation, but at the same time, other genes deteriorate. It is the whole genome, rather than individual genes, that is linked to the fitness of the host. The genetic improvement through beneficial mutation is neutralized by the deterioration through harmful mutations, and cannot be selected for at the level of genome and organism. Because mutations are blind, it is very unlikely that the whole genome is considerably improved by spontaneous dominance of beneficial mutations over harmful ones. The improbability is proportional to the size of genome. The larger the genome, the lower that probability is. Depicted on a landscape, the evolution of asexual genomes is in a deep valley. The depth of the valley is proportional to the size of genome, which is consistent with the Muller's ratchet that asexual reproduction sets limits to the maximum size of genome[160-161]. The barrier is the extremely low probability of spontaneous net beneficial change of the whole genome through blind mutations (Fig. 13). The asexual genomes undergo ineffective thermal-like motion in the deep valley[162]. Therefore, on the one hand, to asexual organisms, their adaptation to novel niches is inversely proportional to the size of their genome; on the other hand, the evolutionary constraint of asexuality is the basis to maintain asexual species (Suppl. Text 5, *Species and speciation: the cause of discrete evolution*).

The conventional view emphasizes fitness but overlooks evolvability. It views asexual evolution from the angle of fitness: that the accumulated mutations in asexual evolution are beneficial or harmful determines the role and weight of asexual reproduction in the history of life. From the angle of evolvability, natural selection is coarse-grained in asexual evolution, while the harmfulness or beneficialness of accumulated mutations is only a specific manifestation of asexual reproduction under a specific condition. The difference in the angle of viewing is not a matter of phrasing but affects our understanding of sex.

## Sex: fine-graining selection at populational level

How can environmental selection resolve the details of the organism without impairing its integrity? If the coarse graining of organisms is not uniform, the masked internal details can be disclosed to environmental selection without affecting the integrity and complexity of the organism. ***The difference exhibited in non-uniform coarse graining indirectly discloses the masked details below the level of coarse graining. This is the indirect fine graining. A typical example is the indirect fine graining of the selection of organism by sex.*** Selection still acts on the whole organism, but the genomes have different combinations of genetic elements through sex (Fig. 13). Therefore, given sufficient time, selection of sexual organisms can resolve the entire genome to the smallest units of combination. The smallest combination units are the limit of indirect fine graining by sex. If coarse graining is uniform, then the coarse-grained cannot be indirectly fine-grained without impairing its integrity, because uniform coarse graining does not exhibit any difference between individual uniformly coarse-grained entities. Sex is a mechanism to fine-grain coupled selection, which is caused by the labor division of internal evolution to pattern formation and functional action. Therefore, ***in order to increase the resolution and efficiency of selection, sex is a necessary result of the genotype-phenotype division.***

The essence of sex is the massive exchange of genetic information between organisms. The information exchange involves syngamy, nuclear fusion and meiosis[163]. Through syngamy and recombination, sex shuffles the genetic entities between the individuals of the same species. Selection still acts on the whole genome, but there are various genomes with different combinations of genetic entities constructed through sex. Selection of these genomes picks out the best combinations. Giving unbiased exchange and sufficient time, it is equivalent in effect to that selection directly acts on individual genetic entities revealed in differential fitness inside a species (Fig. 13). Meiotic recombination extends the information exchange from chromosome to any sequence. Therefore, the resolution of selection could be as small as one nucleotide, and that greatly increases the resolution and efficiency of natural selection. ***As well as the degree of the unbiasedness in configuration space sampling, the resolution and efficiency of natural selection is one of the two essentials of evolvability.*** Only after selection resolves individual nucleotides, can the intragenic structure and intergenic relation gain complexity efficiently. Therefore, in addition to nuclear compartmentation, sex is another important factor in increasing the size and complexity of genome. Sex is more than diversity generation: diversity generation alone cannot explain the evolution of the intragenic structure and intergenic relations. It needs to be emphasized that the diversity generation by sex is the re-organization of the genome at populational level by shuffling genetic components between individual organisms, which is different from the diversity generation by mutation. The fine-graining of natural selection by sex is the basis of sexual species (Suppl. Text 5, *Species and speciation: the cause of discrete evolution*).

Any factor that deviates sex from idealized random shuffling, namely linkage disequilibrium, will decrease the resolution of organismal selection of the genome through sex (Suppl. Text 4, *Linkage disequilibrium decreases the resolution and efficiency of sex*). In addition, sex is an indirect fine-graining at the organismal level; therefore, the details masked by the coarse graining at the molecular level cannot be disclosed by sex, for example the buffering of genetic changes by HSP90. The imperfection of sex is one of the causes of near neutrality in evolution.

From the angle of fitness, the conventional view of sex considers that sex increases fitness by influencing the combination of genes: either promoting beneficial combinations or breaking deleterious ones. However, the opposite, namely breaking beneficial combinations and promoting the deleterious ones cannot be reliably excluded[164-167]. As a process of blind and purposeless evolution, how can sex "know" a combination of genes is beneficial or harmful? Even intelligent humans are unable to know that with certainty. The role of sex in evolution can only be explained from the angle of evolvability. ***According to the present theory, sex per se does not bring a direction to selection: it neither preferentially promotes beneficial genetic combinations nor preferentially breaks deleterious ones. It only facilitates evolution by eliminating evolutionary barriers and thus opening up new niches that are different from the niches of asexual organisms*** (Fig. 13). Hence, sexual organisms acquire more complexity than asexual organisms. In other words, sex provides benefits in evolvability rather than fitness. Evolvability does not directly produce functional output in the struggle for survival, although it is responsible for the acquisition and development of those functions and the consequent increase in fitness. This theory is consistent with Weismann’s theory on sex[166, 168], the mutation theory of evolution[51], and the recent experimental discovery that sex increases the efficacy of natural selection in yeast and water flea populations[162, 169-171].

Some properties of sex can be explained either by coarse graining or by other conventional theories. For example, the long-term advantage of sex can be explained by generating variation[166, 168], resistance to parasites[172-173], DNA repair[174], *etc.*, without resorting to coarse graining. Although both could be equivalent in a specific case, coarse graining has broader and stronger power of explanation: coarse graining is more general and more fundamental, and thus it has much broader and more important application, for example in explaining the role of canalization in genotype-phenotype mapping (2nd section of chapter V, *Coarse graining/canalization reduces the noise in genotype-phenotype mapping for unambiguousness*). The explanation of sex by coarse graining is a natural and harmonious part of a unifying and parsimonious theory of genotype-phenotype division. Other theories are *ad hoc* to address the role of sex and are only particular manifestation of the theory of genotype-phenotype division. Moreover, although the present theory is compatible with many conventional theories of sex, some properties of sex can only be explained by coarse graining. For example, sex increases the resolution and efficiency of Darwinian selection and consequent promotes the evolution of intergenic relations and intragenic structures; a changing environment is not required to realize this advantage of sex, although it makes sex more desirable than a static environment does. In addition, the diversity generated by sex is different from that generated by mutation: the former is the various organizations of genetic elements, while raw genetic elements can only be innovated by mutation. These properties of sex are not addressed in the conventional theories of sex.

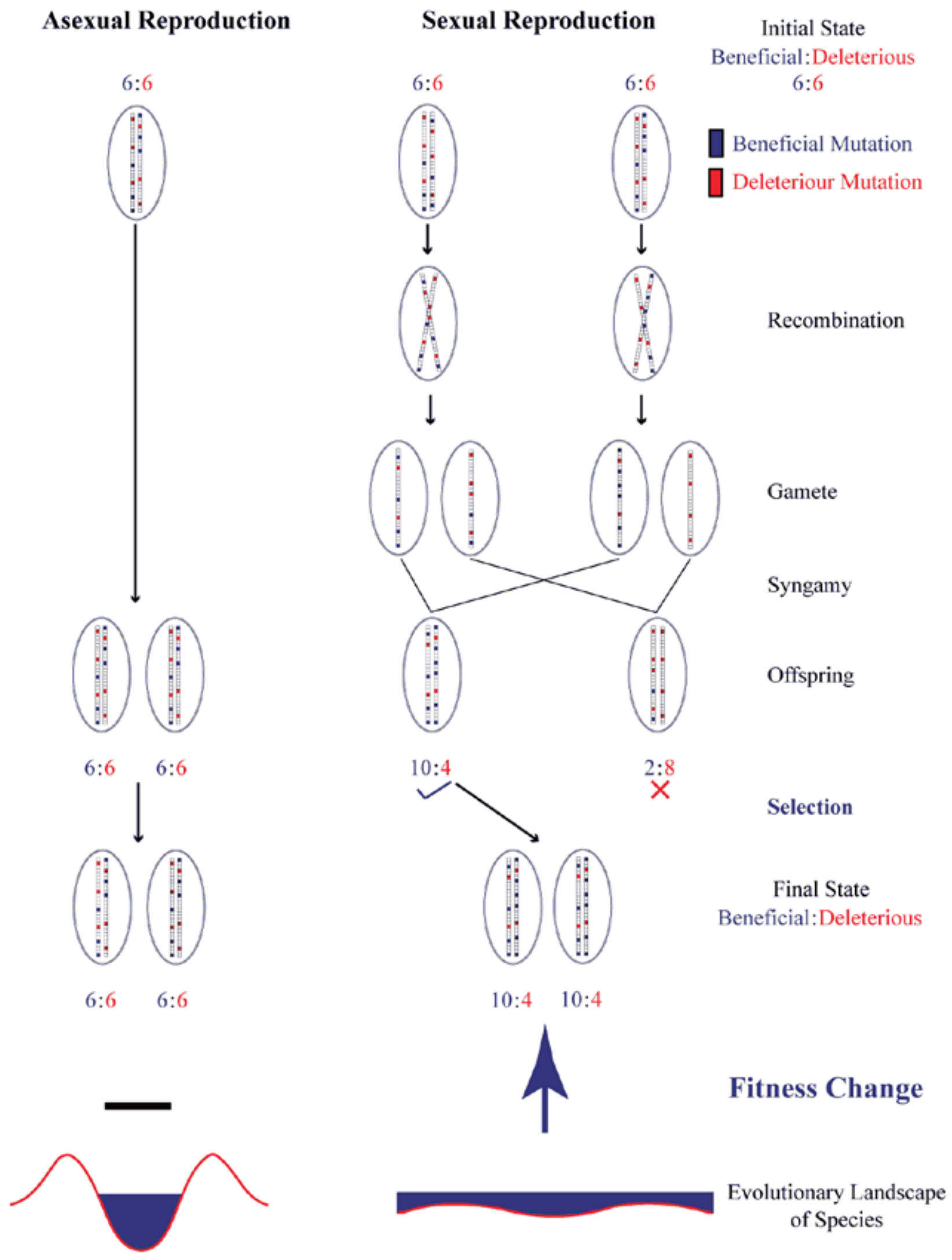

**Fig. 13. The long-term advantage of sex.** Mutations are blind to the fitness of life. In asexual organisms, the whole genome is indivisible and coupled to the host organism. Selection of asexual organisms is at the level of whole genome. To asexual

organisms, beneficial mutations have been neutralized by deleterious mutations because they are inseparable and equally probable. Therefore, These beneficial mutations do not result in any net fitness increase. The genomic evolution of asexual organisms is the ineffective fluctuation restricted in the valley of the evolutionary landscape, where the barrier is the very low probability of spontaneous predominance of beneficial mutations in the whole genome. In contrast, in sexual organisms, genetic elements can be exchanged between organisms through genetic recombination and syngamy. Although mutations are not biased towards beneficial as a whole, beneficial mutations can be enriched in some descendents through the genetic rearrangement that occurs in sexual reproduction. These descendents are selected for while other descendents enriched in deleterious mutations are selected against. In this way, beneficial mutations are preserved and enriched in some individuals through natural selection. In effect, natural selection of sexual organisms can resolve the difference of a single nucleotide, which is the smallest unit of genetic exchange in sexual reproduction. Fine graining of natural selection through sex flattens the barrier hindering fitness increase on the landscape of genomic evolution.

## The short-term advantage of sex

Enhancing the resolving power of selection by sexual reproduction is a long-term advantage at the level of group or species, which provides a maintaining force for sex. However, sexual reproduction brings an immediate disadvantage that sexual organisms reproduce half offspring as asexual organisms, i.e. the twofold cost of sex[175]. Although enhancement of the resolution of selection can provide a long-term advantage to maintain sex, the origin of sex needs an immediate benefit to compensate the twofold cost of sex[175](However, full compensation is not required. See the 5th section on altruism in chapter VI). The immediate benefit of sex is gamete selection. Gamete selection adds a new layer of selection to the selection of organisms, and thus makes the selection of organisms more effective in one generation. Gamete selection has two parts: one part is the environmental selection of gametes for survival; the other part is the competition of gametes for reproduction, mainly the gamete competition for fertilization. Only the latter part is characteristic of sexual selection. The initial benefit of sex is the selection of gametes of better quality and thus breeding better offspring[176-182].

In species of anisogamy, sperm competition is often linked to female polygamy or promiscuity, and sperm competition in strictly monogamous females is ignored. Actually, as well as inter-organism sperm competition, intra-organism sperm competition plays an important role in early evolution. Inter-organism sperm competition selects genetic variations between individual organisms. It mainly reflects the competition between individual organisms. In contrast, intra-organism competition mainly selects the genetic differences in sperms from the same organism. These intra-organism genetic differences are caused by germline mutation and meiotic recombination[177-182]. Therefore, gamete selection selects the gametes of better quality which will result in the full-scale organism of higher fitness. Compared with the selection of individual organisms, gamete selection is more rapid, because it does not require a whole life cycle, and is more economical, because it avoids the waste of resource in the elimination of full-scale organisms.

The advantage of gamete selection is very similar to the advantage of zygote selection proposed in the selection arena hypothesis[183], because the selection of zygotes, the early form of organisms, also saves the resource and time consumed in the selection of full-scale organisms. A prerequisite of the effective gamete selection for reproduction is the asymmetry between the gametes of two genders: one gender has much more gametes which are the active competitor while the other gender has fewer gametes as the target of competition. The greater the ratio of two types of gametes is, the higher the efficiency of gamete selection is, because more male gametes will be eliminated. In isogamous species, the short-term advantage of gamete selection is very weaker than that of anisogamous species. However, full compensation for the two-fold cost of sex is not required to explain the origin of sex, as explained in the 5th section on altruism in chapter VI. Asymmetry between the gametes of two genders in number and behavior increase the short-term advantage of sex through male gamete competition for reproduction. This asymmetry in gamete is the origin of the sexual struggle of organisms and accounts for most organismal sexual differences in morphology, physiology, and behavior, such as the exaggerated secondary sexual characteristics of some males. Moreover, such asymmetry results in different modes of evolution in two genders. For example, the extreme asymmetry in the number of gametes requires more cell divisions of male germline than female, which is the main cause of the male-driven evolution. However, the initial driving force of sex becomes weaker and weaker with the increasing hierarchical complexity. The reason behind this trend is the conflict among different levels of hierarchy, which is discussed in the 3rd section of chapter VI, "Dissect sexual selection to different levels".

# VI. Hierarchization in Genotype-phenotype Mapping – the Conflict and Cooperation in the Multilevel Hierarchy

## What is hierarchization?

All forms of terrestrial life are hierarchical: both the pattern generator and the functional performer are molecules but all forms of life are cellular - a level higher than molecules. Actually, both abiotic and biotic evolution utilize hierarchization repeatedly because of the role of coarse graining. The importance of hierarchy in evolution has been emphasized[184-185]. Here, the author objectively defines the scientific meaning of hierarchy and systematically elucidates its important role in evolution. Hierarchy means that a group of evolutionary entities constitute and subordinate to a novel entity, namely lower-level entities nested within higher-level entities. In coarse graining, interacting component entities lose their independence and constitute a new entity. The component entities are masked in the coarse-grained new organization but determine the evolution of the coarse-grained new organization. Such dependence and masked determining in coarse graining is the evolutionary meaning of subordination. As a consequence of coarse graining, the subordination of masked components to the whole is hierarchization, for example the molecules in cells and the cells in multicellular organisms. Serial coarse grainings constitute a multilevel hierarchy (Fig. 14). As explained in the last chapter, the increase in complexity must result in coarse graining, which in turn results in hierarchy. Conversely, because hierarchization uses a compound entity as a unit for a new evolutionary entity of higher level, hierarchization must consist of coarse graining.

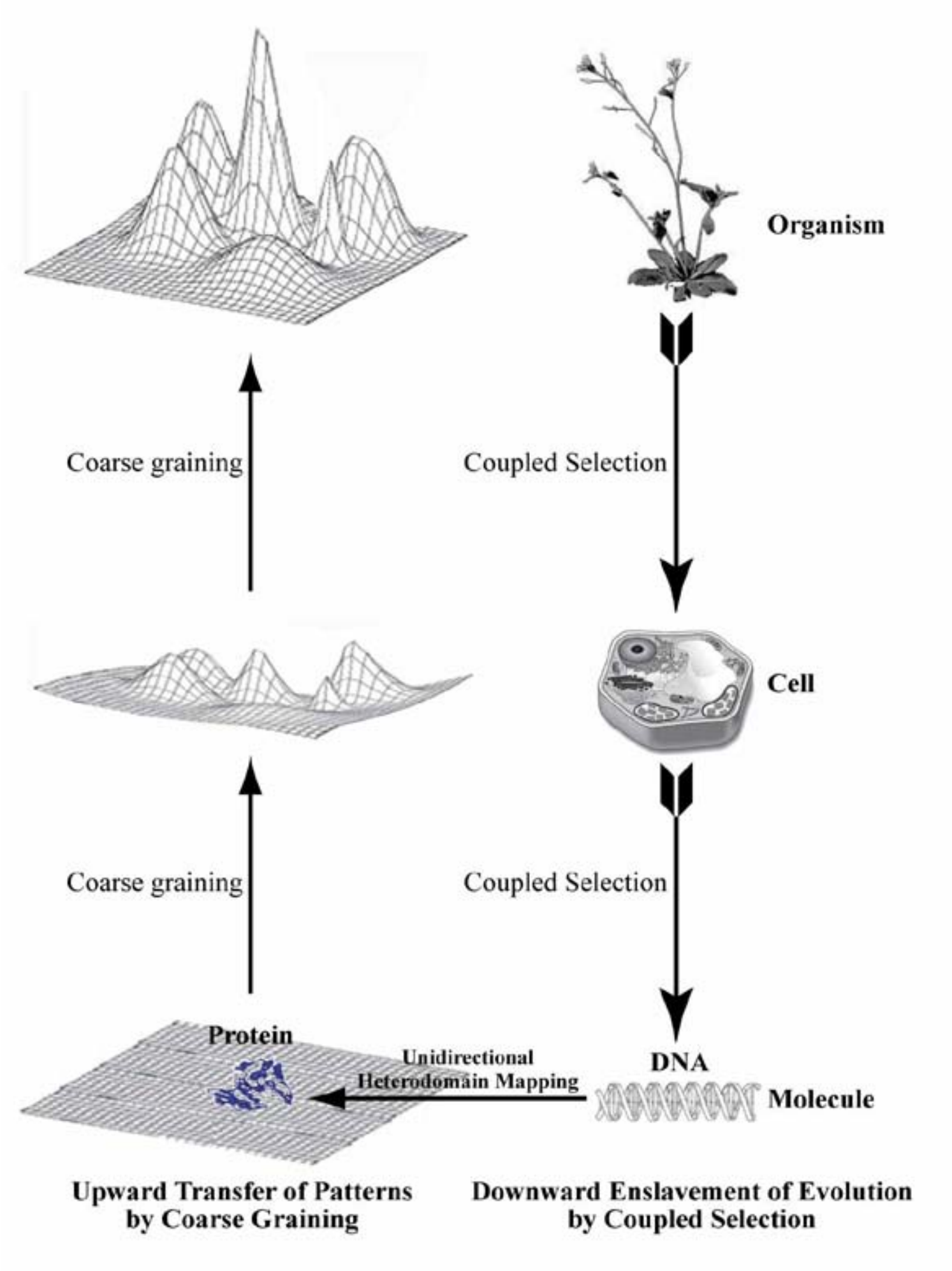

**Fig. 14. The evolution in the hierarchy.** Serial coarse grainings result in the multilevel hierarchy. The evolution of a hierarchy is a balance of the evolution of all levels. On the one hand, the patterns at every level are coarse-grained to form the unit of the pattern at the adjacent higher level. In this way, coarse graining transfers the pattern sequentially upward to the top and serves as an extension of heterodomain mapping. On the other hand, the survival of evolutionary entities at every level depends on the integrity of the adjacent higher level; namely, they are coupled to the host higher-level entity in natural selection. Therefore, every level enslaves the evolution of the lower adjacent level. To every hierarchical level, the lower levels are the pattern supplier while the upper levels are the pattern selector. The asymmetrical upward pattern transfer and downward enslavement are the basic property of hierarchical evolution and explain many problems in the evolution of hierarchical life.

Hierarchization is a synonym for organization: they describe the same process from different angles. A hierarchy can be viewed as serial organizations in an extensive and relatively uniform way, which involves extensive and relative uniform coarse graining. Cells are an example of extensive and relatively uniform coarse graining. As coarse graining is process-specific, the defining of a hierarchical level is purely relative. What matters in evolution is the intensity and evolvability of hierarchy following a specific definition of a level.

The intensity of hierarchization can be measured by the property change after hierarchization. For example, the property of a human is qualitatively very different from that of the cells of a human. The opposite of human is a simple aggregation of bacteria, the property of which may bear a very little qualitative difference from individual bacteria. Between the two extremes are some primitive multicellular organisms, such as *Volvox carteri*[186-188], which has only two types of cell. Vertical property difference in a hierarchy reflects the complexity increase in hierarchization. Another way to measure hierarchization is to examine the dependence of components on the integrity of the hierarchy. In *Volvox carteri*, single cells can live without the multicellular form[186-188], and that is impossible in complex hierarchical life, such as mammals. Dependence reflects the internal labor division of hierarchy, i.e. the horizontal property difference, which is a manifestation of the interaction inside a coarse-grained group. If the survival of the adjacent lower-level entities completely depends on the integrity of hierarchy, it is an **obligate hierarchy**. If the fitness of adjacent lower-level entities only decreases to a nonzero value at the disintegration of the host hierarchy, the hierarchy is a **facultative hierarchy**. The real situation is more complicated: some lower-level entities may require the integrity of their host to survive or exist, and others may not. Cells and proteins are in the former category and many inorganic molecules are in the latter category. Such a difference in dependence affects the evolutionary behavior of lower-level entities.

To biotic hierarchies, the hierarchies with heteromapping mechanism as their principal pattern generator/storage, the key is whether the existence of information carrier requires the integrity of its higher-level host. If it does, then this biotic hierarchy is obligate; if not, then it is facultative. Therefore, a mammal is an obligate biotic hierarchy. In contrast, a group of those mammals is a facultative biotic hierarchy, although a part of this hierarchy is obligate. The evolution of facultative biotic hierarchy is similar to that of obligate biotic hierarchy. The only difference is that the dependence of low-level on higher-level in the facultative biotic hierarchy is weaker than that in the obligate biotic hierarchy. Therefore, hierarchical selection, i.e. the selection coupling lower-level to higher-level, is weaker. Because of the principle of coupled selection explained previously, the complexity of the facultative biotic hierarchy is compromised by the stronger selection at the lower component level. Therefore, only obligate biotic hierarchies can achieve significant complexity. If not specified otherwise, all hierarchies in this paper refer to the obligate biotic hierarchy. ***The coupled selection of genetic information to the host organism is actually a manifestation of obligate hierarchization.***

Hierarchy is not a simple sum of serial coarse grainings. The complicated conflict and cooperation between different levels is an important novel contributor to the complexity of life. Because all forms of life are hierarchical, the conflict and cooperation inside the hierarchy must influence the genotype-phenotype mapping in the hierarchical life.

## The evolution in the obligate hierarchy: inter-level relations

The conflicts in hierarchy can be divided into intra-level and inter-level conflicts. Intra-level conflicts are the same as the ordinary conflicts that do not involve hierarchy. The characteristic conflicts in hierarchy are those between levels. In terrestrial lives, proteins and nucleic acids can be considered as the basal level. Cells, individual organisms, and species are higher levels in rising order. Every level is a coarse-grained extension of the translational output at the basal level and has a different property and evolutionary landscape. Every level provides a basis for the evolution at the adjacent higher level, and imposes selective restrictions on the evolution at the adjacent lower level, because the survival of lower levels is determined by the integrity of higher levels in obligate hierarchies. As a purely eliminative process, Darwinian selection at one level only eliminates the existing individuals at that level and never generates a new substrate for selection at that level, which is the semantic and physical meaning of selection, and is also the essence of Darwinism vs. Lamarckism. However, selection at lower levels, for example molecules and cells, changes the configuration of higher levels and thus generates new phenotypes and genotypes as the substrate for higher-level selection.

The pattern supply in the hierarchy is unidirectionally upward because of the nature of coarse graining organization/hierarchization. A hierarchical level is an organization/hierarchization of its lower-level components, so its changes must be realized through lower-level changes. An organization/hierarchy is a collection of relations between its lower-level components, so it is not an existence independent of its components. Therefore, the lower-level change is the cause and the higher-level change is the coarse-grained effect of cause. As a result, no higher-level change cannot occur without the lower-level change, while the lower-level change can occur without affecting the higher-level. This is the variation reducing by coarse graining. Moreover, genetic information is stored at the bottom of biotic hierarchies, so most of inheritable patterns are transmitted from the bottom. Inheritable patterns at other levels are epigenetic and thus not the principal form of heredity in evolution.

Because of the upward direction of coarse graining and the downward direction of coupled selection, the effects of the evolution at one level on the neighboring higher and lower levels are not symmetrical. The evolution at one level influences the evolution at the adjacent higher level by supplying informational and noninformational configurations/patterns, and influences the evolution at the adjacent lower level by determining the survival of the lower-level entities (Fig. 14). ***The asymmetrical upward pattern transfer and downward selective enslavement are the basic properties of biotic hierarchical evolution and explain many problems in*** ***the evolution of hierarchical life.*** Because one level of the hierarchy can only get patterns from the adjacent lower level, the patterns at the bottom are transformed by coarse graining and passed upward one level by one level. Some patterns at a lower level may be beneficial to a higher level but harmful to that lower level. Selection of the lower levels will eliminate such patterns and make them unavailable for the higher level. For instance, a certain DNA sequence may translate a useful protein for the cell, but such DNA or corresponding RNA is unstable and thus unavailable to the host cell. In other words, the lower-level evolution determines the potential paths and forms of higher-level evolution. This is consistent with the mutationism theory of evolution[50-51, 189]. The influence of mutational bias in the composition of bacterial proteins[83] and the evolution of multiple gene families[51] support that lower-level evolution is the source of the substrate pattern of natural selection. Conversely, some patterns may be beneficial at one level but harmful for the higher levels. Although selection will eliminate the harmful patterns by selecting against the higher-level hosts, such pattern appears repeatedly and decreases the fitness of its host. Trinucleotide disease and cancer are such examples at the DNA and cellular levels, respectively.

Although all patterns available for one level are from the evolution at lower levels, fixation of these patterns in the obligate hierarchy is determined by the evolution of higher levels. This is the essence of selectionism. Therefore, the division to mutation and selection reflects the general environmental action at different levels of the hierarchy: DNA mutation is a type of abiotic selection/transformation of DNA. The difference between hierarchical levels makes the evolution of the hierarchy a balanced result of the selection at all levels. This balanced evolution generates and maintains polymorphism. If selection at one level is very stringent, the evolution of other levels will be constrained by this level: the proper evolution at lower levels will be tightly controlled by the selection at this level, while the evolution at higher levels will be restricted by this level through biased pattern supply. Under this circumstance, information is curtailed to fit the selection at this level and compromises the adaptation to other levels. Sexual selection at cellular level is a typical example of such compromise in adaptation.

## Dissect sexual selection to different levels

No evolutionary entity can be spared from environmental transformation/selection, no matter an independent entity or a dependent part of an entity. Because terrestrial life is hierarchical, the conventional overall selection of organisms can be dissected to different levels, and that clarifies our understanding of evolution. As an important form of natural selection, sexual selection is usually treated as organismal selection. However, sexual selection also occurs at the cellular level. Pre-meiotic germ cell selection, post-meiotic gamete selection, and post-copula gamete competition for fertilization are all sexual selection at cellular level[176-182]. Cellular sexual selection is often simplified as a random genetic drift. Both organismal selection and cellular selection in sex reproduction are evident and have been studied intensively, but no body has paid attention to the conflict between different levels of sexual selection.

As explained in chapter V, sex offers a short-term advantage by increasing the efficiency of selection through gamete selection. To unicellular organisms and primitive multicellular organisms, the functional difference between gamete and organism is small. Therefore, selection of gametes can improve the fitness of full-scale organisms. However, with the increasing complexity and labor division in multicellular organisms, the difference between gamete and organism become greater and greater. Fitness of gametes greatly differs from that of organisms. The conflict between the cellular and organismal levels becomes stronger. Under this condition, excessive gamete competition makes the resulting whole organism less adaptive. The evolutionary conflict between gametes and organisms has actually been raised as early as in 1930s, although the concept of multilevel hierarchy was not explicitly mentioned[190].

Sperm competition is the strongest cellular sexual selection because of its extremely high rate of elimination. Current studies of sperm competition focus on inter-organismal sperm competition, the sperm competition between multiple males mating with one female. The inter-organismal sperm competition is a selection for the fitness of both individual sperms and individual organisms. In contrast, intra-organism sperm competition is stronger at cellular selection because it is a selection of sperms from the same male: the selection is mainly at the cellular level. Therefore, with increasing complexity of hierarchical life, intra-organism sperm competition is weakened through inhibiting post-meiotic gene expression and the intercellular bridges of spermatids[125]. Although pre-meiotic mutations and recombination contribute to the intra-organism variation of gametes, most intra-organism germline variations are meiotic[191-192]. The number of sperms and the volume of ejaculate have to be maintained or even increased for males to compete for fertilization[176], so other mechanisms emerge to weaken the influence of sperm competition on genetic information. The inhibition of post-meiotic gene expression and the intercellular bridges between spermatids make all sperms have the same or a very similar phenotype. These mechanisms decouple genetic information from the phenotype of its cellular host. In this way, the selection of genetic information in intra-organism sperm competition is weakened. Inhibition of cellular selection is so important that post-meiotic gene expression is ubiquitously inhibited in a broad range of animals from Drosophila to human[125, 179]. Inter-organism sperm competition is also inhibited through changes in mating behavior, for example female monogamy, which eliminates the basis of inter-organism sperm competition. Although strict biological monogamy is rare, the inter-organism sperm competition due to polygamy or promiscuity is restrained by non-biological mechanisms, for example social monogamy. Only sexual selection at the organismal level is not weakened at hierarchical complexity increasing, because it is a selection for the fitness of organisms.

Although cellular sexual selection is inhibited, it is not completely shut down because there is a low level of post-meiotic gene expression[176, 178, 182, 193-195]. This is especially important to males because of the extraordinary intensity of sperm competition. Even if the coupled selection in sperm competition is almost completely inhibited, the weak selection for fertilization brings a new level of weakly biased genetic noise, very similar to the weakly biased genetic mutations. Although pre-meiotic DNA replication errors are the principal source of germline mutations, gametic competition for fertilization is an important factor in the selective fixation of those mutations. The bias in fixation is toward the fitness of gametes rather than organisms. One consequence is that the overall fitness of males as a group is lower than that of females because of the much stronger gametic selection in males. Here, the fitness of males, as a fitness of one gender in contrast to the other gender, cannot be measured by the number and survival rate of offspring which results from the sexual reproduction of two genders, but it can be measured as the health and life span of one gender lineage in contrast to the other. Although only Y chromosome is male-specific, males are globally affected by gametic selection more than females. In humans, the spontaneous mutations causing diseases are mostly paternally derived, and some of them bias to male-only origin[192]. The mutations fixed in sperm competition are beneficial to sperms but deleterious to organisms, which is a cause of paternally biased disease genes independent of the more divisions in male germline. Moreover, all male-specific or male-biased genes that are directly or indirectly related to male gamete are more or less tuned by male gamete selection for the fitness of gametes, which compromise male fitness more than female fitness. In other words, male gametic selection preferentially affects male-related genes, which in turn preferentially affects male fitness as compared to females. The female longevity across various animal species supports this theory[196-198].

The female longevity is particularly significant in humans, which may be due to the very large sperm to ovum ratio of humans. The conventional explanation for female longevity has two unexclusive theories. One is the X chromosome hypothesis that the female has one more X chromosome, which has far more genes than the Y chromosome and thus provides advantages for females in longevity[199-203]. However, most genes in the X are inactivated for dosage compensation at the early stage of embryo[204-205]; the expressed genes on the inactive X chromosome are mainly in the pseudoautosomal regions which are homologous to the Y chromosomes[206-207]. Therefore, due to the dosage compensation, the larger size of the X chromosome is effectively trivial and thus cannot account for the female longevity. The other theory proposes that male behavior, such as mate finding and territory defense, put males at higher risk[208]. This theory is actually an organismal version of the above explanation by gametic selection: males have stronger selection at both cellular and organismal levels than females during sexual reproduction. Complementary to each other, both result from the asymmetry between male and female gametes which provides the short-term advantage to sex.

Sperm competition can also influence the evolution of sex chromosomes. Lack of recombination is often considered as the cause of Y chromosome degeneration. Several models have been proposed to explain the Y chromosome degeneration[209]. According to the present theory, the processes in those models are actually different manifestation of the coarse-grained selection under specific contexts, namely the disadvantages of asexual reproduction (chapter V section 3 & 4). However, those models are questioned because the ability and the rate of those processes may not account for the degeneration of Y chromosomes in a short period[209], as in the case of *Drosophila miranda*[210]. The Y chromosome is actually an asexual genetic component in sexual organisms: it never undergoes meiosis, recombination, or genetic fusion in syngamy, the three components of sex[163]. The lower adaptability of asexual evolution is the main cause that the asexual Y chromosome degenerates in contrast to the X chromosomes and autosomes. However, asexuality alone is insufficient to explain the degeneration of Y chromosomes[209-210]. The genome of asexual organisms is much larger than the Y chromosome, but the asexual genome is stable even in long-term evolution. The capacity of asexual genome or chromosome is much larger than the size of Y chromosome. One reason behind the shrinkage of Y chromosome is that Y genes can transfer to other sexual chromosomes while the genes of asexual organisms cannot. Another reason is that the asexual genome is coupled to the organism in natural selection, i.e. the top of the obligate biotic hierarchy, while the Y chromosome is coupled to both the top (organism) and lower (cell) levels of the hierarchy. In the obligate hierarchy, the fate of lower levels (genes and cells) is determined by the highest level (organism), although lower levels have their own evolution. Transfer of Y chromosome genes to X chromosome and autosome weakens the coupling of genetic information to the host gamete and strengthens the coupling to the host organism, because X chromosome and autosome experience less gamete selection. Such transfer is beneficial to the host organism. As a cellular part of sexual selection, gametic selection is an important factor in the Y chromosome degeneration, although it is secondary to the asexuality and thus insufficient to cause the degeneration.

The degeneration of the Y chromosome is similar to the transfer of organellar genome: both are influenced by the conflict between hierarchical levels. However, because endosymbiotic organelles were autonomous and the organellar genes couple to the host organelle all the time, the conflict between the organelles and the cell is the primary force driving the decoupling of organellar genes from the host organelles. In contrast, the Y chromosome only couples to the host sperm in a minor part of its life cycle: most of the time, the Y couples to the organism. Moreover, the sperm selection is weakened through inhibiting post-meiotic gene expression and the intercellular bridges of spermatids[125]. Therefore, hierarchical conflict is only the secondary force for the degeneration of the Y chromosome; in stead, the asexuality of the Y is the primary force for its degeneration.

The subtly difference between asexuality and hierarchical conflict in sex chromosome evolution can be revealed through the detailed analysis of sex chromosome. In the alternation of generations of sexual species, no Y chromosome is recombining, while 2/3 of X chromosomes and all autosomes are recombining. If asexuality is the only force underlying the Y chromosome degeneration, autosomes are the best shelter from the disadvantages of asexuality. In contrast, all Y chromosomes undergo sperm competition, while 1/2 of autosomes and 1/3 of X chromosomes do. As a result, X chromosomes are the best shelter from sperm competition. The X chromosome over-represents genes controlling cognition[211]. Sexual selection of organisms (male competition for mating) has been considered as the reason that cognitive genes are enriched in the X chromosomes[212]. However, sexually selected genes do not have to locate at the X chromosome preferentially, because cognition is beneficial to both genders. A plausible explanation is that cognitive genes locate at the X preferentially to avoid the deleterious effect of gametic selection. As a property of organisms and social groups, cognition is mainly an intercellular activity at organismal level. Although this intercellular activity must be based on specific cellular activities, those cellular activities are not required for gametic struggle. Gametic selection mainly tunes basic cellular mechanisms engaged in gametic struggle. Moreover, as a very late phenotype in the history of life, cognition is not involved in the gametic function, which is very early and primitive, although some cognitive genes have orthologs in very distantly related organisms[211]. As a result, cognition is useless to male gametes in strong competition for fertilization, and thus will deteriorate in strong gametic selection. Because the X chromosome undergoes least gametic selection, cognitive genes tend to emerge at or migrate to the X under the degenerative pressure from gametic selection.

Female longevity, paternally biased disease genes, Y chromosome degeneration, and X chromosome over-representation of cognitive genes have been observed but are never linked to the sexual selection at cellular level in the conventional view of evolution. All these phenomena demonstrate the importance of gamete selection of animals. As a sharp contrast, the consequence of gamete selection in plants is very different from that of animals, although plants have strong gamete selection and the resultant Y chromosome degeneration[213] because of their post-meiotic gene expression[214-215]. The underlying reason is that plant gamete selection takes place as a part of the whole course of cellular selection, because plants do not set aside germline: cellular selection in plants is an intrinsic and thus inseparable part of plant evolution, while sperm competition is only a special period in the life cycle of animals. Cellular selection is the fundamental reason behind the dichotomy of animals and plants (chapter VII).

## Mutationism and selectionism: evolution at different levels of hierarchy

The existence of an entity implies that its configuration is stable throughout its existence. In other words, existence *per se* consists of bias. Similarly, the existence of one hierarchical level must be the result of a stable configuration. Therefore, the configuration of one hierarchical level must bias for its stable existence, rather than evenly distributed in the configuration space. Both Mutational bias and Darwinian selection are the manifestation of this bias at different levels of the hierarchy. In an obligate hierarchy, every level is biased for its own stability/fitness at the cost of other levels. Specifically, lower levels monopolize the supply of patterns to higher levels, while higher levels determine the survival of patterns of lower levels. The bias for one level constraints the exploration of configuration space at other levels, because unbiased exploration of configuration space is crucial for blind evolution to increase stability/fitness and complexity, as explained in chapter I & II. To reduce the bias from lower levels, the landscape of lower levels must be smooth, as the landscape of DNA compared to that of protein. To biotic hierarchy, there is another method to reduce lower-level bias: since DNA is the principal pattern generator/storage, DNA can bypass intermediate levels and directly couple to the top level in natural selection. For example, the germline makes genetic information unaffected by cellular selection, which is why animals are more complex than plants, as will be explained in chapter VII. There are three ways to reduce the bias from higher levels: first, smoothing over the landscape of higher levels; second, reducing the dependence on higher levels for existence/survival; third, weakening the selection at higher levels. This principle applies to all abiotic or biotic hierarchies. Mutationism and selectionism are just the manifestation of this general principle at different levels.

The whole history of life illustrates this general principle of hierarchical evolution. The evolution of genetic information is a balance of the evolution at different hierarchical levels. To increase the complexity and fitness of the top level (the organism), many mechanisms emerge to weaken the bias at lower levels, not only the abiotic evolution of DNA, but also the selection of cells in multiple cellular organisms. However, absolutely unbiased or random evolution is impossible, as the water does not flow on a completely smooth surface. Evolution must be driven by the bias in stability or fitness. Therefore, one of the themes of evolution is to make the landscape of pattern generator/carrier as smooth as possible without affecting the running of pattern formation. First, DNA is chosen as a pattern generator because of it has a smoother landscape than RNA and protein (Fig. 1&4). Second, in heterodomain mapping, the triplet genetic code is degenerative and fault-tolerant, which converts the landscape of DNA to a smoother landscape of genetic information but keeps the dynamics of DNA mutation to drive the evolution of genetic information (Fig. 5&6). Third, as introduced in chapter IV, nuclear compartmentation protects DNA from unsolicited proteinaceous modification(Fig. 10). Fourth, as will be introduced in the next chapter, selection at the cellular level is bypassed through the sequestered germline to reduce the cellular selection-caused bias in the patterns passed upward to the organism.

Strong in early evolution, mutational bias is reduced gradually by these mechanisms. Mutation functions mainly as a driving force for pattern formation, but the pattern of this driving force, i.e. mutational bias, is separated from the generated patterns by the above anti-bias mechanisms. Thus, life approaches unbiased pattern formation and the consequent unbiased exploration of genotype/phenotype space. In other words, ***these mechanisms make mutations to approach a patternless driving force for pattern formation, which reduces the bias in evolution and thus approaches the unbiased exploration of genotype/phenotype space.***

On the other hand, the influence of lower-level bias cannot be completely erased. Trinucleotide repeat diseases are an example of mutational bias[216-217]. "Junk DNA" is also the product of mutational biases. In a general sense, cancer is a cellular bias in multicellular organisms. As nuclear compartmentation and sex underlie the emergence of "junk DNA", cancer also has an evolutionary cause, which will be discussed in the next chapter. Mutational bias is especially strong at the early stage of life when the above anti-bias mechanisms are imperfect or absent. Therefore, the influence of mutation bias on evolution should be more prominent in phylogenetically ancient organisms, such as prokaryotes. To phylogenetically ancient organisms, stasis at low complexity suggests biased exploration of phenotype space due to mutational bias. To modern advanced organisms, the occult influence of mutation bias should be more prominent in phylogenetically ancient and functionally important genes and proteins.

Another more occult influence might be that the mutational bias of DNA/RNA determines the evolutionary path and the developmental form of terrestrial life. All past and present forms of life are based on the patterns generated by mutations. During the origin and early evolution of life, the influence of mutational bias is the greatest. The bias affected the initial direction and path of biotic evolution. Later improvements are only branches of this path. However, life could start from different forms if the mutation was less biased or biased otherwise, as water flows to different sides of the watershed. The essence of life is the complexity and evolvability, not any specific building block, evolutionary path, or developmental form. It would be interesting to design a type of life on the other side of the hill, which has never existed on earth and thus is beyond human imagination so far.

## Altruism: the initiation from the genotype-phenotype division and the fixation by hierarchical conflict

The key to understanding altruism is to distinguish the initiation of altruism and the fixation altruism. Here, altruism refers to strong altruism, an action that decreases the overall fitness of the actor and benefits other individuals. As explained above, every level of the hierarchy has a different property and landscape. The effect of a genotype depends on the specific level of the hierarchy. Altruistic behavior is harmful to the individual actors but beneficial to the higher level. The advantage at the higher level fixes the altruistic phenotype of lower-level individuals, even when the hierarchy is facultative (Fig. 15). The pressure to fix it is partially determined by the intensity of hierarchization, namely the dependence of lower-level components on the hierarchy. In the obligate hierarchy, altruism is intense and conspicuous because the survival of lower-level entities completely depends on higher levels. For example, in animals, all somatic cells abandon their vertical transmission of genetic information. Actually, the molecular division of labor to DNA/RNA pattern formation and protein functional action is the earliest strong altruism which heralded the beginning of life [156].

In the conventional view of evolution, the initiation of altruism is not clearly distinguished from the fixation of altruism. Selective pressure *per se* cannot initiate the pattern change that leads to altruism. As explained above, the inter-level relations in the hierarchy are asymmetrical: enslaving selection is downward while pattern supply is upward. Because of this asymmetry, the conflict between different levels of hierarchy can only explain the fixation of altruism but not its origin. Specifically, the higher level can select for the altruistic phenotype at the lower level, but cannot generate an altruistic phenotype at the lower level for selection, because all patterns are from the lower levels.

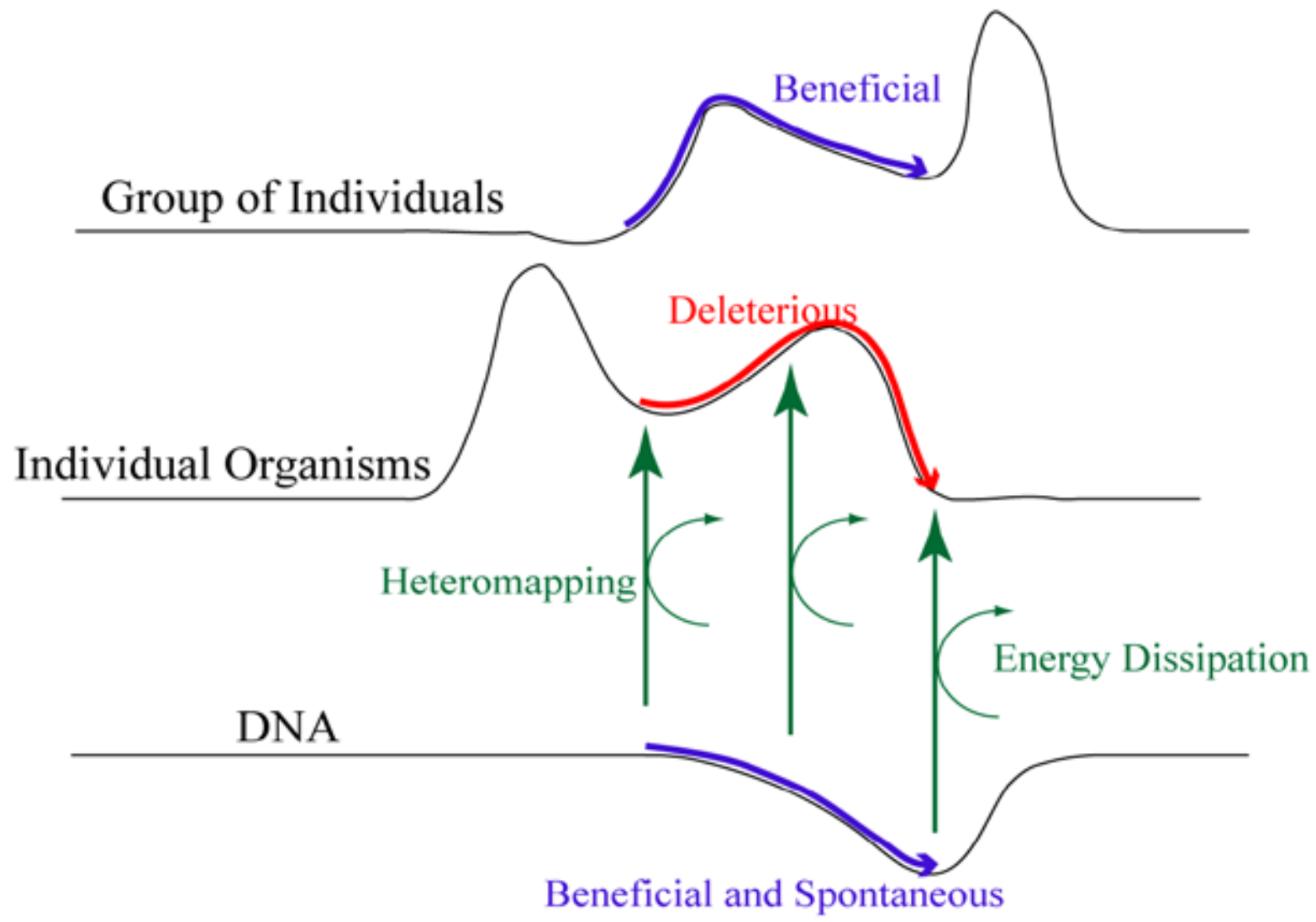

**Fig. 15. The origin of altruism.** Altruism reflects the hierarchical conflicts due to the different evolutionary landscapes at various levels of hierarchy. In hierarchy, higher levels enslave the evolution of low levels, because low-level entities couple to the higher-level host in natural selection. Such downward enslavement explains the fixation of altruism driven by the advantage of the higher levels. However, the origin of altruism is a puzzle: higher levels only select lower levels but cannot change the patterns in lower levels, and thus cannot account for the initiation of altruism. The cause for the origin of altruism is that the energy dissipation in translation enables the component DNA/RNA to enslave the evolution of the higher-level host organism.

Altruism must be disadvantageous to its donor. However, not all disadvantageous changes can initiate altruism: only those crossing the ridge on the landscape can initiate altruism; deleterious lower-level degeneration and environmental change only remove or lower the ridge (*Not all disadvantageous changes can initiate altruism,* Suppl. Text 6). Crossing ridges is the benefit of altruism because it promotes unbiased sampling of configuration despite immediate disadvantages and thus may lead to more complex and stable/fit configurations. The real puzzle of altruism is the origin of altruism: how blind evolution overcomes the immediate disadvantage to establish altruism with distant and unknown long-term benefit to the group of individuals. In other words, environmental action, either transformation or selection, drives evolutionary entities to the stable or fit configuration. In a hierarchical entity, every level can be in conflict with lower and higher levels. How can a level be in conflict with itself? Namely, the paradoxical self-conflict is impossible. However, with a special mechanism, a part (the component) can command the whole (the host entity) and produce a phenotype of the host that is beneficial to that part but deleterious to the whole.

This special mechanism is the labor division of biotic evolution to pattern formation and functional action. As a result, altruism is the characteristic of life. To nonlife, evolution is utterly rigid and “short-sighted”, because pattern formation and functional action are inherently bonded. Although an abiotic entity can be in conflict with its higher-level host, it is never in conflict with itself. As an example of the rigidity of abiotic evolution, water must freeze below the freezing point because its solid state is more stable than its liquid state. An abiotic entity always evolves rigidly and never deviates from the rigid evolutionary path which is determined by its immediate stability/fitness. Energy can change the shape of landscape but never changes the rigidity of abiotic evolution. ***For life, evolution is flexible because pattern formation is separated from functional action; pattern formation and functional action are in separate domains with different evolutionary landscapes; therefore, the evolution at lower levels provides the pattern that instruct higher-level host to deviate from its immediate fitness.*** A genotype may be harmful to the host in the target functional domain but that pattern can be beneficial and thus spontaneous in the source pattern domain. The projection of pattern to functional action is fulfilled by heterodomain mapping, whose fixed energy flow ensures that pattern is generated in the target protein domain according to the pattern in the source DNA/RNA domain (Fig. 15). ***It is the unidirectional energy flow in heterodomain mapping that enables the DNA/RNA to enslave the evolution of the higher-level host, against the general downward enslavement in abiotic hierarchies.*** In other words, with energy dissipation, a part can command the whole. For example, the evolution of DNA can generate a mutation that is directly harmful to its cellular host and initiate the altruistic phenotype of the host; this deleterious mutation must be beneficial to DNA (more stable) so it occurs spontaneously. In this way, the evolution of host is not blocked by its immediate disadvantage. When combined with the selection from higher levels, unbiased configuration sampling finally results in "flexibility" and “foresight”: altruism is only one of manifestations. ***Energy dissipation is the ultimate driving force for all upward evolution on the evolutionary landscape, including the macroscopic upward evolution, such as the altruism of life.***

Another prerequisite for altruism is that the immediate disadvantage against altruism is not prohibitive to all carriers. The disadvantage may eliminate some but not all carriers. If the immediate disadvantage is completely prohibitive, the disadvantaged state eliminates all carriers, and the pattern underlying altruism becomes extinct; the road to altruism is blocked completely. In evolution, disadvantage generates a low fitness niche while advantage generates a high fitness niche. Low fitness does not necessarily mean zero fitness. As long as there is a non-zero fitness for altruism carrier, advantages from other levels will help to overcome the ridge (Fig. 15). A lineage with low fitness but high evolvability still can thrive in future, because it can gain higher and higher fitness through its high evolvability. In this sense, evolvability is more fundamental and thus more important than fitness. That is why the watershed between nonlife and life is a mechanism of evolvability rather than fitness. As a metaphor, fitness is the first-order derivative of evolution while evolvability is the second-order derivative of evolution, as the velocity is the first-order (time) derivative of position while acceleration is the second-order derivative of position.

The significance of altruism is beyond the unselfishness in social behavior. Labor division to pattern formation and functional action results in altruism, which librates evolution from the rigid environmental constraint. Unbiased sampling of configuration space is crucial for the increase of complexity and fitness because of the blindness of evolution. Unbiased sampling of configuration space often needs to overcome immediate disadvantages and not to be trapped in immediate advantages which lead to a dead end. For instance, somatic cells of animal give up the transmission of their own genes; sex has an immediate disadvantage in the number of offspring, i.e. the two-fold cost of sex. The immediate benefit may be absent or insufficient to compensate for the immediate disadvantage. Rigid abiotic evolution cannot cross these ridges to increase complexity and fitness. Only the labor division of internal evolution to pattern formation and functional action can overcome this ridge. ***That is why molecular altruism, the labor division to DNA/RNA pattern formation and protein functional action, is the starting point of life.*** When combined with natural selection, the flexibility results in the “foresight” of biotic evolution, which finally leads to the miraculous complexity of life. The so-called intelligence of human is actually a neural crystallization of that “foresight” of biotic evolution. From the angle of individual humans, the initiation of altruism is in conflict with the short-term benefit. Therefore, such initiation appears irrational or erroneous to the intelligent human, such as “trembling hands”, “fuzzy minds”, or the “defective genotype”[218-221]. If the subjectiveness and teleology of human consciousness are removed, altruism is actually as common as the selfishness in biotic evolution from the beginning of life. It could be misleading to apply the game theory to the study of evolution[6, 54], because the game theory assumes intelligence and rationalness, which are absent in most time of life history. The prisoner’s dilemma does not exist in the life history before the emergence of intelligence; it is only human’s retrospective analysis of life history. Even humans are not perfectly intelligent and rational. It is anthropocentric to use an evolutionarily very late derivative of human to explain earlier events of evolution. The so-called kin selection is another anthropocentric interpretation of kinship-related altruism.

### Genetic kinship enhances the evolvability of altruism but kin selection is unreal

If the disadvantage of altruism cannot be prohibitive, how is it possible that some altruists are completely sterile, for example worker ants or somatic cells of an animal? The answer is that although they are sterile and cannot pass their altruistic patterns, other individuals from the same ancestor, namely their kin, carry altruistic patterns and thus are able to pass altruism. However, such continuance of altruism is driven by hierarchical selection, namely group selection, not by the so-called kin selection[222-224]. In order to understand this, the role of genetic heredity in the emergence and evolution of altruism needs to be analyzed.

Although the initiation of altruistic phenotype requires genetic mapping, its operation and further development do not require genetic heredity. Organisms from different species can form an ecological group as a higher-level evolutionary entity[223-224]. The relational patterns between these organism lineages are ecological rather than genetic. In other words, the altruistic relationship between altruistic donors and recipients can be noninformational. Because of the low evolvability of noninformation (chapter III and IV) such altruism is unstable and of low complexity. In contrast, the relationship between organisms from a common ancestor can be encoded as genetic information, which has much better evolvability than noninformation (chapter III and IV). The altruism that is encoded and inherited as information can acquire more complexity and stability/fitness in evolution than noninformational altruism. The genetically related altruistic donors and recipients are actually the performers of the genetic strategy of the common ancestor. Both cooperation and conflict are a part of this genetic strategy. Moreover, ***the higher percentage of genetic information in the relational patterns between the individual of an altruistic group, the higher evolvability the altruism in this group has, and, as a result, the more complex and stronger the altruism is.*** When the donors and recipients are more closely related, the percentage of genetic information in their relationship increases: the interaction between them is encoded as genetic information more than it is between those loosely related or non-related. Therefore, the altruism between closer relatives has better evolvability and higher complexity than that between less close relatives. ***That is why complex and stable altruism usually occurs in a genetically related group and why altruism is apparently stronger with the increasing kinship.*** However, at the same time, other relations between closer relatives, such as competition, also have better evolvability and higher complexity. The interactions between kin are not uniformly beneficial.

The concept of kin selection was raised as early as 1930s[225]. It has been widely accepted since the term "kin selection" was created and defined in 1964: "By kin selection I mean the evolution of characteristics which favor the survival of close relatives of the affected individual, by processes which do not require any discontinuities in the population breeding structure"[226]. However, kin selection is not a concrete form of selection as sexual selection, although kin altruism indeed exists. Selection, either conventional natural selection or sexual selection, is the destruction of low stability/fitness entities in the fluctuating process of existence. The entities escaped from the destruction can have further evolution, for example survival and reproduction, but such further evolution is the consequence of selection, not the process of selection *per se*. Therefore, selection must be negative elimination, which is not only the scientific and semantic meaning of selection but also the essence of Darwinian selection vs. Lamarckian transformation: evolution by selecting existing diversity vs. evolution by *de novo* producing diversity. Any form of positive "selection" that favors or promotes the survival/existence or reproduction/replication of evolutionary entities is not a concrete form of selection. Therefore, kin "selection" by favoring the survival of relatives according to the degree in genetic relatedness cannot be true.

Moreover, selection only recognizes stability/fitness. Other properties are only indirectly selected through their influence on stability/fitness. To a defined entity in a defined environment, a property has a defined value for stability/fitness and is selected accordingly. However, kinship is not such a property: it is only a relationship between the individuals of a group; the property behind the kinship varies to different individual and in different environment. To individuals, the property of a kinship could be beneficial (altruism) or harmful (competition)[224, 227-230]. To the group, kinship is actually one of the forms of complex organization of the group, so it has both altruistic and competitive and other more complicated properties, definitely not uniformly altruistic[224, 227-230]. Organisms have to survive and reproduce to avoid the elimination of their lineage, and that is why natural selection and sexual selection exist concretely. However, organisms do not have to save their relatives in order to survive and reproduce. As well as competition and selfishness, kin altruism is only one type of the inter-organismal relations encoded as genetic information. Kinship is not fundamentally different from other types of inter-organismal relations, and kin interaction is not fundamentally different from other inter-organismal interactions. The concept of kin selection is an anthropocentric sorting of altruistic kin relations from various kin relations. Natural selection recognizes fitness and sexual selection recognizes capability for reproduction, because those selections are actually a physical process of destruction according to the stability/fitness of an organism or a lineage (a tautology!). In contrast, there is no physical basis for kin recognition and selection, because kinship is an abstract relation rather a physical property and cannot be recognized by blind evolution. Only intelligent life recognizes kinship. The proposed mechanisms of kin selection, including kin recognition, are an anthropocentric interpretation of the common interactions between kin as kinship-specific, although these interactions are not fundamentally different from the interactions between non-kin. In short, the concept of kin selection is an anthropocentric interpretation of the importance of genetic information in the evolvability of inter-organismal relations and organization.

Although complex and stable altruism can only develop inside a species, or more strictly, an interbreeding group because of the role of genetic heredity in the development of altruism, it does not mean that altruism only occurs inside a species. Altruism between genetically unrelated entities has also been demonstrated[223-224]. The complex relations inside an ecological group, either intraspecies or interspecies, often consist of altruism. A deleterious phenotypic change of a species is beneficial to the competitors and predators of this species. Such relation is a strong altruism because the adverse phenotype is purely harmful to its donor/carrier and beneficial to the recipient. This type of altruism still evolves in noninformational (non-genetic and thus unstable) ecological relations, instead of genetic relations. The initiation of such deleterious (thus altruistic) changes still require the labor division, but the relation between the host species of deleterious change (altruistic donor) and the beneficiary species (altruistic recipient) are purely non-genetic. Moreover, such inter-species altruism develops under the selective pressure at the level of ecological group of species, a level higher than individual species, and confers stability/evolvability to the host ecological group. The evolvability and complexity of ecological altruism is much lower than genetic altruism (kin altruism) because the altruistic pattern is not information, which could be a reason why ecological altruism is overlooked.

Therefore, genetic relatedness is not required for altruism to evolve. The (genetic) kin altruism is only a special and advanced form of general altruism. Genetic kinship only increases the evolvability of altruism, but it is neither the cause nor the prerequisite for altruism. Genetic kinship is only a special and advanced form of patterns encoding altruism. With the emergence and evolution of consciousness and neural information, the pattern of altruism is extended from genetic relations to cultural relations and the corresponding kin altruism is extended to individuals that are culturally but not genetically related.

## The neutral and nearly neutral theories

If a change is neutral at one level, then its evolution at that level, either occurrence or fixation, will be completely random and unbiased. In other words, neutral evolution will be completely random, and vice versa. However, no completely random process has been found so far, even at the level of elementary particles[231]. A change must be due to an uneven landscape, namely a biased evolution at the corresponding level. A mutation must be due to the differential stability of DNA and a selection must be due to the differential fitness of organisms. Completely smooth landscape or complete randomness does not exist. Most “random” processes are nearly or practically random processes, which are made up of numerous non-random sub-processes. These “random” processes are still non-random, but are impervious to precise analysis. Occurrence of a genetic change at the level of DNA/RNA is a motion at an unstable position on the landscape of DNA/RNA. In consideration of our current capability of modeling and computation, the hypothesis of randomness or neutrality may be a necessary simplification for quantitative study of complex phenomena, for example biological evolution. However, such simplification could bring undesirable effects together with the resulting convenience. Although the bias in DNA/RNA mutation may be very weak, the subtle effect of mutational bias is no longer negligible in long-term evolution or at early stage of life when anti-bias mechanisms are absent. In the modern evolutionary synthesis, evolution is treated as “shifting gene frequencies” driven by Darwinian selection of organism, while evolution at lower levels, for example mutational bias, is not fully considered[7-9, 49-51]. Although mutation is considered as a source of novelty, the bias of mutation is not taken into consideration. The modern synthesis may explain the short-term evolution at the late stage when the effect of changes in intergenic relation is much greater than the innovation of genes by mutation[75-78], but cannot explain the origin and early evolution of life or the long-term phenomena such as the emergence and dichotomy of animals and plants.

As the first theory introducing lower-level evolution, the neutral theory[232-233] could have brought a much greater change in life sciences if people recognize the importance of internal lower-level evolution through this theory. However, because the conventional view overemphasizes organismal selection, the diverse and complex molecular evolution has been reduced to a random and neutral process, as a null hypothesis of organismal selection! The modification by the nearly neutral theory, although a progress in replacing the strict neutrality at organismal level with the bias in organismal selection[234-236], also only emphasizes the organismal effect of molecular evolution. The importance of molecular evolution *per se* is still overlooked. In the conventional view, evolution is just the passive gene frequency change in the organismal selection by the external environment. The diverse and complex lower-level evolution is simplified to random changes. All deviations from randomness are attributed to organismal selection, including those actually caused by lower-level evolution. As a consequence, the current theory of evolution has problems in explaining molecular evolution, such as the low level of polymorphism, intolerable death of organism, and the great variation of the molecular clock[232, 236-237].

The polymorphism of genetic information is the result of the balance between various evolution at different levels, as well as the environmental changes[21, 238]. In the neutral theory, the polymorphism resulting from mutation-selection balance depends on the effective population size and mutation rate. However, the level of protein polymorphism/heterozygosity is smaller than that predicted by the neutral theory[236, 239-240]. The traditional notion of balance selection can explain the polymorphism, but the consequent genetic load due to the strong selection is intolerable, namely intolerable organismal death[241]. Similarly, the explanation provided by the nearly neutral theory also incurs a similar excessive genetic load[237]. Actually, focusing only at the organismal effect of molecular evolution brings intolerable death of organism in explaining molecular evolution, not only about polymorphism but also about the rate of molecular evolution[232]. The reason is that lower-level selection/transformation influences molecular and cellular evolution without causing organismal death. The "excessive" genetic load is on lower-level molecules and cells rather than host organisms, so it does not cause death of individual organisms: it only affects the diversity of patterns available to the organism. For example, because some genes are disadvantaged in gametic or pre-gametic selection, for example gamete killers and some B chromosomes, those genes are eliminated together with the host germ cell[176-178, 182, 242]. Such elimination, however, does not cause death in the downstream organisms.

At the molecular level, the diversity of biased mutation is less than that of random mutation. All causes of mutation have a specific pattern in the mutations they bring, and mutations are thus not evenly distributed. For example, deamination has a C to U and 5-methylcytosine to T pattern; replication slippage usually occurs at repetitive sequences[21]; the GC mutational pressure even influences the amino acid composition of proteins in bacteria[83]; the mutational bias to transition over transversion reduces the rate of nonsynonymous mutation and the proportion of polar changes among nonsynonymous mutations[243]. These observations evidence that mutation is limited in some patterns instead of being evenly distributed in all possible patterns as the hypothetic random mutation does. Therefore, molecular evolution *per se* has reduced polymorphism without organismal death. As a result of sub-organismal evolution, polymorphism is reduced without incurring the death of organism. The conventional theories only consider the selective elimination at the level of organism and overlook the elimination at lower levels. That is why conventional theories cannot explain the low range of polymorphism. The relatively invariant level of polymorphism across various species with greatly different population size[236, 239-240] suggests that the balance between hierarchical levels, not population size or environmental effect, is the principal determinant of polymorphism. Another cause for the smaller polymorphism is hitchhiking[236, 244-246], because it deviates sexual reproduction from idealized random genetic shuffling. Actually, any process that deviates sex from random genetic shuffling decreases the resolution of Darwinian selection on genetic entities through sex, as explained in chapter V. As a result, the genome shaped by Darwinian selection has fewer details and hence is coarser than in the idealized sex of random shuffling. Although non-random genetic exchange contributes to the spatial variance of polymorphism inside genome, it is largely unknown how important it is in the inter-species variance of polymorphism. Such decrease in polymorphism results from the coarse graining of Darwinian selection at the organismal level. Actually, coarse graining at various levels of the hierarchy not only extends the range of near neutrality[247] but also accounts for the role of effective neutrality in evolution: it smoothes out the biases in evolution and the developmental noise for the fitness and complexity of the host organism (chapter V).

Similarly, the molecular clock, the evolutionary rate of genetic information, is influenced by all levels of hierarchical life. The neutral theory only considers the levels of genes and organisms. Selection/transformation at other levels also contributes to the variation of clock, such as cellular selection. Moreover, selection/transformation at lower levels is biased and thus does not result in a Poisson distribution[36]. Thus, the variation of the molecular clock is higher than the calculation according to the neutral theory. One of the prerequisites of the molecular clock is the constant and random occurrence of neutral or nearly neutral mutation, which only holds approximately at current stage of life with many anti-bias mechanisms. It is risky to extrapolate this prerequisite to the early stage, because the emergence of some important anti-bias mechanisms may have changed the rate of molecular evolution significantly. First, nuclear compartmentation reduces the mutational bias and increases the evolutionary rate of noncoding sequence, as explained in chapter IV. Second, sex fine-grains the organismal selection of genome and thus transforms one uniform rate of the whole genome to various rates of individual genetic elements, as explained in chapter V. This transformation may dramatically change the rate of (effectively) neutral substitution and accelerate that of fixation/elimination of non-neutral substitutions. Third, mass extinction may reduce the selective pressure of the species competing with the dying species. As a result, the organismal selection of genome is relaxed and the effect of lower-level selection is magnified, as a result of loss of equilibrium between different levels. The molecular clock in the evolution of early organisms couldn't be represented by the evolution of modern analogues and thus need to be reframed.

Molecular evolution is as diverse and colorful as organismal evolution. Selective pressure from the host organism is neither constant nor uniform to a gene[240, 248-250]; similarly, at the molecular level, mutational rate and bias are neither temporally constant nor spatially uniform[243]. The unit of molecular evolution is the nucleotide, not the gene. Molecular evolution should be individualized in our study of molecular evolution, as individualizing organisms in our study of organismal evolution. After all, the principal part of the diversity and complexity of terrestrial life is encoded in the genome. In consideration of our current capability of modeling and computation, simplification may be a necessary compromise for quantitative study of complex phenomena, for example biological evolution. However, it is unlikely to simplify evolution without adverse consequence[251], especially in the study of early or long-term evolution.

Although the proportion of neutral/nearly neutral evolution could be lower than what the theories originally proposed, effectively neutral evolution is still a valid mode of evolution, because organismal selection cannot discern very weakly beneficial or harmful changes[240, 248]. A better interpretation is that the so-called neutral/nearly neutral evolution is actually the evolution of one of the lower levels of the hierarchical life. The evolution of hierarchical entities is a balance of all levels. Lower-level evolution provides pattern for higher levels and thus is the basis of the evolution of the entire hierarchy, and higher levels enslave the lower levels through coupled selection. For example, the minor allele frequency of some functionally important polymorphic sites is about 1% - 10%, significantly lower than the frequency of classic Mendelian disease genes[9, 252-253]. This observation suggests that such type of polymorphism is the variation of lower-level substrate rather than the result of Darwinian selection. Although the pattern in such polymorphism is not optimal for organismal fitness, it is the only building block available for evolutionary tinkering[5].

Because of the universality of unidirectional translation and the paucity of retrotranscription, the evolution of all molecules but DNA only indirectly influences genomic evolution through changing organismal fitness, and never directly modifies genomic evolution as mutational bias does. Moreover, the weak bias in mutation is further reduced by the smoothing effect of genetic coding and the protective effect of nuclear compartmentation. All these mechanisms reduce mutational bias and increase the evolvability of biotic evolution, but also make it very difficult to find a negative control to prove the proposed role of molecular bias in biotic evolution. However, at the cellular level, there is an almost all-or-none contrast to prove the role of lower-level cellular selection in organismal evolution: the absence and the early-specification of germline in plants and animals, respectively, demonstrate that cellular selection drives the dichotomy of animals and plants. Such effect of lower-level cellular evolution is actually a manifestation of the extended central dogma in the hierarchy.

## VII. The Extended Central Dogma in Hierarchy – the Integration of Development and Ecology into the General Theory of Evolution

The early-specified germline in animals is sequestered from functional differentiation and cellular selection. According to the principle of coupled selection, such suppression of cellar selection should have influenced the evolution of multicellular organisms. The role of germline in evolution has been overlooked for long time, because conventional theories haven't clearly recognized the principle of coupled selection of genetic information in the multilevel hierarchy. Only after the four elements of biotic evolution, heteromapping, coupled selection, coarse graining, and hierarchization, are elucidated in previous chapters as necessary preparations, can it be revealed that the early-specified germline accounts for the dichotomy of animals and plants and the differences between them in nourishment, motility, cell fate, development, and oncogenesis.

## The upward extension of the central dogma in the multilevel hierarchy

The upward extension of the central dogma applies to the complexity increase in the hierarchy, which is a noninformational extension of heteromapping and gives full play to genetic evolution. In life with more than one level higher than the basal level, for example the multicellular life, the selection of genetic information can be coupled to any one of higher levels. When information couples to a specific level of informational output, the complexity of that level will increase at the cost of other levels. In order for the organism to be as complex as possible, the coupling of genetic information to other levels should be minimized. This extended central dogma in the multilevel hierarchy involves the complicated inter-level relations.

If not specifically suppressed, natural selection occurs at every level of the hierarchy. The evolution of such a hierarchy is an average of all levels, namely the average of the whole and the components. Because of this character, this type of hierarchy is a distributed hierarchy with compromised complexity. Because a level of the hierarchy can only obtain patterns from the adjacent lower level, the informational patterns are transformed by coarse graining and passed upward one level by one level. The patterns generated by heterodomain mapping are selected at every level in this type of hierarchy. The patterns at the peaks of the landscape of a level are unstable. Therefore, the pattern that is at the peak of any level is sifted out (Fig. 16). The informational patterns from the bottom are filtered by the selection at every level until they reach the top level – the entire hierarchy. The screening of the patterns is unfavorable for unbiased sampling of the entire phenotype space. Therefore, ***in distributed hierarchies, the complexity increase of the top level, i.e. the whole hierarchy, is compromised by the pattern selection at intermediate levels. In order to reduce the screening of patterns, the landscapes of the intermediate levels between the bottom and the top should be as flat as possible, and that is another characteristic of the distributed hierarchy.*** When the intermediate levels are relatively flat, the intermediate entities can switch between different configurations. To cells, such a property is pluripotency. Another consequence of the flat landscape is the plasticity and the responsiveness of intermediate entities to the external environment, because there is no barrier in the response to the environment (Fig. 16). Consistently, the plasticity of intermediate entities reduces the pattern screening by intermediate levels, and thus promotes the unbiased sampling of configuration space. The distributed hierarchies that keep a rugged landscape at intermediate levels fail to acquire complexity as much as those hierarchies that keep smooth landscape at intermediate levels.

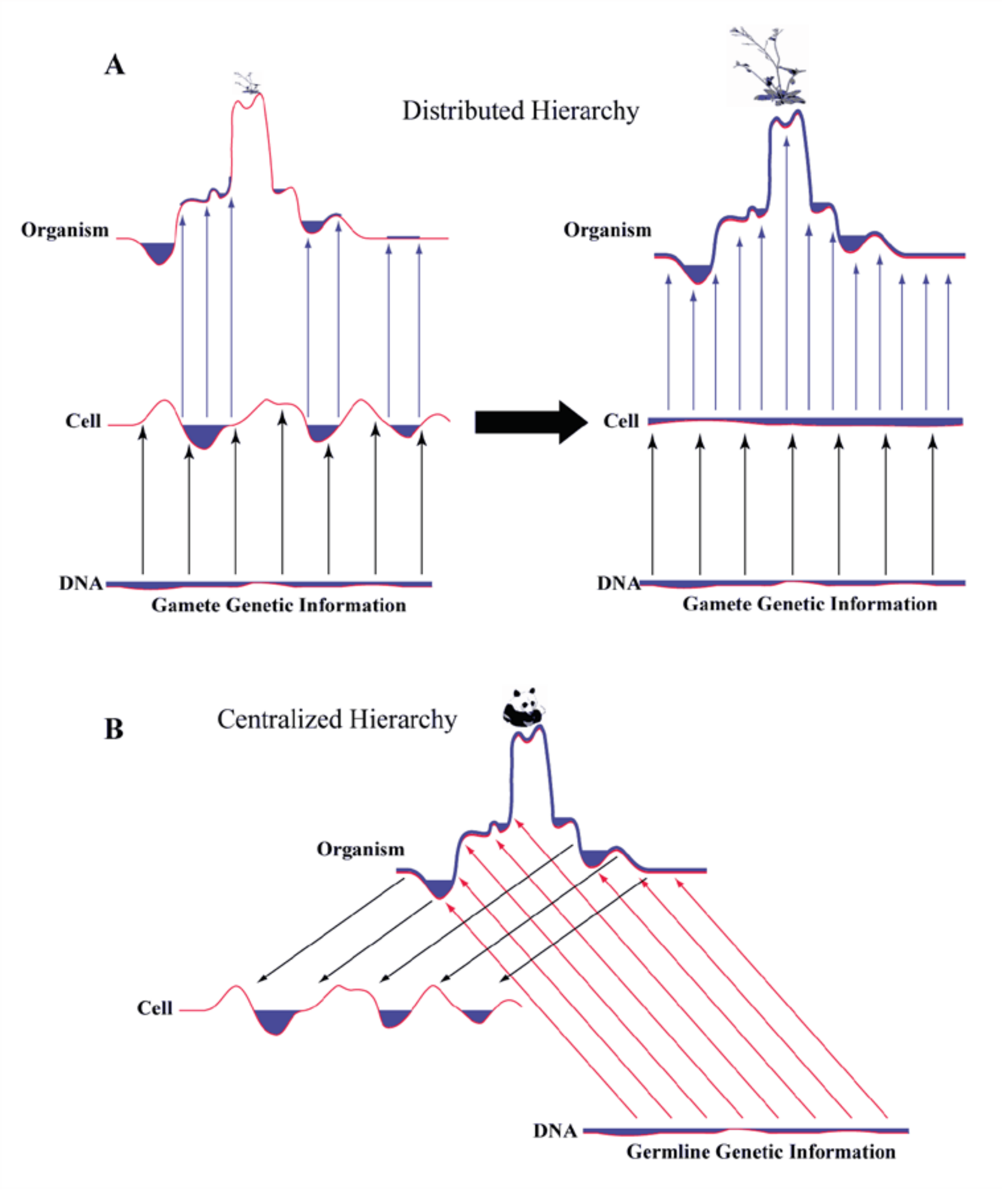

**Fig. 16. The selection in the hierarchy influences the evolution and development of the hierarchy. A.** In the distributed hierarchy, because patterns are transferred upwardly one level by one level through coarse graining, the patterns from the bottom are selected at every level. The intermediate levels between the top and the bottom have to be smooth to reduce the screening of patterns. Otherwise, the patterns are screened at the intermediate levels and the patterns transferred to the top are severely biased. Such bias is harmful to the unbiased probing of phenotype space and thus constrains the complexity increase. The smooth landscape of intermediate levels leads to the pluripotency, plasticity, responsiveness to the environment, and autotrophy in intermediate entities. The plant is an example of the distributed hierarchy. **B.** In the centralized hierarchy, selection at intermediate levels is inhibited. The patterns from the bottom are directly transferred to the top without screening. Therefore, intermediate levels can have a rugged landscape, which is beneficial to the functional action and fitness of hierarchy. The rugged landscape leads to the determined cell fate, autonomy, and heterotrophy in intermediate entities. An example of the centralized hierarchy is the animal, which inhibits intermediate cellular selection of genetic information through the early specification of germline.

In contrast, in a centralized hierarchy, the genetic information is directly projected to the top of hierarchy without intermediate passing, because genetic information only couples to the top level in natural selection (Fig. 16). ***Selection of genetic information at intermediated levels is weakened or stopped.*** Therefore, the complexity of the top level, i.e. the whole hierarchy, is efficiently increased in evolution. ***The centralized hierarchy evolves as an indivisible whole: its components do not have an independent position in evolution. Moreover, because intermediate levels do not screen patterns, the intermediate entities can have a rugged landscape.*** Because rugged landscapes have better functional activity than smooth ones, centralized hierarchies tend to have rugged landscapes for their intermediate levels under selective pressure. Rugged landscapes make the intermediate levels less responsive to the environment, and more autonomous because of the barriers. It is more difficult for the intermediate entities of centralized hierarchy to switch between various configurations than it is for the distributed hierarchy (Fig. 16).

Plants are the distributed hierarchy while animals are the centralized hierarchy. The cause for this division is that animal have an early-specified germline while plants do not have a specified germline[254-255]. The early-specified germline of animals suppresses the selection at the level of cell and couples genetic information to the top level of hierarchy, namely the host organism, because the germline is sequestered from functional differentiation and selection. The early-specified germline explains not only the greater complexity of animals than that of plants, but also the differences between them in motility, cell fate, development, nutrition, and oncogenesis (Table 1). Under the framework of the extended central dogma, not only the development but also the ecology of life are united with the evolution of life.

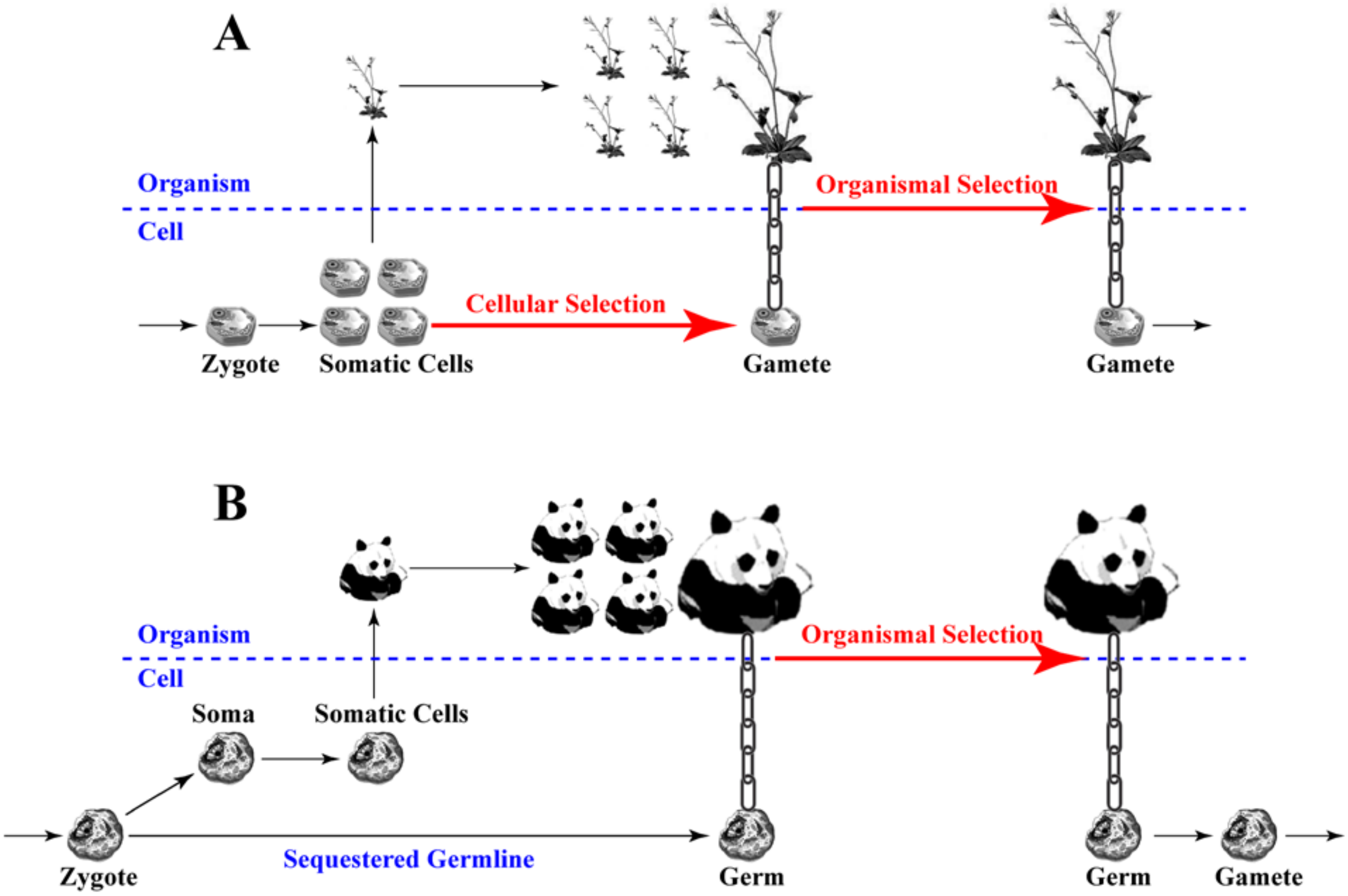

**Fig. 17. The role of germline. A.** In plants, there is no specified germline: gametes differentiate from somatic cells, and thus they undergo many divisions and various differentiations and perform various somatic functions. During these processes, future gametes are subjected to various somatic mutations and selections. Therefore, genetic information undergoes cellular selection as well as organismal selection. The genetic information must be trimmed to fit the cellular niche as well as the organismal niche. **B.** In animals, the early-specified germline is determined at the embryonic stage. The germline undergoes much fewer divisions than somatic cells, remains undifferentiated, and does not participate in any somatic function. Consequently, the germline is sequestered and only subjected to organismal selection. The early-specified germline explains not only the much greater complexity of animals than that of plants, but also the differences between them in motility, cell fate, development, nutrition, and oncogenesis.

## The labor division of cells into pattern generator and functional performer: the germ-soma division explains the difference between plant and animal

To the multicellular forms of hierarchical life, the germ-soma division is the labor division of cells to pattern formation and functional action, while genotype-phenotype division is the labor division of molecules. Specifically, the selection of individual cells constrains the evolution of a multicellular organism as a whole. The evolutionary advantage of germline is to suppress the selection at the cellular level and thus strengthen the selection at the organismal level. Most animals have the early-specified germline that is determined at the embryonic stage[256-257]. This early-specified germline undergoes much fewer divisions than somatic cells, remains undifferentiated, and does not participate in any somatic function. ***Consequently, the germline is sequestered and*** ***only subjected to the selection at the level of whole organism*** (Fig. 17). Because only the germline can pass genetic information to the offspring, it is a generator and preserver of genetic information, while soma is a purely functional performer. Therefore, the intermediate level of cells does not have any significant role in the evolution of genetic information. Since the selection at the intermediate cellular level does not affect germline information, animals evolve as a centralized hierarchical entity, as defined in the last section. Similar to the transfer of organellar genes to the nuclear genome and the degeneration of the Y chromosome, the specified germline is an embodiment of the principle of coupled selection in the hierarchical life: because the cell is the only residence of genes, the genetic information of a multicellular organism cannot be kept outside of the cell; therefore, specification of a line of sequestered cells to keep genetic information is the only feasible method to shift the coupling of genetic information from the cell to the multicellular organism.

In contrast, plants do not have a specified germline. Plant gametes derive from somatic (vegetative) cells, which undergo many divisions and various differentiations and perform a multitude of somatic functions[254-255]. During these processes, future gametes are subjected to various somatic mutations and selections, and thus become adapted to various somatic niches[254-255]. Because selection at the intermediate cellular level affects the genetic information passed to the offspring, plants are the distributed hierarchy. During long-term evolution, the selection at the level of cell averages out the selection at the level of whole organism with which the gametes are coupled. According to the extended central dogma, coupling of genetic information to intermediate levels decreases the complexity of the host organism in evolution. Therefore, the evolvability and complexity of plants is less than those of animals. Animals have more types of cell and organ; animal metabolism is more complex than that of plant, although autotrophic plant metabolism has a more basic position in food chain; plants have larger genome than animals only because of the increase in ploidy[*] rather than in the number of genes and regulatory elements; the most convincing reason is that the neural system and consciousness emerge in animals instead of plants, which is an compelling evidence that animals are more evolvable and complex.

[*] Polyploidy of plants results from its smooth landscape of the whole genome, because smooth landscapes are linearly addable. The smooth landscape of plant genome maps to the smooth landscape of plant cells. This hypothesis will be discussed in another paper.

**Table 1. A comparison between plants and animals***

| | Plant | Animal |
|---|---|---|
| **Germline** | No Germ–Soma Division | Early Specified Germline |
| **Development** | Continuous, Postembryonic, and Plastic | Brief, Embryonic, and Predetermined |
| **Cell Fate** | Determined by Extrinsic Positional Information | Predetermined by/Committed to Intrinsic Lineage |
| **Cell Growth** | Indeterminate Growth with Indefinite Lifetime | Limited in Replicative Capacity due to Telomere |
| **Cell Differentiation** | Reversible and thus Pluripotential | Irreversible and thus Committed |
| **Defence & Repair** | Developmental Response | Highly Specialized Immune Response |
| **Cell Motility** | No Cell Movement on Solid Surface, No Myosin II | Amoeboid Movement on Solid Surface with Myosin II |
| **Cytokinesis** | *de novo* Cell Wall Formation | Contractile |
| **Nutrition** | Autotrophy | Heterotrophy |
| **Complexity** | Fewer Organs and Cell Types | A Greater Variety of Organs and Cell Types |
| **Tumor** | Benign and Non-metastatic | Malignant and Metastatic |

*According to the references [254-255, 258-263].

The consequences of the specific germline are not only in the degree of complexity but also in the development of complexity. Because plants are a distributed hierarchy, plant cells, the intermediate entities of plant, have pluripotency, plasticity, responsiveness to the environment, and autotrophy* due to their smooth landscape, which reduces the pattern screening at the intermediate level and thus ensures the unbiased sampling of

* Autotrophy is a nutritional manifestation of the responsiveness to the environment, because the responsiveness favors utilizing inorganic molecules and energy from the environment. In contrast, autonomy, namely low responsiveness, is unfavorable for utilizing environmental materials and energy to synthesize organic compounds. Therefore, autonomic organisms take energy and raw organics from autotrophs. This requires autonomic phenotypes that break the restraint by environment, such as motility and phagotrophy.

configuration space. In contrast, in animals, because selection at the intermediate cellular level does not affect the germline information, animal cells have a rugged landscape, which results in terminal differentiation, intrinsic lineage of cells, autonomy, and heterotrophy[*].

The same conclusion can be drawn from an angle different from hierarchical analysis. In long-term evolution, the genetic information in plants is an averaged mixture of the information about the whole plant and its various parts. If we track a plant gamete backward through many generations, we will find that this gamete has experienced various types of cell fate. Therefore, plant cells have retained information for differentiating to other cell types. A plant cell is the temporal average of its historical fates. Any intrinsic barriers between various cell types, if they exist, have been smoothed by this temporal averaging.  Plant cells can easily convert to other cell types or even grow to a whole plant [254-255]. In this sense, every plant cell is a stem cell all the time. Due to this property, plant development can occur anytime, and does not require cell migration. Because of the flat landscape of plant cells, the proliferation and differentiation of plant cells are determined by extrinsic signals. Positional information instead of lineage (clonal history), is the primary determinant of cell fate in plants[254-255]. Moreover, the pluripotency and relatively smooth landscape results in high responsiveness to environment, and that makes plant cells less autonomous and thus less amenable to oncogenesis [264-265]. Pluripotent plant cells do not require migration to develop the organism. The absence of cell migration makes plant tumor cells motionless and thus less malignant. All of these can explain why most plant tumors are extrinsic and benign[254-255, 258, 263].

In contrast, due to the early-specified germline specification, animals are a centralized hierarchy: animal germ cells do not experience any somatic cell fate. The genetic information in germline cells represents the entire organism. Somatic cells only affect the evolution of genetic information as a dependent functional part of the whole organism. Because evolving as a whole, animals develop as a whole. The germline information about various somatic cell types is an inseparable whole. The intrinsic path, rather than the extrinsic signal, is the principle determinant of somatic proliferation and differentiation. Differentiations and commitment of animal cells are downward paths separated by barriers on the rugged landscape. Only cells at the branching point have the potential to be committed to different paths on the landscape. These properties can explain why animals undergo organogenesis only once, and why animal cells are more autonomous and less responsive to the environment. Consequently, animal cells tend to be more susceptible to oncogenesis and the resultant tumor is more malignant than plant tumors. The structure of animals is discontinuous: a type of cell cannot be differentiated from adjacent cells; adjacent cells belong to different lineages and perform distinct functions. The discontinuous structure is a necessary result of the complexity increase. Because of the predetermined lineage, cell migration is beneficial for embryogenesis to form discontinuous structures by cells from different embryonic locations. Ideally, discontinuous structures can form through programmed cell death (apoptosis) and/or shape change of cell mass without involving cell migration. For example, programmed cell death can disconnect one group of cell from the parental cells, which makes that group of cells discontinuous to adjacent cells. However, such a way requires the growth of the cell group into the target area, which affects the structures and function of the target area. Therefore, it can only form simple and limited discontinuous structures, but cannot generate delicate and/or extensive discontinuous structures. Both cell mass migration and individual cell migration are very common in animal development and function[266-268]. As an inevitable result, the ability to migrate makes animal cancer invasive and thus more malignant.

These properties of animals and plants relate to one another and form an indivisible network. Which property is the initiating factor in the emergence of this network? The answer is cellular motility. Flagellation, the motility in liquid, drives the emergence of multicellularity in protistan ancestors, while amoeboid crawling, the motility on solid surface, underlies the emergence of germline. Animals and plants are both multicellular forms of hierarchical life, but chose different ways to construct the hierarchy. Such bifurcation is not haphazard. The common choice of multicellularity is due to the flagellation constraint on the protistan ancestors of both animals and plants: simultaneous flagellation and mitosis are prohibited. Multicellularity with the labor division is the common way for ancestral animals and plants to solve the flagellation constraint. Other unicellular organisms choosing different ways to solve the flagellation constraint go to the dead end of complexity increase. On the other hand, ancestral animals and plants have different flagellation constraint: that of ancestral animals is due to the single microtubule-organizing center (MTOC), while that of ancestral plants is due to the cell wall. This seemingly minute difference leads to different strategies to construct hierarchical multicellularity: early-specified germline or no germline, which finally results in the bifurcation of animal and plant.

## The flagellation constraint drives the emergence of multicellularity

According to phylogenetic studies, unikonts* and bikonts are the protistan ancestor of animal and plant, respectively[104, 269]. At an early stage of flagellar evolution, unikonts have a single flagellum with one centriole, while bikonts have two flagella. According to phylogenetic studies, the unikont-bikont bifurcation is a very early, if not the earliest, diversification of known eukaryotes[260, 269-271]. It is not accidental that metazoan, namely multicellular animals, originated in the branch of unikont.

The origin and early evolution of flagella is closely related to mitosis, because both mitosis and flagellation need microtubule-mediated motility[272-276]. They both need the microtubule-organizing center (MTOC) for anchorage, positioning, and orientation. It is believed that at the early stage of the evolution of microtubule-based structures, there is only one MTOC; mitosis and flagellation compete for the MTOC[277-278]. Simultaneous mitosis and flagellation is prohibited, and that imposes severe disadvantage to all flagellates, because flagellation is very important for phototaxis[186, 279] and predation[274] while mitosis is required for reproduction. This type of constraint is named the flagellation constraint by MTOC. Several paths of evolution can overcome this constraint (Fig. 18). First, for ciliates, atypical mitosis or amitotic division are used and thus the MTOCs are not required in division[277, 280]. Second, the MTOC develops special structures that enable it to fulfil flagellation and mitosis simultaneously. For instance, the MTOCs in *Barbulanympha* are long filiform structures with one end as the anchorage of flagella and the other end as the spindle pole of mitosis[277, 281]. Third, many flagellates develop multiple MTOCs, and can have flagellation and mitosis simultaneously [277-278]. Bikonts take the third path. In contrast, the ancestors of animals take the fourth path: multicellularity with the labor division, which results in the emergence of germline and animal from unikonts.

---

* -kont: Greek: κοντ•ς *(kontos)* = "pole" i.e. flagellum.

Multicellularity with the labor division is one way to achieve simultaneous flagellation and mitosis[282-283]. A part of cells give up mitosis temporarily and maintain functional flagella, while the remaining cells give up flagella but keep the function of mitosis. Although multicellularity has many long-term advantages over unicellularity, these advantages cannot provide direct and immediate selective pressure for the development of multicellularity. In the evolution of multicellularity and labor division, especially germ-soma division, the flagellation constraint by MTOC is not only the initiating force but also the maintaining force at the early stage. Therefore, the flagellation constraint is incorporated into the regulation of cell division from the beginning of metazoan. When multiple MTOCs develop in animals at a late stage, additional MTOCs are available for spindle assembly; the MTOC no longer imposes constraints on cell division any more[284-285]. At this stage, due to their long-term advantages over unicellularity, multicellularity and the labor division are maintained and consolidated by long-term advantages through genetic mechanisms in advanced animals. The role of the flagellation constraint by MTOC in maintaining multicellularity and the labor division is gradually lost.

The flagellation constraint by MOTC has left imprints in modern animals. The cell division, differentiation, and movement in the initiation of multicellularity were driven by the flagellation constraint and developed to the blastrula formation and gastrulation in primitive ancestors of animal[286]. These early events of multicellularity have been fixed as the very conserved early ontogeny of embryonic development across animal kingdom[286]. The role of the flagellation constraint in animal evolution and development is so fundamental that some relics without any fitness value have been left in all modern animal: ciliary resorption is coordinated with cell cycle and the centrosome serves as a scaffold to anchor cell cycle regulatory proteins[287-288], although the centrosome is not required for spindle formation during mitosis in modern animal cells[284]. Neither animal cells nor their ancestral protists can divide while retaining flagella or any other derived structures, such as the axons and dendrites of neurons, the kinocilia of cells in vertebrate ear, and the tail of spermatids[276-277]. This phenomenon is a puzzle, because many other modern flagellated protists can divide[186, 276, 279]. The conventional theories of evolution cannot solve this puzzle. In retrospect, this phenomenon is the relic of a then cumbersome solution to the flagellation constraint by MTOC. This strategy later brings complexity and prosperity to its users. A different flagellation constraint occurs in the bikont protistan ancestors of plants (Fig. 18). Bikonts have developed two flagella and MTOCs. Therefore, the number of MTOC does not prohibit simultaneous flagellation and mitosis any more. However, there is a new flagellation constraint in the walled bikonts. Flagella are anchored to the cell through their basal bodies. At interphase, two basal bodies are connected and placed close to each other. During mitosis, they migrate and take positions near the spindle poles, behaving like centrioles. In naked flagellates, the basal bodies can migrate while remaining attached to their flagella. In walled flagellates, the rigid cell wall prevents any lateral movement of flagella. In most walled unicellular flagellates, the flagella are resorbed before mitosis to allow basal body migration and cell division[279]. The flagellation constraint creates a dilemma: a walled unicellular flagellate cannot fulfil both flagellation and mitosis simultaneously. In order to discriminate it from the flagellation constraint by MTOC, the constraint in walled bikonts is named as the flagellation constraint by wall.

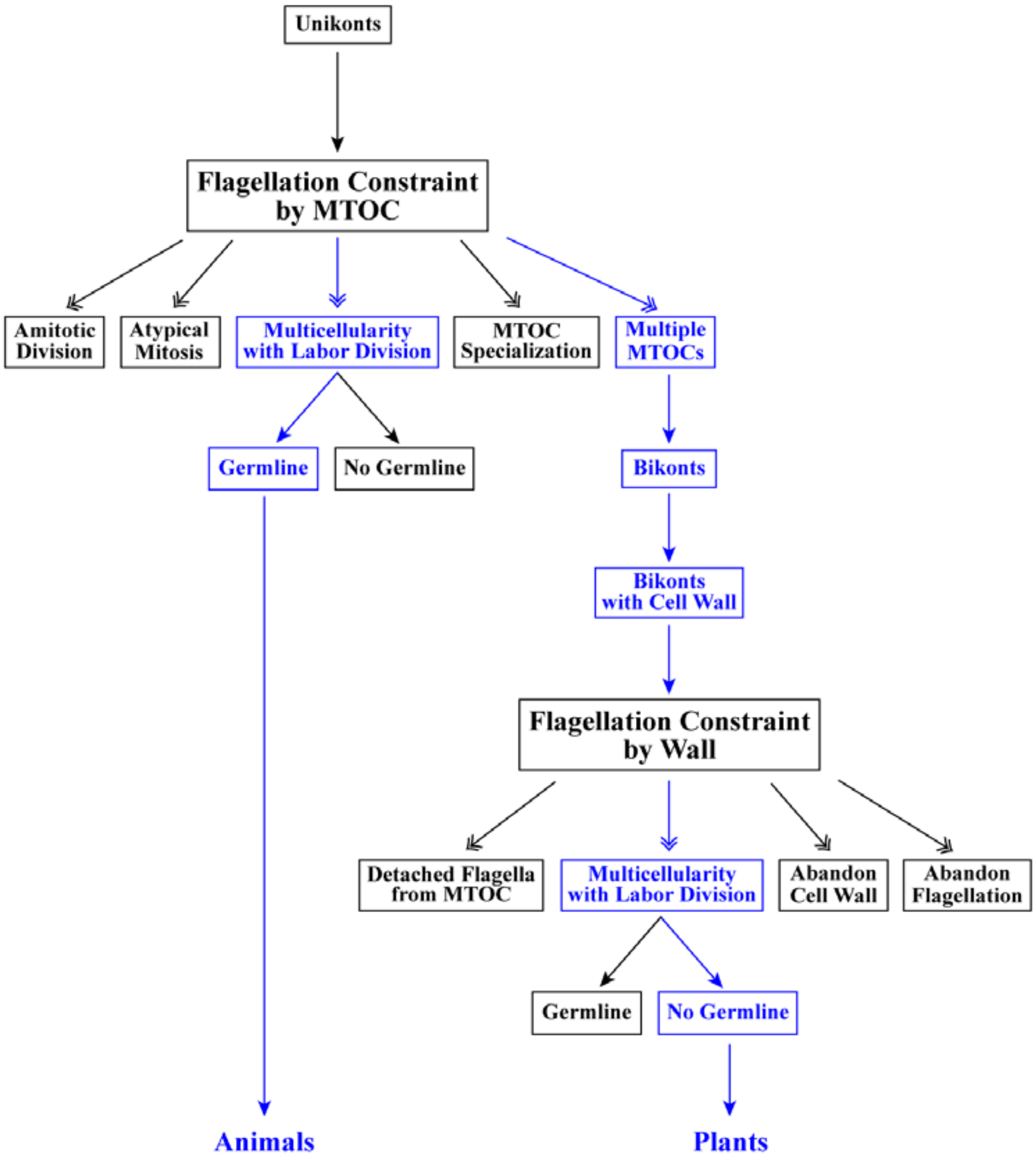

**Fig. 18. The bifurcation of animal and plant.** The flagellation constraint by MTOC (Microtubule-organizing center) is the inability to divide and flagellate simultaneously, because division and flagellation both require exclusive using of MTOC. The flagellation constraint by wall is because the cell wall prohibits the lateral movement of flagella and centrioles, which is required by mitosis. Various solutions to these constraints lead to the emergence of multicellularity, germ-soma differentiation, and the dichotomy of animal and plant.

The pathways in black color are all the dead end of complexity increase. The pathways in blue color represent the evolution to the complex form of multicellular life – animal and plant.

Cell wall is important for osmoregulation, and mitosis is required for reproduction, while flagellation is important for both phototaxis[186, 279] and predation[274]. Similarly, one solution is multicellularity and labor division. The flagellation constraint by wall promotes the assembly of cells to form a multicellular organism[186]. Some cells abandon mitosis and keep flagellation, while other cells abandon flagellation but keep mitosis. Other solutions include abandonment of flagella in the asexual phase, the abandonment of cell wall in the sexual stage, or detachment of flagella from basal bodies. However, only multicellularity with the labor division allows significant complexity increase (Fig. 18)[186, 188, 279].

## Germ-soma division: the selected strategy for amoeboid multicellularity

Due to the difference in flagellation constraints, two types of multicellularity develop: amoeboid multicellularity develops in unikonts due to the flagellation constraint by MTOC, and walled multicellularity in bikonts due to the flagellation constraint by cell wall. Different multicellularity has distinct mechanisms to increase its complexity and fitness. The effectiveness of these mechanisms is mainly determined by whether they establish and improve the specialized pattern generator from the general functional labor division.

To the walled multicellular bikonts, there are several possible paths for evolution (Fig. 18). The first is to acquire early germ-soma division with a simple structure: flagellar cells lose their fertility and become terminal cells. Mitotic non-flagellar germ cells are the cellular source for various structures, including the soma and the offspring. In the walled multicellular bikonts, the cell wall prohibits amoeboid cell migration. Early germ-soma division must suppress complexity increase, because the combination of immotility and germ-soma division forbids the formation of discontinuous cellular structures in multicellular organisms. This path is a dead end of complexity increase. This is the case of *Volvox carteri*, which has a continuous and simple structure with only two types of cells: fertile germ and sterile soma[187]. Because it has only two types of cells, the mitotic cell is the only fertile candidate for the germline. During embryogenesis, the non-flagellar germ cells undergo asymmetrical divisions to produce large and small cells. The large cells become the germ of the juvenile, and the small cells become the flagellar soma of the juvenile[187].

The second path is to abandon the cell wall. Even if this path is not lethal, it removes the flagellation constraint. At the early stage of multicellularity, the flagellation constraint is not only the main driving force but also the major maintaining force for multicellularity. Loss of this force makes the early multicellular organism retrogress to a unicellular state.

The third path is to keep fertility in both flagellar and mitotic cells and specify gametes at a very late stage. The universal fertility with smooth landscape develops to pluripotency. The resultant pluripotency allows for discontinuous structures despite the cell wall and leads to the emergence of complex plants. Land plants originate from aquatic green algae[289-291], which is the walled descendent of bikont. It is reasonable to propose that the enhanced evolvability of multicellularity with the general functional division leads to the evolution from bikonts to advanced land plants.

The situation of amoeboid multicellularity is subtly different: the unikonts with the flagellation constraint by MTOC can still keep their amoeboid cellular motility all the time, while walled bikonts never acquire the motility on solid surface. This explains why myosin II, the principal force generator for amoeboid crawling[292-295], arose only in unikonts after the divergence of eukaryotes into unikonts and bikonts[260]. This subtle difference at the early stage of eukaryote evolution results in the great difference between animals and plants. Pluripotency is not required to form discontinuous structures, because amoeboid migration of predetermined cells can form discontinuous structures. Once the mandate of pluripotency is removed, various cellular divisions of labor with rough landscape emerge, because the roughness of evolutionary landscape brings advantages in functional action of cells and organisms. The most important labor division is the germ-soma division, which is actually the division of internal evolution to pattern formation and functional action at the cellular level. As in *Volvox carteri*, flagellar cells become functional soma; mitotic non-flagellar cells become germline as the cellular source for soma and offspring. During animal evolution, soma becomes more and more complex; somatic function is expanded from flagellation to many other activities, but the relics of flagellation constraint are left. In this way, the survival of the silent germline only couples to the organism in natural selection. ***The importance of the flagellation constraint in the germ-soma division is that it provides an initiating mechanism to sequester a specific type of cell from function, differentiation, and the attendant selection.*** According to this theory, the genetic mechanism that sequesters the germline should be a derivative of the historical flagellation constraint by MTOC.

***All important aspects of either amoeboid multicellularity or walled multicellularity are consistent with their strategy in the germ-soma division.*** To unikonts, amoeboidy in the absence of cell wall facilitates phagotrophy, promotes multicellularity and the labor division with the aid of the flagellation constraint by MTOC, and allows discontinuous structure through cellular migration in development. All of these features lead to heterotrophy, development of cellular motility to muscle contraction, embryonic development, determined cell fate, cellular autonomy, and predisposition to cancer. In contrast, to bikonts with cell wall, phagotrophy and cellular motility on solid surface are prohibited. After the emergence of multicellularity driven by the flagellation constraint by wall, pluripotency and the late specification of gametes are the only choice for complexity increase, and that accounts for the autotrophy, postembryonic development, indeterminate cell growth, plasticity, and resistance to oncogenesis. ***None of the organisms deviating from these two paths acquire significant complexity***. For instance, with both specified germline and cell wall, *Volvox carteri* has only continuous and simple structure with

two types of cell, fertile germ and sterile soma[187]; as walled unikonts without germline, fungi are heterotrophic and thus fail to acquire advanced complexity.

Obligate hierarchies are either distributed or centralized hierarchies. In distributed hierarchies, intermediate levels require a smooth landscape for the host organism's unbiased sampling of the phenotype space. Pluripotency, plasticity, responsiveness, and autotrophy are the necessary results of a smooth landscape. In centralized hierarchies, intermediate levels tend to acquire a rough landscape in order to maximize their functional complexity and fitness. Cell lineage, autonomy, cellular diversification, and heterotrophy are the necessary results. Therefore, on the one hand, bifurcation of animals and plants are the necessary result of hierarchical biotic evolution; on the other hand, animals and plants are the only two strategies for hierarchical lives to acquire complexity.

The labor division of internal evolution into pattern formation (configuration sampling) and functional action is the only way to break the universal polarity of evolution. Heterodomain mapping and coupled selection are the component mechanisms of the labor division of internal evolution, while coarse graining and hierarchization are the extension of these component mechanisms. As a molecular extension of the conventional theories of evolution (Suppl. Table, *A Comparison of the Conventional View and the Present Theory*), this general theory of evolution unitarily and parsimoniously explains the essence of life.